\documentclass{cernyrep}
\usepackage{amsmath}
\usepackage{wrapfig}
\usepackage{caption}
\usepackage{mdframed}
\usepackage{enumitem}
\usepackage{afterpage}
\usepackage{lipsum}
\usepackage{array}
\usepackage{xcolor, soul}
\usepackage{colortbl}
\usepackage{subcaption} 
\usepackage{mathtools} 
\usepackage{pdfpages}
\usepackage{siunitx}

\usepackage[OT2,OT1]{fontenc}

\DeclareSymbolFont{cyrletters}{OT2}{wncyr}{m}{n}
\DeclareMathSymbol{\Sha}{\mathalpha}{cyrletters}{"58}

\newmdenv[backgroundcolor=lightgray]{importantbox}
\fancypagestyle{ARTTITLE}{%
\fancyhf{} 
\lhead{\small{Proceedings of the CERN--Accelerator--School course:\\ \it{Introduction to Accelerator Physics}, courses 2024 and beyond}}
\lfoot{Available online at \url{https://cas.web.cern.ch/previous-schools}}
\rfoot{\thepage\hspace*{3mm}}

}
\providecommand*{\Eq}[1][s]{\ifthenelse{\equal{#1}{b}}{Equation}{Eq.}}
\providecommand*{\Eqs}[1][s]{\ifthenelse{\equal{#1}{b}}{Equations}{Eqs.}}
\providecommand*{\Figure}[1][s]{\ifthenelse
                                      {\equal{#1}{b}}{Figure}{Fig.}}
\providecommand*{\Figures}[1][s]{\ifthenelse
                                      {\equal{#1}{b}}{Figures}{Figs.}}
\providecommand*{\Ref}[1][s]{\ifthenelse{\equal{#1}{b}}{Reference}{Ref.}}
\providecommand*{\Refs}[1][s]{\ifthenelse{\equal{#1}{b}}{References}{Refs.}}
\providecommand\Table{Table}
\providecommand\Tables{Tables}

\providecommand\Section{Section}
\providecommand\Sections{Sections}

\providecommand*\Eref[2][s]{\Eq[#1]~(\ref{#2})}

\providecommand*\Fref[2][s]{\Figure[#1]~\ref{#2}}

\providecommand*\Sref[1]{\Section~\ref{#1}}

\providecommand*\Tref[1]{\Table~\ref{#1}}

\begin{document}

\title{Time-domain and frequency-domain signals in accelerators and their analysis}
\author{H.~Schmickler}
\institute{CERN, Geneva, Switzerland}

\begin{abstract}
Depending on their application people may use time-domain or frequency-domain signals in order to measure or describe processes. First we will look at the definition of these terms, produce some mathematical background and then apply the tools to measurements made in accelerators. We will first look at signals produced by a single bunch passing once through a detector (transfer line, linac), then periodic single bunch passages (circular accelerator) and at the end multi-bunch passages in a circular accelerator.
The bunches themselves are considered rigid; oscillations within the bunch are treated in another CAS course (i.e. general advanced CAS course \cite{bib:TFD-advanced-2013})
\end{abstract}


\maketitle
\thispagestyle{ARTTITLE}

\section{Introduction and Recapitulation}

\subsection{Definition of time- and frequency-domain}

Objects move or evolve, electrical signals change along a continuous time-line.
The observation of processes along this time-line we call the time-domain observation of that process. The principal sensors of the human body for time- domain observation are the eyes, which are well know to be very much bandwidth limited. Processes faster than about 10 Hz humans cannot resolve with their eyes and for very slow processes our memory cannot retain long enough the visual information. So we construct a variety of instruments, which allow us to observe processes over years (photographs) down to very short time-scales. Modern pump-probe laser techniques allow the observation of biochemical processes in the fs-domain. Recent publications in this domain are \cite{bib:TFD-pumpprobe-2007}\cite{bib:TFD-pumpprobe-2019}.

\begin{figure}[ht]
    \centering
    \begin{minipage}{0.45\textwidth}
        \centering
        \includegraphics[width=0.8\textwidth]{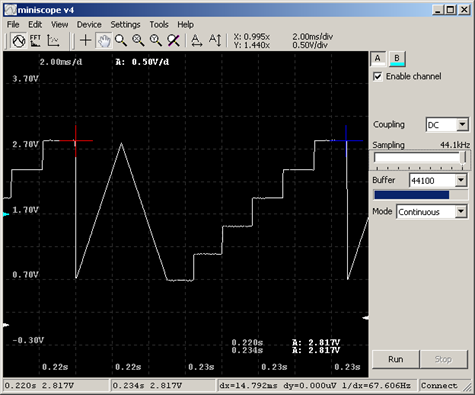}
        \caption{time trace of an electrical signal on an oscilloscope}
        \label{fig:TFD-oscilloscope}
    \end{minipage}\hfill
    \begin{minipage}{0.45\textwidth}
        \centering
        \includegraphics[width=0.8\textwidth]{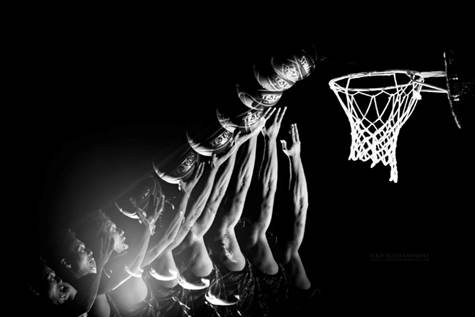}
        \caption{movement of a basketball player photographed at regular intervals}
        \label{fig:TFD-basketball}
    \end{minipage}
\end{figure}

\Fref[b]{fig:TFD-oscilloscope} and \Fref{fig:TFD-basketball} are put as illustrations of instruments to extend our bandwidth of time-domain observations by eye.

Things get a bit more complicated when we extend the "instruments" of our human body to the ears. \Fref[b]{fig:TFD-musictimetrace} shows the time trace of a piece of music. The data represented is the amplitude of the musical signal sampled at regular intervals at 44.1 kHz sampling frequency. Simply by looking at this data, it is impossible to determine what song it is unless you play it and listen (as we do in the course). 
\vspace{1cm}
\begin{figure}[ht]
  \centering
  \includegraphics[width=0.9\textwidth]{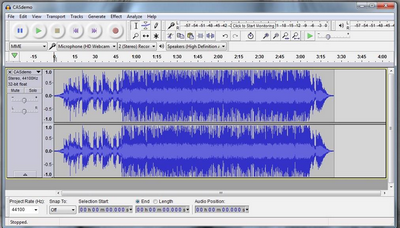}
  \caption{"Soldier of Fortune" of "Deep Purple" in time-domain representation}
  \label{fig:TFD-musictimetrace}
\end{figure} 
\vspace{1cm}

One actually needs the ears and the connected brain (hopefully) in order to recognize the music. The brain will use the time domain information for the rhythm, but the frequency decomposition in order to disentangle the different instruments and voices.
\newpage

\subsection{Basics of Fourier Transforms (FT)}

The Fourier-Series and -Transform go back to the mathematical work of J.~Fourier at the end of the 18th century. This is all well documented and can be found in many textbooks \cite{bib:TFD-fouriertextbook} and on-line for example in \cite{bib:TFD-fourier}. 
Here we summarize the qualities of Fourier transforms, which are relevant for the rest of the lecture.\\

\subsubsection{Fourier-Transform formalism}
\label{sec:FT}

The FT is a mathematical operation, which allows us to transform time-domain data of an observable into frequency-domain and vice versa (\textbf{inverse Fourier transform}). Depending on the use case we will have continuous time-domain functions, potentially even periodic time domain functions, or like in most cases of signal measurements, a limited number of measured samples $x[n], n=0...N$, which have been acquired at regular time intervals $\Delta t$. \Tref{tab:TFD-FT} shows these various use-cases for Fourier transforms and the corresponding mathematical operation, which is needed to compute the transform. The FT of $x[n]$ is depicted as $\mathcal{F}\{ x\} [k]$. Often we find the term \textbf{FFT = Fast Fourier transform}. FFT  simply depicts a DFT with the number of samples $N$ being a power of 2 ($N= 2^m$). This condition allows a faster computation of the Fourier transform, but apart from the computing speed a DFT or a FFT are the same operations.
\begin{table}[h]

\centering
\begin{tabular}{|l|l|l|} 
\hline
\multicolumn{3}{|c|}{Time Duration}                                                                                                                                     \\ 
\hline
\multicolumn{1}{|c|}{{\cellcolor[rgb]{0.753,0.753,0.753}}Finite} & \multicolumn{1}{|c|}{{\cellcolor[rgb]{0.753,0.753,0.753}}Infinite}                 &                  \\ 
\hline
Discrete-FT (DFT)                       & Discrete Time FT (DTFT)                                                           & discrete         \\
$\mathcal{F}\{ x\} [k]\ = \  \sum_{n=0}^{N-1} x[n]\cdot e^{-j\frac{2\pi}{N}nk}$                    & $\mathcal{F}\{ x\} (\omega)\ = \  \sum_{n=0}^{N-1} x[n]\cdot e^{-j\omega n}$                                                          & time index n     \\
\multicolumn{1}{|c|}{{\cellcolor[rgb]{0.753,0.753,0.753}}$k=0,1,...N-1$ }              & \multicolumn{1}{|c|}{{\cellcolor[rgb]{0.753,0.753,0.753}}$\omega \in (-\pi,+\pi)$   } &                  \\ 
\hline
Fourier Series (FS)                              & Fourier Transform (FT)                                                            & continuous        \\
$\mathcal{F}\{ x\} [k]\ = \  \int_{0}^{P} x(t)\cdot e^{-j\frac{2\pi k}{N}t}dt$                                                 & $\mathcal{F}\{ x\} (\omega)\ = \  \int_{-\infty }^{+\infty } x(t)\cdot e^{-j\omega t}dt$                                                          & time variable t  \\
\multicolumn{1}{|c|}{{\cellcolor[rgb]{0.753,0.753,0.753}}$k=-\infty ,...+\infty $ } & \multicolumn{1}{|c|}{{\cellcolor[rgb]{0.753,0.753,0.753}}$\omega \in (-\infty , +\infty )$}                       &                  \\ 
\hline
\multicolumn{1}{|c|}{{\cellcolor[rgb]{0.753,0.753,0.753}}discrete frequency k  }       &
\multicolumn{1}{|c|}{{\cellcolor[rgb]{0.753,0.753,0.753}}continuous frequency $\omega$} &                  \\
\hline
\end{tabular}
\caption{Different Use-Cases for Fourier transforms}
\label{tab:TFD-FT}
\end{table}

\begin{figure}[h]
  \centering
  \includegraphics[width=0.7\textwidth]{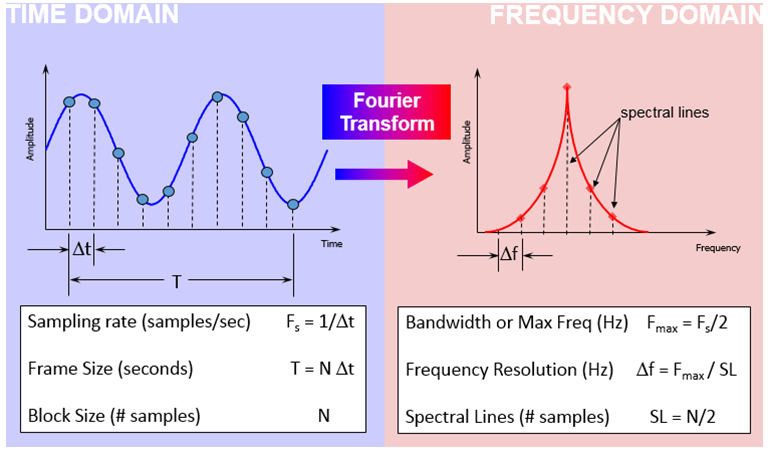}
  \caption{Illustration of a DFT}
  \label{fig:TFD-DFT}
\end{figure} 

\subsubsection{Usable Spectral Information (Nyquist-Shannon Theorem)}
\label{sec:aliasing}
The result of any DFT of $N$ observables $x[n]$ is a set of $N$ complex numbers. In most cases  we use as power spectrum the sum of the squares of the real- and imaginary part or as \textbf{amplitude spectrum the root of the power spectrum}. \Fref[b]{fig:TFD-DFT} illustrates the situation and explains the most important parameters. One should note that  only $N/2$ spectral lines are available in the frequency range $(0 < f < f_{s}/2)$. The result of any DFT delivers a symmetrical amplitude spectrum for negative frequencies, which is of no importance for the description of physical phenomena. The frequency resolution measured by the distance of neighbouring bins is $\Delta f = 2f_s/N$.

\begin{importantbox}
An important  limitation of spectral information when using sampled information is expressed by the Nyquist-Shannon-theorem \cite{bib:TFD-nyquist}.
Only signals, which have frequencies below half the sampling frequency $f_s$, give unambiguous results in frequency domain.
Signals of higher frequency $f$ are wrongly interpreted at a mirror-frequency ($f_s-f$). This phenomenon is called "\textbf{aliasing}".
\end{importantbox}

\Fref[b]{fig:TFD-aliasing} illustrates the effect of aliasing in time-domain. Wave(a) represents a continuous wave at 7 kHz. Wave(b) shows the same wave sampled at 10 kHz (blue points), which is a sampling well below the Nyquist-Shannon limit. (c) shows the corresponding 3 kHz wave (10 kHz - 7 kHz) as which the Fourier transform of the blue sampled points will be interpreted. In order to avoid such ambiguities designers have to take care that before sampling through filtering techniques any frequency components above the Nyquist-Shannon limit are suppressed.

\begin{figure}[h]
    \centering
    \begin{minipage}{0.45\textwidth}
   \includegraphics[width=0.9\textwidth]{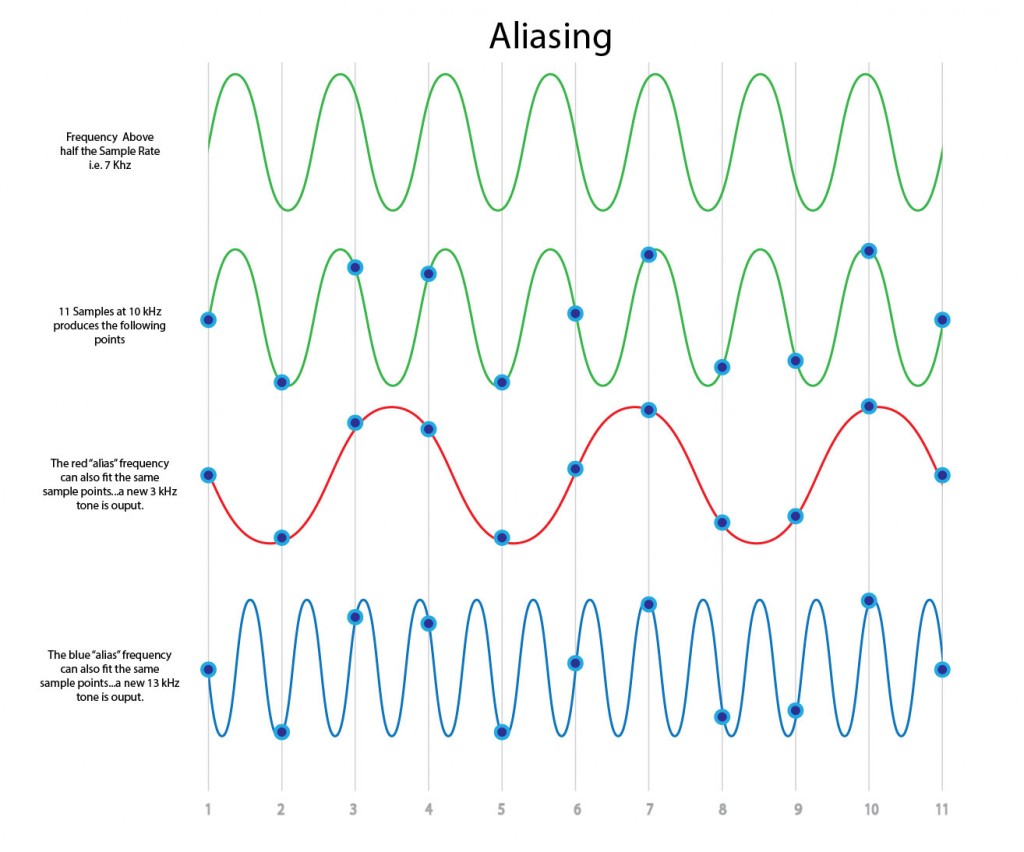}
  \caption{(a) continuous 7 kHz wave, (b) sampled at 10 kHz (blue points), (c) interpretation of the same samples point as a (10kHz -7kHz) = 3 kHz wave}
  \label{fig:TFD-aliasing}
    \end{minipage}\hfill
    \begin{minipage}{0.45\textwidth}
        \centering
  \includegraphics[width=0.9\textwidth]{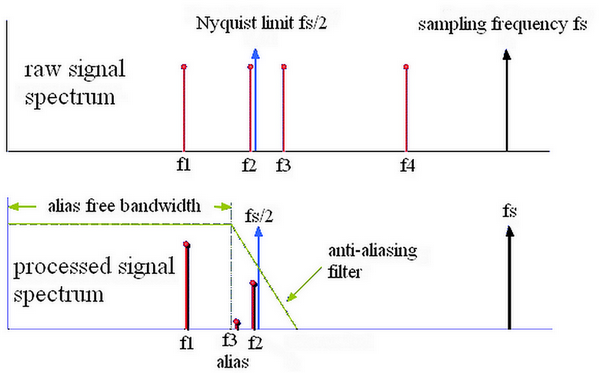}
  \caption{Illustration of aliasing effects in frequency domain }
  \label{fig:TFD-aliasing2}
    \end{minipage}
\end{figure}

\Fref[b]{fig:TFD-aliasing2} explains the effect of aliasing in frequency domain. Four different signals at frequencies f1...f4 are present in a raw signal which is measured with a sampling frequency fs. Frequencies below half the sampling frequency appear at the correct place in the spectrum of the processed data (f1 and f2). With the help of anti-aliasing filters engineers have to make sure that potential frequency components above half the sampling frequency are effectively suppressed. The signal at f3 is attenuated, but since it is still in the passband of the anti-aliasing filter, a small signal appears at (fs-f3) in the processed spectrum. The signal at f4 is effectively suppressed.
The transfer function of the anti-aliasing filter is indicated in light green.
\newpage

\subsubsection{Convolution theorem and amplitude modulation (AM)}
\label{sec:TFD-convolution}

The Fourier transform of a time-domain function $h(t)$, which itself is the convolution of two functions $f$ and $g,\ h(t)=f(t)\circledast g(t)\ =\ \int{f(t)g(z-t)dz}$ is the point wise product of the Fourier transforms of $f(t)$ and $g(t)$. Or:\\

\noindent let $\mathcal{F}\{ f\} $ be the Fourier transform of $f(t)$, \\
and $\mathcal{F}\{ g\} $ be the Fourier transform of $g(t)$
then\\
\begin{equation}
\mathcal{F}\{ h(t)\}\  =\ 
\mathcal{F} \{f(t)\circledast g(t)\}\ =\ \mathcal{F}\{ f(t)\}  \cdot \mathcal{F}\{ g(t)\} 
\label{eq:TFD-convolution}
\end{equation}
More details on the convolution theorem under \cite{bib:TFD-convolution}.

\begin{figure}[h]
    \centering
    \begin{minipage}{0.5\textwidth}
        \centering
        \includegraphics[width=0.8\textwidth]{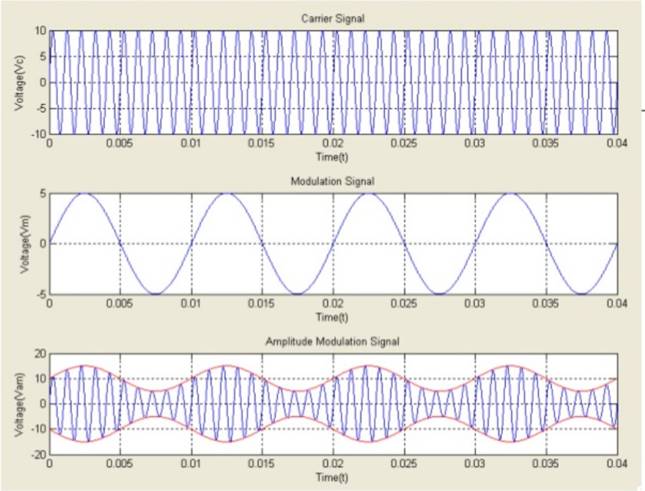}
        \caption{\label{fig:TFD-specmod}Time domain representation of a 500 kHz sin-wave carrier with amplitude-modulation by a sin-wave of 1 kHz }
    \end{minipage}\hfill
    \begin{minipage}{0.45\textwidth}
        \centering

        \includegraphics[width=0.8\textwidth]{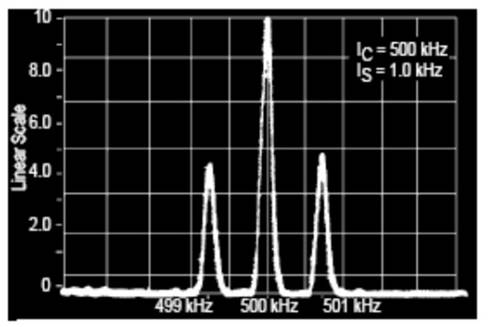}
        \caption{\label{fig:TFD-P26}Measured frequency domain spectrum of \Fref{fig:TFD-specmod}}
    \end{minipage}
\end{figure}

A simple application of the above is the Fourier transform of a sine-wave (so called carrier), which is modulated (:= multiplied with) in amplitude by another sine-wave of lower frequency. This technique is known from old radio communication as AM (amplitude modulation) technique.
The resulting Fourier transform is a single spectral line at the frequency of the carrier with two side-bands left and right of the carrier-frequency. The \textbf{amplitude} of the side-bands is given by the modulation depths of the carrier (see \Fref{fig:TFD-specmod}). 

\noindent So using the above convolution theorem, the Fourier transform of the carrier($f(t)$) is the middle spectral line in \Fref{fig:TFD-specmod}, the Fourier transform of the modulating sine-wave $g(t)$ is again a single line, but the convolution in frequency domain places the modulation lines instead around the "zero" frequency around the carrier frequency. So in this case also the negative frequency of the Fourier transform gets a physical meaning, namely the left (lower) sideband of the carrier. The positive frequency yields the right (upper) sideband of the carrier.

\noindent If one replaces the low frequency sine-wave by a music-signal one obtains the music spectrum left and right around the carrier. The carrier with its sine wave can be propagated as a radio wave and in this way former music or speech transmission was achieved.

For this simple case of an AM modulation of a sinusoidal carrier we do not need a the convolution theorem in order to understand the frequency domain result:
We write
\begin{equation}
V_{AM}\ =\ V_c\cdot (1+m \sin{2\pi f_m t})\cdot\sin{2\pi f_c t}
\end{equation}
with ($0 \le m \le 1$) as modulation index, $f_m$ as modulation frequency and ($f_c / V_c$) as carrier frequency/voltage.\\[5mm]
Using the trigonometric identity
\begin{equation}
(\sin a)(\sin b) = 1/2\left[ \cos{(a-b)} - \cos{(a+b)} \right]
\end{equation}
we get
\begin{equation}
V_{AM}\ =\ V_c\sin{2\pi f_c t} +
\frac{V_m}{2}\cos{2\pi (f_c -f_m)t} -
\frac{V_m}{2}\cos{2\pi (f_c +f_m)t} \quad \rm{with:}\ V_m =\frac{mV_c}{2}
\end{equation}
The first term clearly describes the carrier signal, the second term the lower side-band with the amplitude depending on the modulation index and the third term the equivalent upper side-band.

\subsubsection{Fourier transform of various signals}
\label{sec:TFD-varsig}
\begin{figure}[ht]
\begin{center}
  \includegraphics[width=0.6\textwidth]{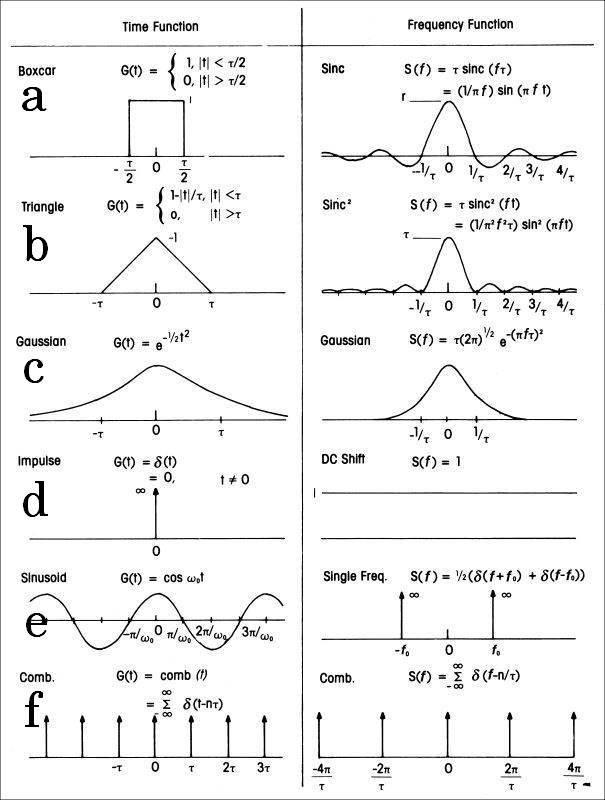}
  \caption{Fourier transform of various time domain functions a-f. Figure taken from \cite{bib:FTs}}
       \label{fig:FTs}
\end{center}
\end{figure}

\noindent \Fref{fig:FTs} shows several examples of time-domain functions and the corresponding Fourier transforms. Most of the selected functions have an important application in describing particle beams:
\begin{itemize}
\item{
function "e", a continuous sinewave describes well a linear transverse or longitudinal beam oscillations. In frequency domain, the resulting obvious single discrete frequency line we will identify later on as transverse or longitudinal fractional tune.
In all physical interpretations we neglect negative frequencies.
}
\item{
The longitudinal shape of particle bunches can at higher energies be well described by a Gaussian distribution (function "c"). The Fourier transform is also a Gaussian distribution (more details in \Sref{sec:bulen}).
Important to note that the width of the curve in frequency domain is the inverse of the width in time domain. So short bunches contain a frequency spectrum that extends to higher frequencies and vice versa.
}
\item{
Function "d" represents the limit of function "c" for an infinitesimal short bunch.
The transform indicates that the frequency content of such a bunch will extend to infinite high frequencies.
}
\item{
Function "a" or "b" we will use as so called "window-functions" to describe the finite observation length of any real measurement (see \Sref{sec:windowing}).
}
\item{
In most accelerators the particle beam is concentrated in several bunches, which means a dense concentration of the particles in the longitudinal plane. This is necessary for the acceleration of the beam by an alternating RF-field. The time distance between bunches is a multiple of the RF-period and often in mathematical terms the beam is represented by a series of dirac pulses or comb function "f" (neglecting the finite bunch length).
A series of Dirac pulses with distance\ $\tau $\ is often referred to as \textbf{Shah-function}, possibly because its graph resembles the shape of the Cyrillic letter sha ($\Sha$ ).
It turns out that the Fourier-Transform of the Shah-function is also a Shah-function
(see for example \cite{bib:TFD-shah}).
\begin{equation}
\Sha(t)=\sum_{n=-\infty}^{\infty} \delta (t-n\cdot\tau)
\end{equation}
\begin{equation}
\mathcal{F}\{ \Sha(t)\} =\mathcal{F}\{ \sum_{n=-\infty}^{\infty} \delta (t-n\cdot\tau)\}\ =\ \frac{1}{\tau}\sum_{n=-\infty}^{\infty}\delta (f-\frac{n}{\tau} )
\label{eq:TFD-shah}
\end{equation}
Rewriting \Eref{eq:TFD-shah} with the definition of a fundamental frequency $\omega_0$  yields in frequency domain for k=1 a line at the fundamental frequency followed by higher harmonics for k $\geq $2. Much more details in \Sref{sec:sbmp}.
\begin{equation}
\Sha(t)=\sum_{n=-\infty}^{\infty} \delta (t-n\cdot\tau)\quad\Longrightarrow\quad  
\mathcal{F}\{ x(t)\} =\sum_{k=-\infty}^{\infty} \delta(\omega - k\omega_{0})
\qquad \omega_{0}=\frac{2\pi}{\tau}
\label{eq:TFD-shah2}
\end{equation}
}
\end{itemize}

\subsection{Accelerator bunch filling patterns}
\label{sec:bunchpatterns}

In most particle accelerators the particle beams are generated in bunches and kept in this shape by the RF-system throughout the operational cycle. In most cases it is sufficient to assume a Gaussian-like shape in time-domain for the longitudinal particle distribution. In that case we characterize the longitudinal distribution by the so called 
\textbf{bunch length $\sigma_l$}\ (:= one sigma width of the Gaussian distribution).

\begin{figure}[ht]
  \begin{center}
    \includegraphics[width=0.5\textwidth]{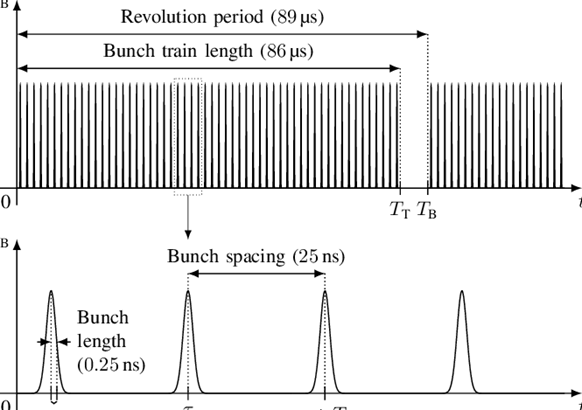}
 	\end{center}
  	\caption{Illustration of the bunch distribution along the perimeter of the LHC}
    \label{fig:TFD-bunchpattern}
\end{figure}

There are only few cases where all particles are accumulated in one bunch, usually the particles are distributed in several bunches, which are placed and kept at their longitudinal position by the RF system. The distribution of bunches on the perimeter of a circular accelerator is called the filling scheme of the accelerator. In a linac, the bunches are usually arranged in bunch trains with the same distance between individual bunches. A few definitions, valid for circular accelerators,  are useful for the following sections.
\begin{itemize}
\item{ \textbf{the revolution frequency $f_{rev}$} or its inverse the \textbf{revolution time} $T_{rev} = 1/f_{rev}$ is the time a bunch needs to travel once around the perimeter of the accelerator. Since this frequency is linked to the geometry of the ring (and the velocity of the particles), it is independent of the filling scheme and hence often referred to as $f_0$ or fundamental frequency of the accelerator.}
\item{\textbf{ The RF-frequency} $f_{RF}$. This frequency is always a multiple of the revolution frequency. The proportionality factor $h$ is called the \textbf{harmonic number} of the accelerator $f_{RF} = h\cdot\ f_{rev}$. The harmonic number of the accelerator represents the maximum number of bunches, which could be filled into the accelerator (In that case all RF buckets would be filled with particle bunches). }
\item{\textbf{the bunch repetition frequency} $f_{rev} = f_{RF}/n$. In many cases, not all consecutive RF buckets are filled with bunches; then $n$ depicts the number of RF cycles between adjacent bunches. $M=h/n$ is the maximum number of bunches that can be filled into the accelerator with even bunch spacing (no bunch gaps).}
\end{itemize}
\Fref[b]{fig:TFD-bunchpattern} illustrates these definitions for the LHC. A revolution time of 89 $\mu sec$ corresponding to a circumference of about 27 km, a RF frequency of 400 MHz and a bunch repetition frequency of 40 MHz. Hence $n$ in the above definitions has a value of 10: A bucket distance of 2.5 nsec and a bunch distance of 25 nsec. The harmonic number of the LHC $h$ is 35640, which makes $M=3564$.
\pagebreak

\section{Signals of stable bunched beams}

\subsection{Signal description}
\label{sec:bulen}

Any beam pickup or detector located in an accelerator monitors the passing charged particle beam,
thus receives a \emph{beam signal} as a function of time. 
As the charged particles are accelerated by gaining energy from time varying fields, they are not randomly distributed along the
circumference of a storage ring or along a linear accelerator, but the periodic RF accelerating field defines \emph{buckets} 
in the longitudinal phase space that result in the formation of \emph{beam bunches}.
Regardless of the operational principle and function of the beam pickup, e.g.\ directly utilizing the EM-field of the primary beam, using it's synchrotron light or other types of radiation, or detecting secondaries from a target made out of a specific material,
in all cases the signal of the primary beam current $i_{beam}(t)$  is transformed into a voltage signal $v_{out}(t)$, 
which is digitized and further processed.
Following that concept, we may characterize any beam pickup by it's \emph{transfer impedance}
\begin{equation}
Z_{PU}(\omega)=\frac{V_{out}(\omega)}{I_{beam}(\omega)}
\label{eq:PUtrans}
\end{equation} 

\begin{figure}[h]
  \centering
  \includegraphics[width=0.8\linewidth]{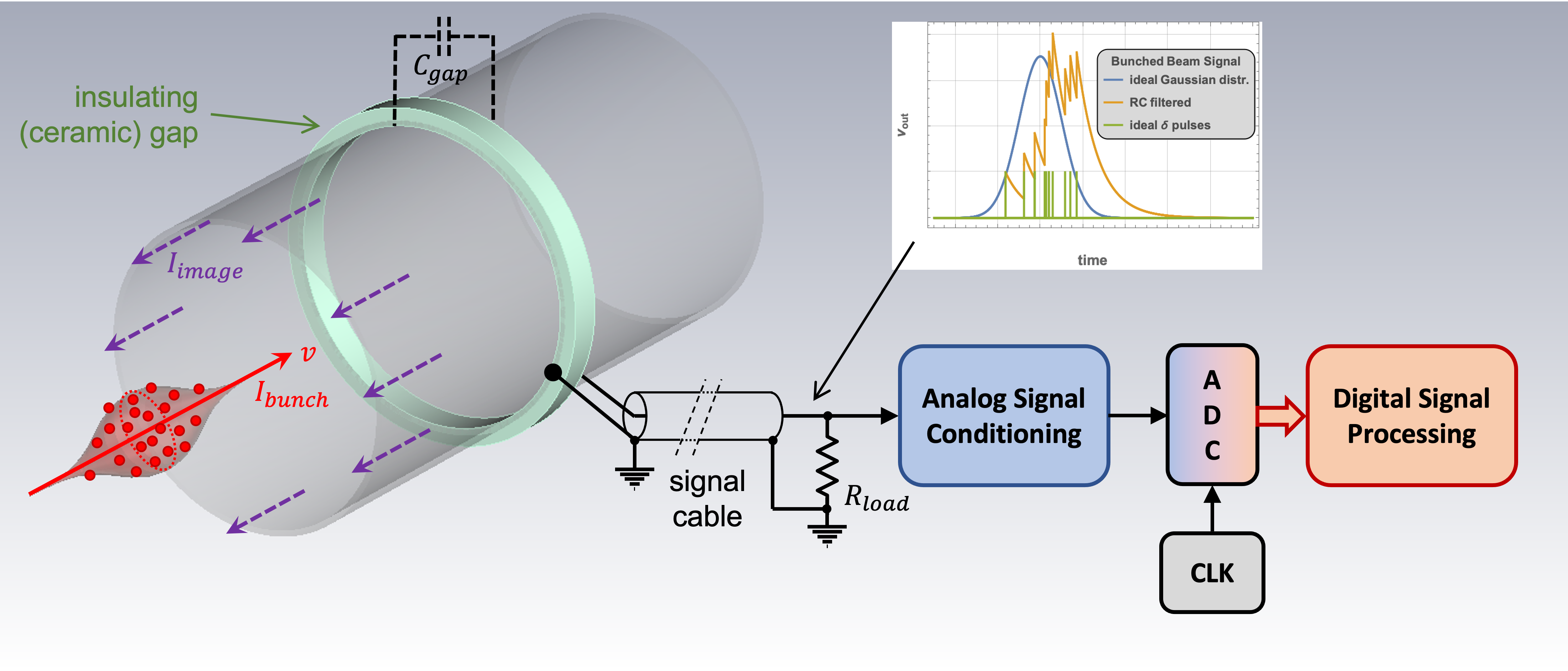}
  \caption{Bunch beam signal form a wall current monitor (WCM).}
  \label{fig:WCMsig}
\end{figure}
Fig.~\ref{fig:WCMsig} illustrates the principle for the example of a \emph{wall current monitor} (WCM), with a bunched beam
traveling at a velocity $v$ passing the ceramic gap of this simple beam pickup.
Lets first assume the beam bunch is just a single point charge $q=ze$, with $z$ being the charges state in case of ions, 
and $e=\SI{1.602e-19}{C}$ being the elementary charge, 
and is traveling at relativistic velocity $\beta=v/c=1$ in a 
perfectly conducting vacuum chamber. 
It has the beam current:
\begin{equation}
i_{beam}(t)=\frac{\Delta Q}{\Delta t}=q\,\delta(t)=-i_{\text{image}}(t)
\label{eq:iSingle}
\end{equation}
which is cancelled by the image (or \emph{wall}) current $i_{\text{image}}=i_w(t)$, 
originated  by the image (wall) charges $q_w=-q$ distributed around
the azimuth of the beam pipe wall.
The \textit{Dirac} delta function in Eq.~(\ref{eq:iSingle}) is defined as
\begin{equation}
\delta (x) =
 \begin{cases}
 +\infty , & x=0 \\
 0 & x\neq 0
 \end{cases},\qquad \int_{-\infty}^{+\infty} \!\!\delta(x) \, dx = 1
\label{eq:DiracDelta}
\end{equation}

For many point charges $q_i$, all 
traveling at a relativistic velocity $v\approx c$ and distributed at locations $\vec{r_i}=(x_i,y_i,z_i)$ only the
longitudinal $z$-coordinate $z_i=\tau_i c$ in the direction of motion has to be considered, and
 the beam current becomes a train of $\delta$-signals.
\begin{equation}
i_{beam}(t)=\frac{\Delta Q}{\Delta t}=\sum_i q_i\,\delta_i(t-\tau_i)
\label{eq:iMulti}
\end{equation}
This also is the case for a bunched beam, however, now the distribution $\tau_i=z_i/c=s_i/c$ \footnote{%
Please note the longitudinal $z$-coordinate is often denoted as $s$-coordinate.}
of the $\delta$-signals follows a 
specific distribution function given by the accelerating RF fields, in many practical cases 
it is the Normal (\textit{Gaussian}) distribution.

\begin{figure}[t]
\begin{subfigure}{0.69\textwidth}
  \centering
  \includegraphics[width=0.99\linewidth]{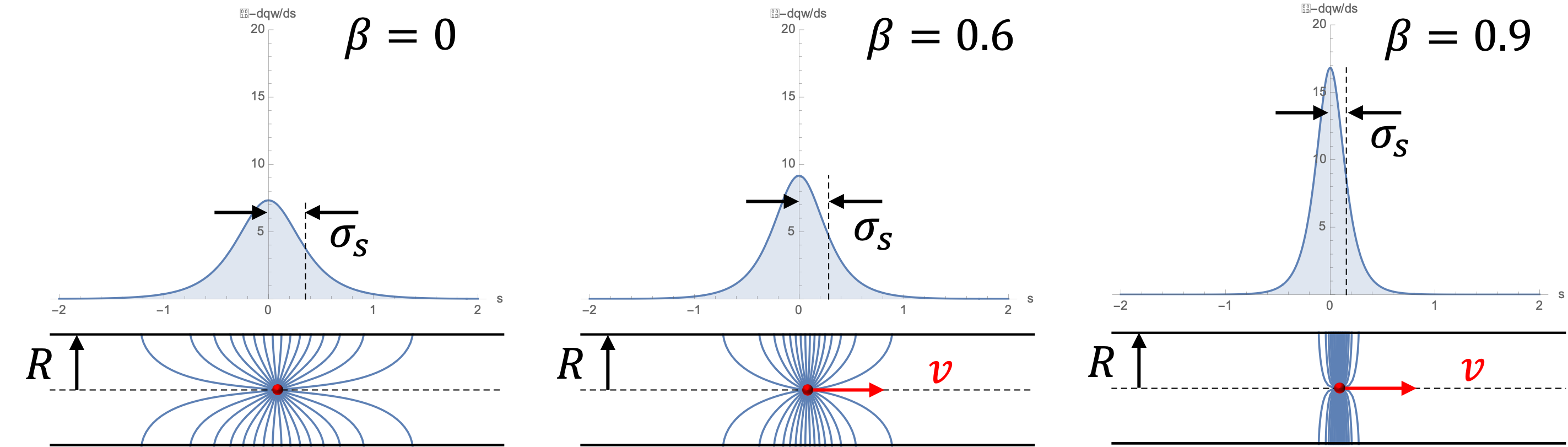}
\end{subfigure}
\hfill
\begin{subfigure}{0.29\textwidth}
  \centering
  \includegraphics[width=0.99\linewidth]{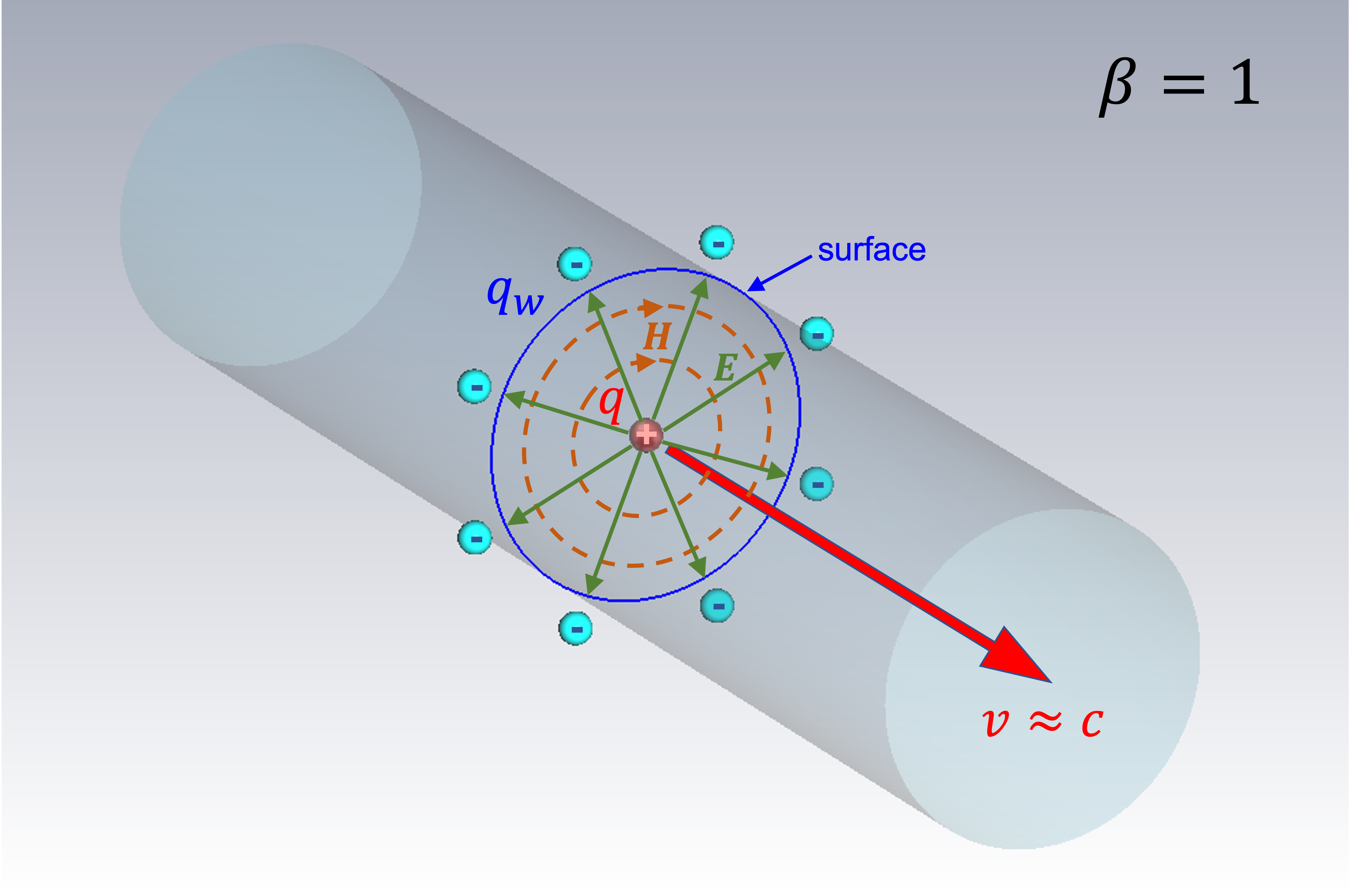}
\end{subfigure}
  \caption{EM-field of a point charge $q$, traveling in the center of a beam pipe at different velocities $\beta=v/c$, 
  and the corresponding distribution of wall (image) charges $q_w$.}
  \label{fig:qfield}
\end{figure}
Figure \ref{fig:WCMsig} shows the voltage output signal of an ``ideal'' wall-current monitor (WCM), indicated by the green
$\delta$-pulses for $N=10$ \textit{Gaussian} distributed point charges.
In practice, beside other ``limitations'' in time resolution, the gap of the WCM has some unavoidable 
capacitance $C_{\text{gap}}$, thus the $\delta$-pulses experience
some integration effects, the result is presented as orange trace.

Finally, the distribution of the electromagnetic field of the moving charges itself present a limitation for their monitoring as
the EM-field has an opening angle that is proportional to  $1/\gamma$ (\textit{Lorentz} factor: $\gamma=1/\sqrt(1-\beta^2$
and normalized, relative velocity: $\beta=v/c$).
For an electromagnetic beam pickup, like the WCM sketched in Fig.~\ref{fig:WCMsig}, this results in a widening
of the detected image charge distribution $-dq/ds$ at the surface of the beam pipe wall of radius $R$:
\begin{equation}
\sigma_s=\frac{R}{\sqrt{2}\gamma}
\label{eq:sigmaqw}
\end{equation}
as illustrated in Fig.~\ref{fig:qfield}.

\begin{figure}[t]
\begin{subfigure}{0.45\textwidth}
  \centering
  \includegraphics[width=0.9\linewidth]{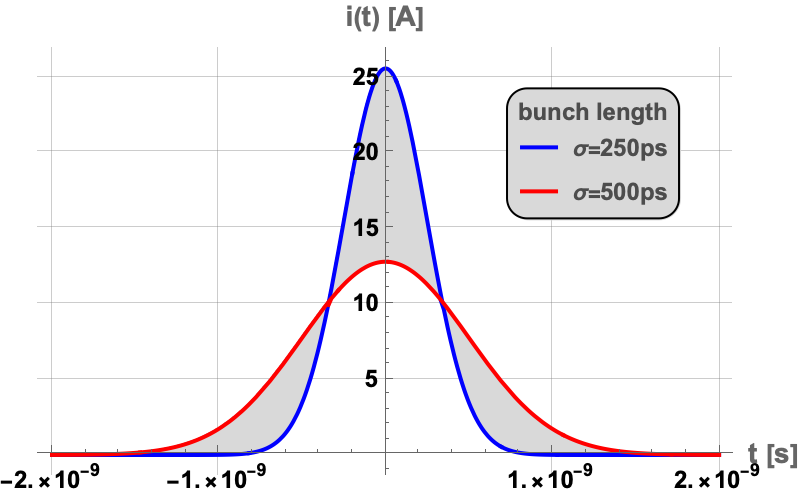}
  \caption{\textit{Gaussian} bunch current.}
  \label{fig:gaussTD}
\end{subfigure}
\hfill
\begin{subfigure}{0.45\textwidth}
  \centering
  \includegraphics[width=0.95\linewidth]{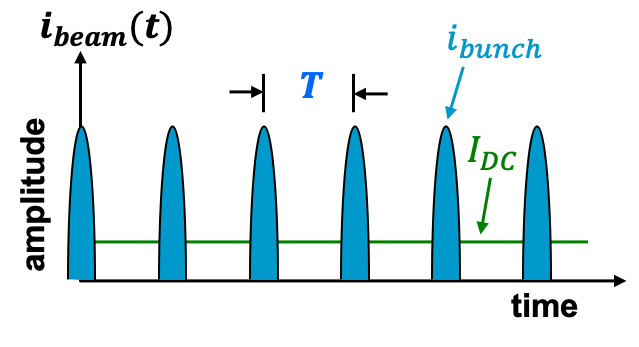}
  \caption{Beam current made out of bunches of same intensity and shape.}
  \label{fig:bTrain}
\end{subfigure}
\caption{Bunch and beam current signals in the time-domain.}
\label{fig:BeamBunchCurrent}
\end{figure}
Both factors, the limited time resolution of the beam pickup and the EM-field configuration of the moving charged
particles to be detected, together with the fact that typically the number of particles $N$ in a beam bunch is large result
in a continuous bunch signal, which in many cases can be approximated by the \textit{Gaussian} normal distribution
(see also Fig.~\ref{fig:gaussTD}) 
\begin{equation}
i_{\mathrm{bunch}}(t)=\frac{zeN}{\sqrt{2\pi}\sigma_t}e^{-\frac{t^2}{2\sigma_t^2}}
\label{eq:gaussTD}
\end{equation}
with $\sigma=\sigma_t v$ being the bunch length.

For many beam signal processing tasks it makes sense to understand the 
bunched beam signal contents in the frequency-domain by applying the \textit{Fourier} transformation; see also \Sref{sec:FT}.
\begin{equation}
F(f)=\int_{-\infty}^{+\infty} \!\! f(t)e^{-j2\pi ft}
\label{eq:FT}
\end{equation}
While the \textit{Gaussian} distribution function Eq.~(\ref{eq:gaussTD}) is most popular for lepton machines, in some
hadron accelerators the longitudinal particle distribution can better be approximated by other functions,
such as the $\cos^2$-function or the generalized \textit{Tsallis} q-\textit{Gaussian} distribution.
\small
\begin{align}
&\text{time-domain}    &\qquad   \text{freq}&\text{uency-domain} \nonumber \\ 
i_{\textit{Gauss}}(t)&=\frac{zeN}{\sqrt{2\pi}\sigma_t}e^{-\frac{t^2}{2\sigma_t^2}} &
I_{\textit{Gauss}}(f )&=zeNe^{-2(\pi f\sigma_t)^2} \label{eq:GaussTDFD} \\[1em]
i_{\cos^2}(t)&=
\smash{\left\{\begin{array}{c@{}c@{}}
\frac{zeN}{t_b} \left ( 1+\cos\frac{2\pi t}{t_b} \right ), \; -t_b/2<t<t_b/2 \\ [\jot] 
0,  \qquad\qquad\text{elsewhere} 
\end{array}\right.}
& 
I_{\cos^2}(f)&=\frac{zeN\sin\pi f t_b}{\pi f t_b \left [1-(f t_b)^2\right ]} \label{eq:cos2TDFD} \\[1em]
i_{\text{q-}\textit{Gauss}}(t)&=
\frac{zeN\sqrt{1-q}\left (1+\frac{(q-1)t^2}{2\beta_t^2}   \right )^{\frac{1}{1-q}}}{\sqrt{2\pi}\beta_t 
\Gamma\left ( 1+\frac{1}{1-q} \right )}, \;\; q<1 \nonumber 
\end{align}
\vspace{-3mm}
\begin{equation}
\hspace{10000pt minus 1fil}
I_{\text{q-}\textit{Gauss}}(f)= \frac{zeN(1-q)\left ( \frac{1-q}{2}  \right )^{\left | \frac{1}{2}-\frac{q}{1-q}  \right |/2} 
J_{\frac{1}{1-q}+\frac{1}{2}} \left ( 2\pi f \beta_t\sqrt{\frac{2}{1-q}}  \right ) \Gamma \left ( \frac{1}{1-q}+\frac{3}{2} \right )}
{2f^{\frac{1}{1-q}+\frac{1}{2}}\pi^{\frac{1}{1-q}+\frac{1}{2}}\beta_t^{\frac{1}{1-q}+\frac{1}{2}}} 
\hfilneg
\label{eq:qGaussTDFD}
\end{equation}
\normalsize
\begin{figure}[htb]
\begin{subfigure}{0.32\textwidth}
  \centering
  \includegraphics[width=0.99\linewidth]{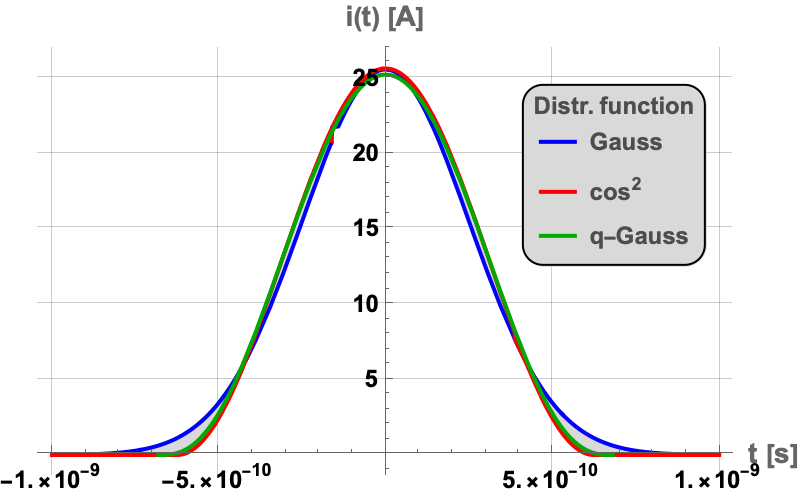}
  \caption{Bunch current \\ in the time-domain.}
  \label{fig:gaussCos2qGaussTD}
\end{subfigure}
\hfill
\begin{subfigure}{0.32\textwidth}
  \centering
  \includegraphics[width=0.99\linewidth]{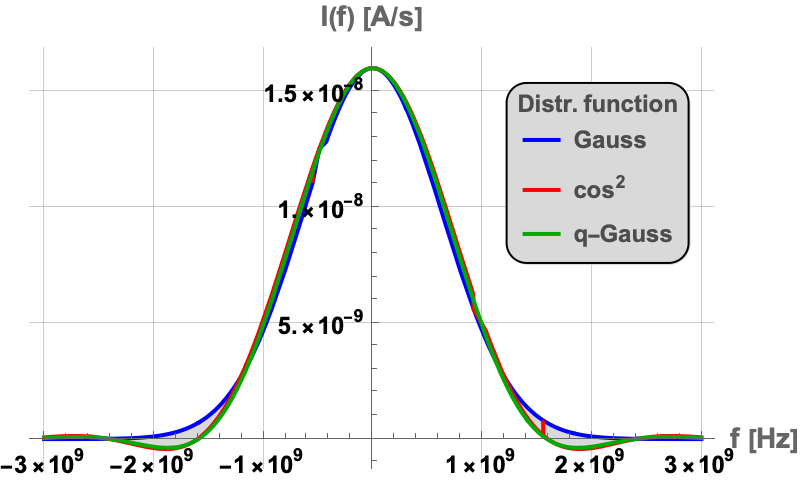}
  \caption{Bunch current \\ in the frequency-domain.}
  \label{fig:gaussCos2qGaussFD}
\end{subfigure}
\hfill
\begin{subfigure}{0.32\textwidth}
  \centering
  \includegraphics[width=0.99\linewidth]{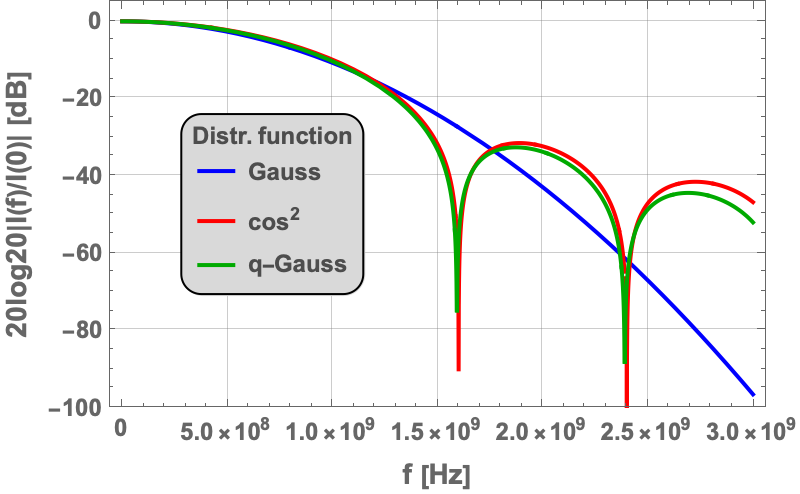}
  \caption{Bunch current in the frequency-domain, normalized and $\log_{10}$ scaled.}
  \label{fig:gaussCos2qGaussFDdB}
\end{subfigure}
\hspace*{5mm}
\caption{Examples of bunched beam current signals for different particle distributions.}
\label{fig:bunchTDFD}
\end{figure}
Equations (\ref{eq:GaussTDFD}),  (\ref{eq:cos2TDFD}) and (\ref{eq:qGaussTDFD})
express the longitudinal distribution functions in time and frequency-domain for the \textit{Gaussian},
$\cos^2$ and the \textit{Tsallis} q-\textit{Gaussian} functions.
The latter is only given for $q<1$, as for $q=1$ Eq.~(\ref{eq:qGaussTDFD}) merges into 
Eq.~(\ref{eq:GaussTDFD}) with $\beta_t=\sigma_t$,
and the case for $1<q<3$ is off less relevance.
Figure \ref{fig:bunchTDFD} shows an example of the beam bunch current signal in time and frequency domain
for the discussed three different distribution functions.
The bunch intensity is identically for three cases, $N=\num{1e11}$, 
and also the bunch length is similar, $\sigma_t=\SI{250}{ps}$ for the \textit{Gaussian} distribution,
$t_b=\SI{1.25}{ns}$ for the $\cos^2$ distribution, and $\beta_t=\SI{285}{ps},\, q=0.65$ for
the \textit{Tsallis} q-\textit{Gaussian} distribution.
With these parameters the peak current, $i(t=0)\approx\SI{25}{A}$, is almost identical, also the bunch shape is
similar, in particular the $\cos^2$ and \textit{Tsallis} q-\textit{Gaussian} distributions are almost identically.
Only in the graph of the normalized, logarithmically scaled bunch spectra 
$20\log_{10}[I(f)/I(0)]$, Fig.~\ref{fig:gaussCos2qGaussFDdB},
little differences between the two can be identified.

Please note that the definition of the \emph{bunch length} varies not only if different distribution functions are in the
play, but also for the same \textit{Gaussian} distribution function the definition has to be stated, e.g.\
as $\sigma$, $4\,\sigma$, or $FWHM$ (full-width, half maximum), with the latter given as
\begin{equation}
t_{FWHM}=2\sqrt{-2\ln{(1/2)}}\, \sigma_t\approx 2.355\, \sigma_t
\label{eq:fwhm}
\end{equation}

Strictly speaking, each bunch passing a beam pickup is different, and even the same bunch
circling in a storage ring is a bit different each times it appears at the beam pickup location,
the intensity may have dropped due to particle losses, the bunch shape has changed due to the various forces
on the particles throughout a revolution.
Still, for some time period we can assume a bunch in a ring accelerator of circumference $l_c=vT_{rev}=v/f_{rev}$
as being time invariant.
The RF frequency defines \emph{RF-buckets} at equidistant time intervals
$$
T=\frac{2\pi}{\omega}=\frac{1}{h\,f_{rev}}
$$
where $h=f_{RF}/f_{rev}$ is the \emph{harmonic number}.
In many cases each RF-bucket is filled with the same number of charged particles, such that all beam bunches
stored in the ring accelerator are basically identically of equal intensity and shape, spaced by an
equidistant time $T$, see also the illustration in Fig.~\ref{fig:bTrain}.
In this case the resulting beam current can be expressed in terms of a \textit{Fourier} series expansion
\begin{equation}
i_{\mathrm{beam}}(t)=\langle I_{DC}\rangle + 2 \langle I_{DC}\rangle \sum_{m=1}^{\infty} A_m \cos (m\omega t)
\label{eq:iTDfseries}
\end{equation}
with the average beam current $\langle I_{DC}\rangle=zeN/T$ and frequency harmonics spaced by $\omega = 2\pi f$.
The bunch shape in Eq.~(\ref{eq:iTDfseries}) is defined by a harmonic amplitude factor. 
Table~\ref{tab:Am} lists the harmonic amplitude factor $A_m$ for some typical,
and also for some less common bunch shape functions,
with $t_b$ defined as the bunch length (in time) at the base of the function (except for the \textit{Gaussian}),
and a normalization of $A_m\rightarrow 1$ for
$\omega\rightarrow 0$.
\begin{table}[b!]
\begin{center}
\caption{Harmonic amplitude factors for various bunch shapes.}
\label{tab:Am}
\begin{tabular}{ccc}
\hline\hline
\textbf{bunch shape}             & \textbf{harmonic amplitude factor} \boldmath$A_m$
                                                & \textbf{comments}\\
\hline
$\delta$-function   & 1             & for all harmonics\\
\textit{\textbf{Gaussian}}     & \boldmath$\exp \left [\frac{(m\omega\sigma_t)^2}{2} \right ] $           
 & \boldmath$\sigma_t=$ \textbf{RMS bunch length} \\
 parabolic & $3 \left ( \frac{\sin\alpha}{\alpha^3}-\frac{\cos\alpha}{\alpha^2}\right )$ 
 & $\alpha=m\pi t_b/T$ \\
(cos)$^2$ & $\frac{\sin(\alpha -2)\frac{\pi}{2}}{(\alpha -2)\pi}+\frac{\sin\frac{\alpha\pi}{2}}{\frac{\alpha\pi}{2}}+
\frac{\cos(\alpha +2)\frac{\pi}{2}}{(\alpha +2)\pi}$ 
 & $\alpha=2m t_b/T$ \\
 triangular & $\frac{2(1-\cos\alpha )}{\alpha^2}$  & $\alpha=m\pi t_b/T$ \\
 square & $\frac{\sin\alpha}{\alpha}$  & $\alpha=m\pi t_b/T$ \\
\hline\hline
\end{tabular}
\end{center}
\end{table}

\begin{figure}[t]
\begin{subfigure}{0.45\textwidth}
  \centering
  \includegraphics[width=0.8\linewidth]{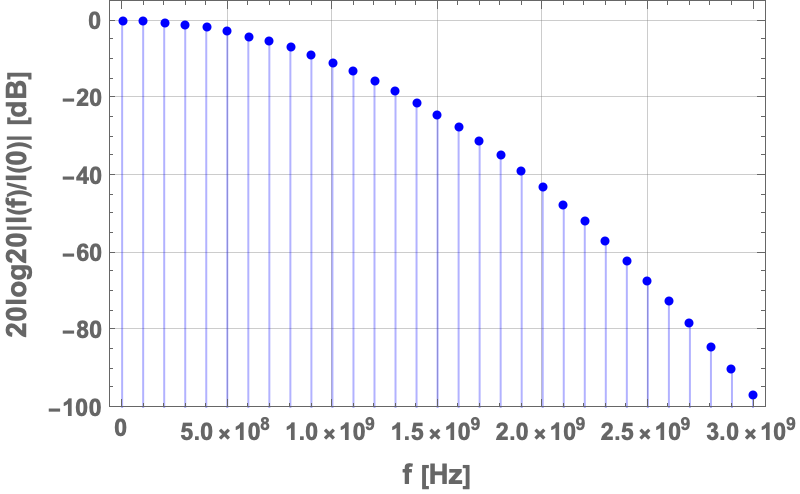}
  \caption{Line spectrum of a \textit{Gaussian} bunch.}
  \label{fig:gaussSpectr}
\end{subfigure}
\hfill
\begin{subfigure}{0.45\textwidth}
  \centering
  \includegraphics[width=0.8\linewidth]{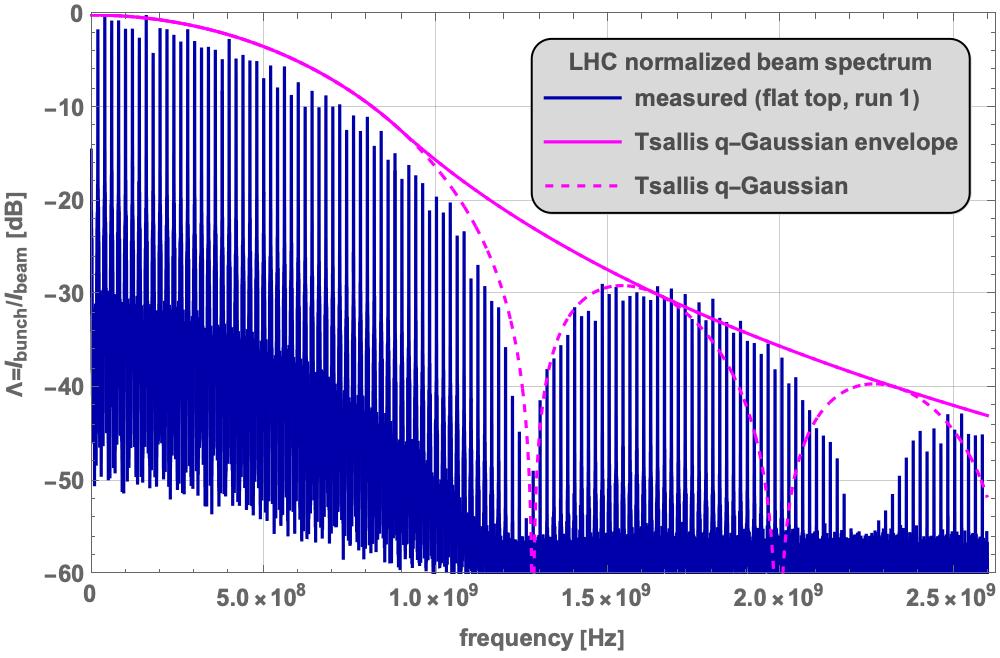}
  \caption{Multi-bunch beam spectrum measured in the LHC.}
  \label{fig:LHCspectr}
\end{subfigure}
\caption{Normalized beam spectra, logarithmically scaled.}
\label{fig:BeamSpectr}
\end{figure}
Figure \ref{fig:gaussSpectr} illustrates the normalized bunched beam spectrum for \textit{Gaussian} bunches
equidistantly spaced by $T=\SI{10}{ns}$, which results in the shown line spectrum with the harmonics spaced by
$\Delta f=1/T=\SI{100}{MHz}$, e.g.\ for a storage ring $l_c\approx\SI{30}{\meter}$,
the bunched beam traveling at $v\approx c$, $h=10$, and
all RF-buckets filled with bunches of same intensity.

\begin{figure}[hbt]
\begin{subfigure}{0.45\textwidth}
  \centering
  \includegraphics[width=0.8\linewidth]{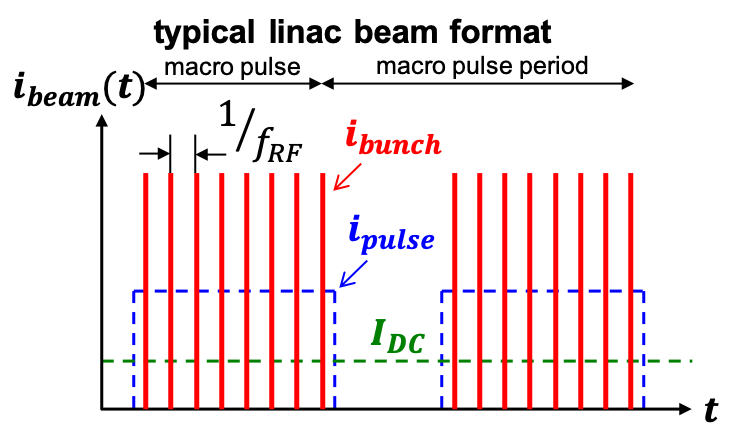}
  \caption{Ideal beam format in a linac.}
  \label{fig:linacBeam}
\end{subfigure}
\hfill
\begin{subfigure}{0.45\textwidth}
  \centering
  \includegraphics[width=0.8\linewidth]{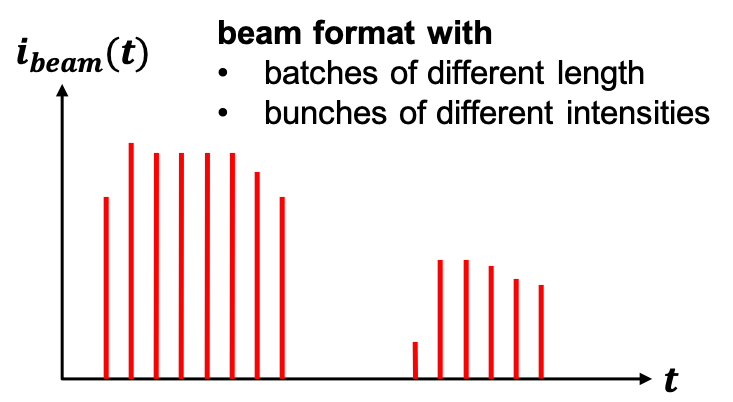}
  \caption{Multi-bunch beam format.}
  \label{fig:mBunch}
\end{subfigure}
\caption{Multi-bunch beam formats.}
\label{fig:BeamFormats}
\end{figure}
Even with similar or same particle distributions for the bunches, the \emph{beam format} in an accelerator can be very 
complicated (see also Fig.~\ref{fig:mBunch}), 
e.g.\ due to filling patterns and RF frequencies from pre-accelerators, kicker gaps, bunch
intensity variations, etc., which then result in a complicated beam spectrum, not covered by Eq.~(\ref{eq:iTDfseries}).
Figure \ref{fig:LHCspectr} shows the normalized line spectrum for a multi-bunch beam, measured at the
\emph{Large Hadron Collider} (LHC) at CERN, and approximated by the single bunch spectrum of
a \textit{Tsallis} q-\textit{Gaussian} distribution.
For completeness, Fig.~\ref{fig:linacBeam} illustrates the ideal bunched beam format as it typically appears in pulsed
linear accelerators (linac).

In practice, the response of the beam pickup -- as defined by the transfer impedance Eq.~(\ref{eq:PUtrans}) -- 
to a single bunch, with the related single-bunch spectra examples given by Eqs.~(\ref{eq:GaussTDFD}),  
(\ref{eq:cos2TDFD}) and (\ref{eq:qGaussTDFD}) is sufficient for the beam signal analysis.
Most beam pickups are linear, time-invariant systems, and the response to a complicated beam format can be
evaluated by superposition of several single bunch stimulus signals with the appropriate parameters (time delay,
intensity, length, shape, etc.).
\clearpage\newpage
 
\subsection{Bunch length measurement in time- and frequency domain}

We shall round-up the treatment of a beam signal for a single beam passage with a bunch length measurement in time-domain and in frequency-domain. The intention to show the two measurement examples of the same physical quantity is to demonstrate again, that time- and frequency domain are not only useful mathematical descriptions of a physical process, but also that a physical quantity can be measured in either domain.

\subsubsection{time-domain measurement}

\begin{figure}[ht]
    \centering
    \begin{minipage}{0.48\textwidth}
  \begin{center}
    \includegraphics[width=0.7\textwidth]{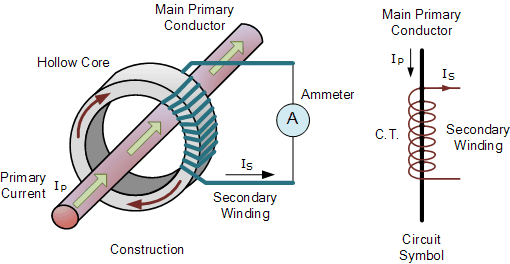}
 	\end{center}
  	\caption{Measurement of the longitudinal bunch shape with a so called bunch current transformer}
    \label{fig:TFD-BCT}
    \end{minipage}\hfill
    \begin{minipage}{0.48\textwidth}
  \begin{center}
    \includegraphics[width=0.7\textwidth]{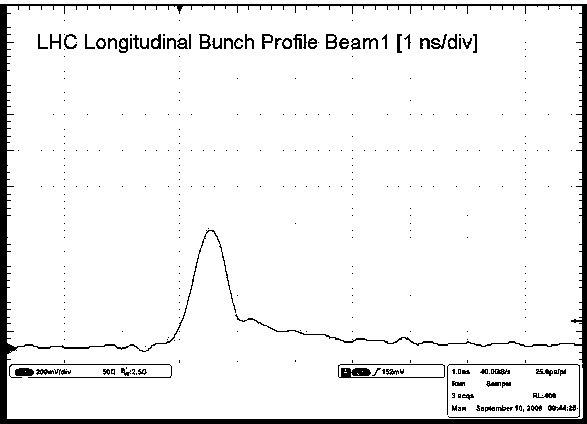}
 	\end{center}
  	\caption{bunch length signal on an oscilloscope with a time base of 1 nsec/div showing the LHC beam\cite{bib:TFD-valuch}}
    \label{fig:TFD-bunchlength}
    \end{minipage}
\end{figure}

\Fref[b]{fig:TFD-BCT} shows the principle of a bunch current measurement with a so called bunch current transformer (see for example \cite{bib:TFD-rhodri1}, a device similar to the wall-current monitor described before in \Sref{sec:bulen}. The device intercepts the image current of the bunch charges and produces a proportional electrical signal. For relativistic particle beams not only the bunch charge, but also the longitudinal profile can be measured with this electrical signal. Below the nsec-scale the measurement of the bunch shape using time-domain information reaches its limits due to bandwidth limitations in the transformer and also in the measurement electronics. \Fref[b]{fig:TFD-bunchlength} shows such a measurement for LHC bunches of around 300 ps bunch length.

\subsubsection{frequency-domain measurement}

For very short bunches (well below the ns-scale) direct measurements in time-domain are less accurate if not impossible. Instead one might use frequency-domain measurements. \Fref[b]{fig:TFD-13} shows an experimental setup of such a bunch length measurement in the CLIC test facility \cite{bib:TFD-CTF3} for ps-long bunches. 

\begin{wrapfigure}{r}{0.6\textwidth}
  \begin{center}
    \includegraphics[width=0.3\textwidth]{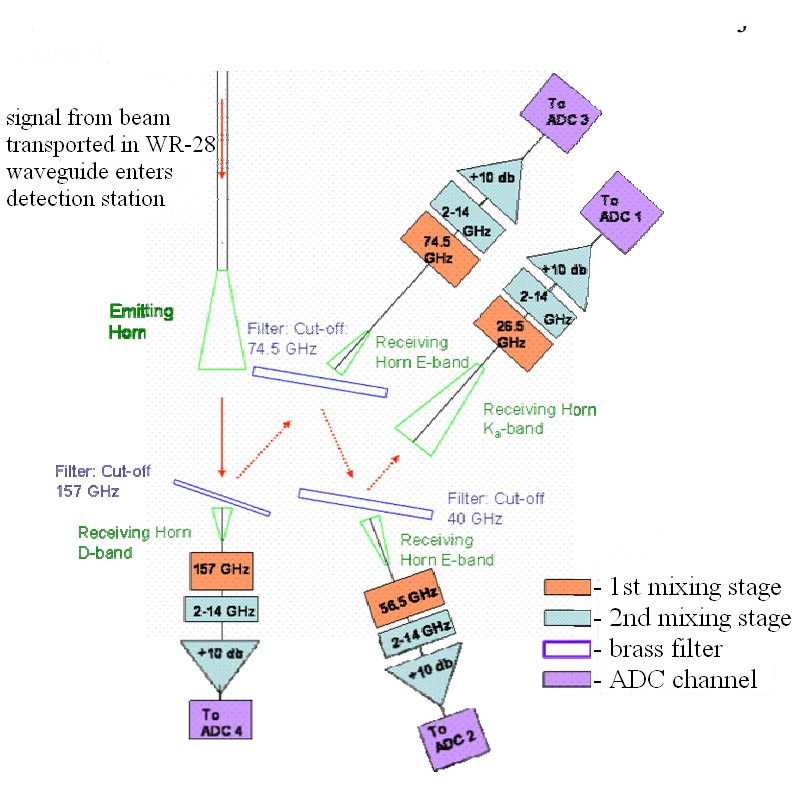}
   \end{center}
  	\caption{Experimental setup in CTF3 (CLIC test facility CERN) for a bunchlength measurement in frequency-domain}
    \label{fig:TFD-13}
    \end{wrapfigure}
\noindent For the measurement  the electromagnetic radiation of the beam was used, which was emitted with a horn antenna directly onto a setup of four micro-wave receivers. By choosing the appropriate geometry and by choosing absorbers, these receivers were made sensitive to different frequency ranges of the electromagnetic radiation. As we have seen in \ref{sec:bulen}, the radiation spectrum of short bunches will contain more signal in the higher frequency bands than the spectrum of long bunches. 
\clearpage

\begin{figure}[!ht]
    \centering
    \begin{minipage}{0.45\textwidth}
        \centering
        \includegraphics[width=0.9\textwidth]{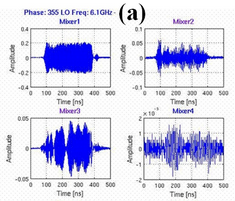} 
        \caption{\label{fig:TFD-14}time-domain signals of the beam pulse recorded in the 4 frequency bands }
    \end{minipage}\hfill
    \begin{minipage}{0.45\textwidth}
        \centering
        \includegraphics[width=0.9\textwidth]{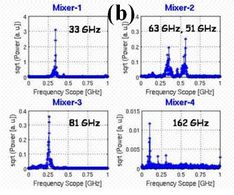} 
        \caption{\label{fig:TFD-15}Spectrum of the 4 time-domain signals in \Fref{fig:TFD-14} after down conversion and Fourier-transform}
    \end{minipage}
\end{figure}

\begin{figure}[!ht]
    \centering
    \begin{minipage}{0.45\textwidth}
        \centering
        \includegraphics[width=0.9\textwidth]{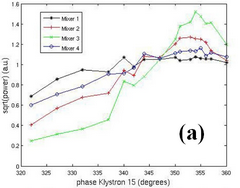} 
        \caption{\label{fig:TFD-16}amplitudes of the highest peaks in the spectra of \Fref{fig:TFD-15} as a function of a klystron phase setting}
    \end{minipage}\hfill
    \begin{minipage}{0.45\textwidth}
        \centering
        \includegraphics[width=0.9\textwidth]{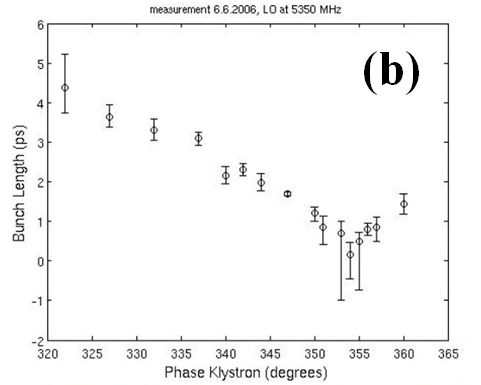} 
        \caption{\label{fig:TFD-17}Computed bunch-length from data in \Fref{fig:TFD-16} as a function of a klystron phase setting}
    \end{minipage}
\end{figure}

In the following figures measurement examples of the CTF3 test setup are given from an experiment, which used the so measured bunch-length in order to optimize a klystron setting.
\Fref[b]{fig:TFD-14} shows the signals of the four micro-wave receivers in time-domain.
\Fref[b]{fig:TFD-15} shows the same signals after down conversion and Fourier transform.
In \Fref{fig:TFD-16} the amplitude of the highest peak in \Fref{fig:TFD-15} is extracted and plotted as a function of a klystron setting (phase). One clearly sees how the ratio of lower to higher frequencies changes. Finally \Fref{fig:TFD-17} shows the same data, but the peak-amplitudes have been converted into the bunch-length by using a mathematical model.
The klystron setting with the smallest bunch length was be the optimal setting.
    
\clearpage\newpage

\subsection{Spectra of regular equidistantly bunched beams}
\label{sec:sbmp}

\begin{wrapfigure}[12]{r}{0.65\textwidth}
  \begin{center}
    \includegraphics[width=0.62\textwidth]{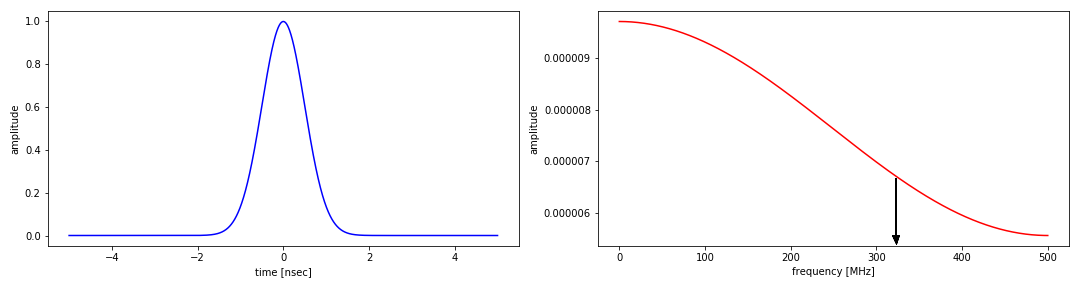}
 	\end{center}
  	\caption{Time domain representation of a bunch with 0.5 nsec bunch length($\ \sigma_{s}$) (left) and frequency-domain information of the same bunch (right) passing only \textbf{once} through the sensor. The width in frequency domain  $\sigma_{f} =\frac{1}{2\pi \sigma_s}$ = 318 MHz is indicated as black arrow.}
    \label{fig:TFD-Onebunch}
\end{wrapfigure}

We recapitulate the information of \Sref{sec:bulen} by drawing the simple \Fref{fig:TFD-Onebunch}. 

\noindent Assuming a Gaussian longitudinal particle distribution we have also in frequency domain (the positive half of) a Gaussian distribution. This type of observation is typical for a beam transport line or for a linear accelerator. It is important to note that the frequency spectrum is a continuum, i.e. \textbf{all frequencies} within the given envelope are present in the bunch spectrum.
\vspace*{1cm}
    
\noindent
The same bunched beam passing continuously through the sensor represents a continuous periodic signal. 
\begin{wrapfigure}[14]{r}{0.65\textwidth}
  \begin{center}
    \includegraphics[width=0.62\textwidth]{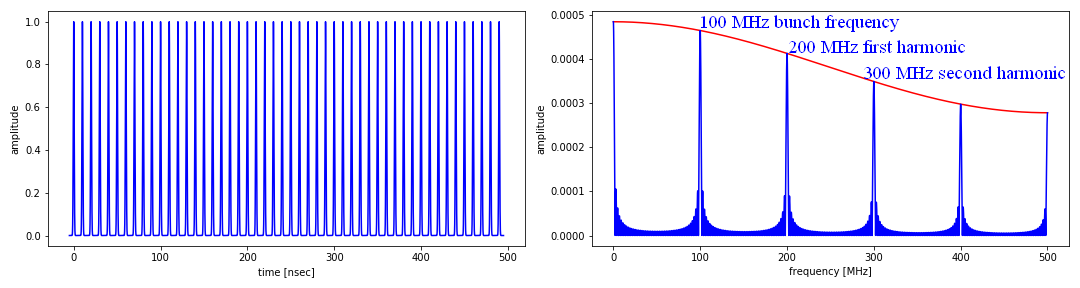}
 	\end{center}
  	\caption{Time domain representation of a continuous sequence of bunches with 0.5 ns bunch length($\ \sigma_{s}$) and 10 ns  bunch distance (left) and frequency-domain information of the same (right). The red amplitude envelope is the same as obtained from a single bunch of 0.5 ns bunchlength in \Fref{fig:TFD-Onebunch}}
    \label{fig:TFD-infinitebunches}
\end{wrapfigure}

\noindent Please compare the spectra in \Fref{fig:TFD-Onebunch} and \Fref{fig:TFD-infinitebunches}. A somewhat surprising result: The continuous spectrum of a bunch at single passage becomes a line spectrum with discrete frequencies if the bunch signal appears periodically.
We see a first line at "zero" frequency, which corresponds to the total intensity of the bunch. Then we see at 100 MHz the bunch frequency, which corresponds to the 10 ns bunch distance. Furthermore we see an (infinite) sequence of harmonics of the bunch frequency, but the envelope of these harmonics corresponds to the frequency content in the initial distribution ( see \Fref{fig:TFD-Onebunch}(left)).

\noindent The above bunch pattern could originate from the following accelerator types:
\begin {itemize}
\item{
a Continuous Wave (CW) linac or in an associated transport line of a CW linac}
\item{
In a very small circular accelerator with 10 ns revolution time filled with a single bunch}
\item{
In a larger circular accelerator with many bunches at 10 ns distance and with no time gap for the injection or extraction of bunches}
\end{itemize}

\noindent The last example is made up in a somewhat artificial way in order to demonstrate that with a single pick-up and by looking at beam spectra it is not obvious to tell on what accelerator one is working. It is in particular curious to see that using a single observation sensor, one can not tell if one works on a CW linac or at a circular accelerator with one or many bunches! 

\noindent The appearance of such a line spectrum is so fundamental, that we want to understand this in two rather different ways:

\subsubsection{Understanding the line spectrum by doing the Fourier transform}

First we can entrust the understanding of the appearance of a line spectrum entirely to our mathematical skills. We take the mathematical description of a continuous bunched beam as a sequence of dirac pulses (see \Sref{sec:TFD-varsig}) and then give the bunches a finite length by expressing them as Gaussian profiles in time.

The resulting spectral information results from a Fourier-transform by using the convolution theorem (\Sref{sec:TFD-convolution}).

\begin{equation}
\Sha(t)\ =\ \sum_{n=-\inf}^{\inf} \delta (t-n\Delta t)
\end{equation}
A single Gaussian pulse is mathematically expressed as:
\begin{equation}
g(t)\ =\ \frac{A_0}{\sqrt{2\pi}\sigma_t}\cdot e^{-\frac{t^2}{2 \sigma_t^2}}
\end{equation}
a sequence of \textbf{Gaussian pulses} inter-spaced by an interval $\Delta t$ can be written as the convolution of the above functions
\begin{equation}
f(t)\ =g(t)\circledast\ \Sha(t)\ =\  A_0\cdot\sum_{n=1}^{\inf} e^{-\frac{(t-nT)^2}{2 \sigma_t^2}}
\end{equation}
From the convolution theorem (see \Eref{eq:TFD-convolution}) we know the Fourier transform of $f(t)$ will be the simple product of the Fourier transforms of $g(t) \rm{and}\ \Sha(t)$.
Hence we can write:
\begin{equation}
\mathcal{F}[f(t)]\ =\mathcal{F}[g(t)]\cdot\ \mathcal{F}[\Sha(t]\ =\  
 e^{-\frac{\omega_{0}^2}{2 (1/\sigma_t)^2}} \cdot
\sum_{k=-\inf}^{\inf} \omega_{0}\delta(\omega - k\omega_{0})
\qquad \omega_{0}=\frac{2\pi}{\Delta t}
\end{equation}
The second term of the above Fourier transform gives an infinite sequence of frequency lines, whereas the first term modulates the amplitudes of these frequency lines depending on the bunch length.

\subsubsection{Understanding the line spectrum by the superposition of individual bunch signals}

\begin{figure}[!ht]
  \begin{center}
    \includegraphics[width=0.99\textwidth]{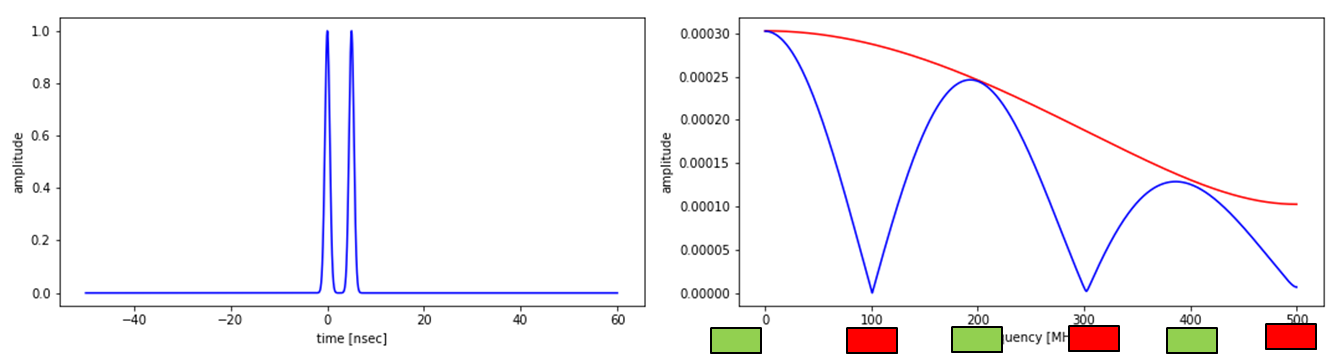}
 	\end{center}
  	\caption{Time domain representation of two bunches with 0.5 nsec bunch length($\ \sigma_{s}$) and 5 nsec bunch distance (left) and frequency-domain information of the same (right). }
    \label{fig:TFD-twobunches1}
\end{figure}

The mathematical description above does not give us directly an understanding of the physics. So complementary to the Fourier transform we explain the forming of spectral lines using a phenomenological description.
As first step we add only a second bunch to the first bunch.
Both bunches shall have the same characteristics and the bunch distance is 5 nsec.

In order to understand the resulting frequency-domain spectrum we simply have to recall that each individual bunch contains a continuous frequency spectrum with an envelope given by its length and that now all frequency components of the two bunches will interfere, since they are received in our sensor with a 5 nsec time delay.
All frequency components in the regions indicated with green boxes in \Fref{fig:TFD-twobunches1} have roughly the same phase, so they will interfere constructively (for example at 200 MHz a full wave length just corresponds to the time delay of 5 nsec, at 400 MHz the time delay corresponds to two full wavelength). Please note also that the vertical scale on the right of \Fref{fig:TFD-twobunches1} is twice as large as of \Fref{fig:TFD-Onebunch}. 

For frequencies in the ranges indicated by the red boxes, the time difference of the two bunches corresponds to (n+1/2) wave length, so the signals of both bunches extinct each other.

\begin{figure}[!ht]
  \begin{center}
    \includegraphics[width=0.9\textwidth]{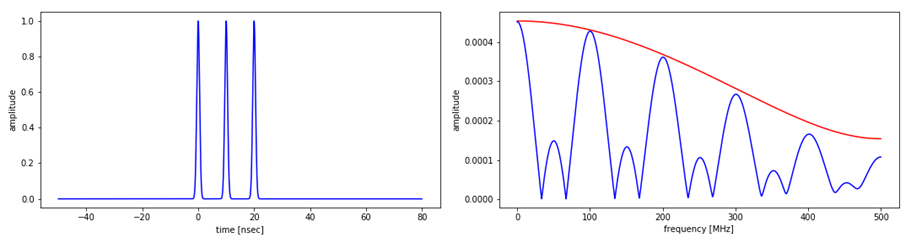}
    \includegraphics[width=0.9\textwidth]{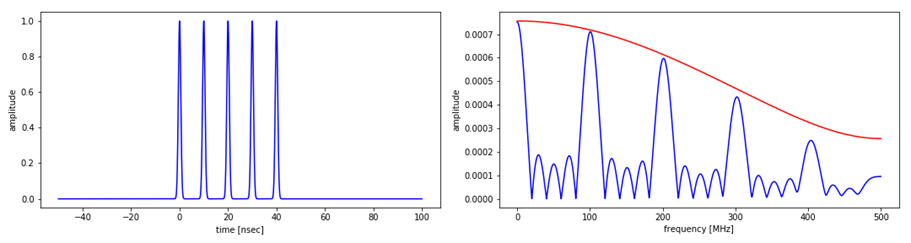} 
    \includegraphics[width=0.9\textwidth]{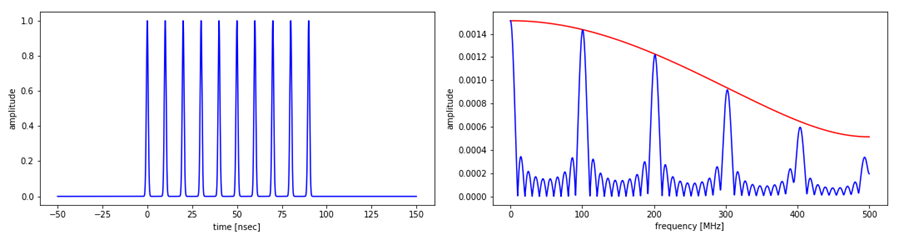}  
    \includegraphics[width=0.9\textwidth]{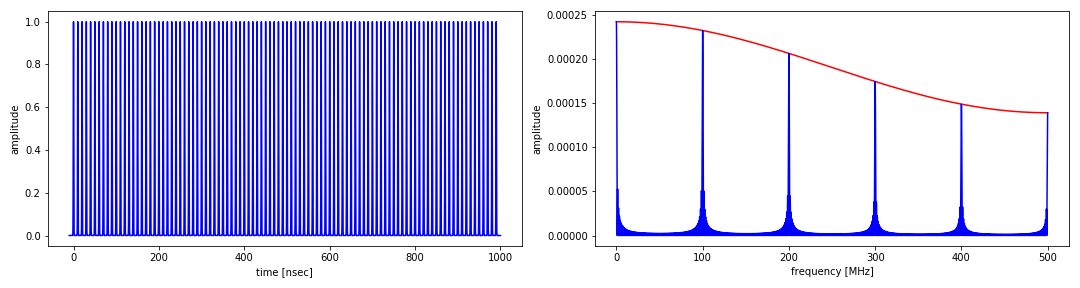} 
 \end{center}
  	\caption{Time domain representation of 3(top trace)5, 10 and 100 bunches (bottom trace) with 0.5 nsec bunch length($\ \sigma_{s}$) and 5 nsec bunch distance and frequency-domain information of the same. }
    \label{fig:TFD-multibunches1}
\end{figure}

Still very far away from single spectral lines, we already see the formation of frequency bands due to constructive or de-constructive interference. But by increasing the number of bunches and hence the number of combinations for constructive and destructive interference, we finally build up discrete residual spectral lines (bottom part of 
\Fref{fig:TFD-multibunches1}.
\clearpage

\subsubsection{Illustration of spectra for different bunch spacings}

As a complementary illustration we go back to two bunches and we vary the bunch spacing.
\Fref[b]{fig:TFD-multibunches2} shows the case for 5, 10 or 20 nsec bunch time distance.
One clearly sees how the frequency bands with constructive and destructive interference move due to the varying bunch distance.
\begin{figure}[!ht]
  \begin{center}
    \includegraphics[width=0.9\textwidth]{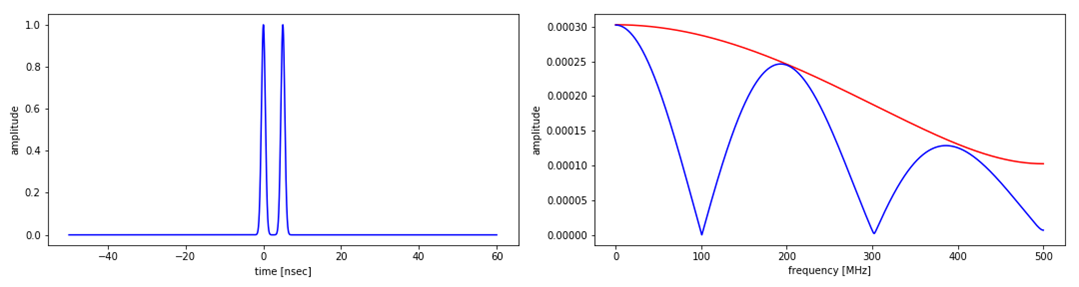}
    \includegraphics[width=0.9\textwidth]{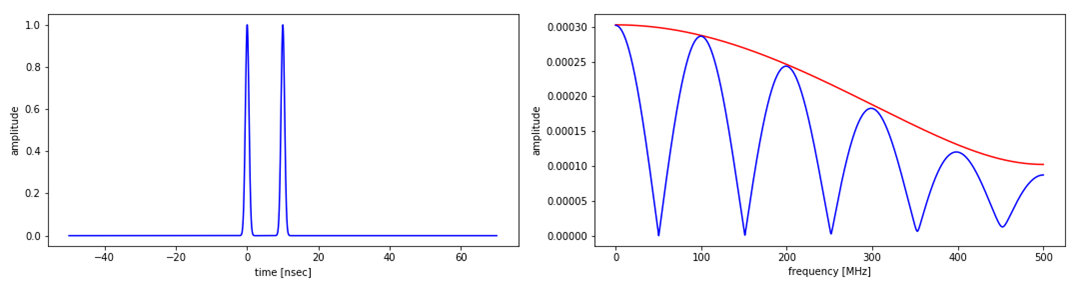} 
    \includegraphics[width=0.9\textwidth]{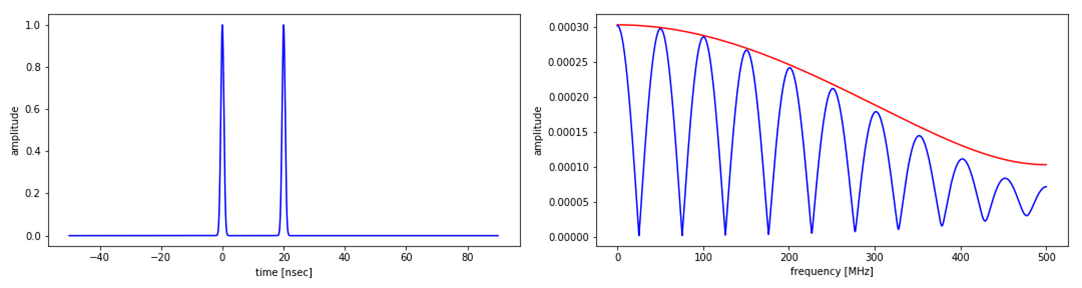}  
 \end{center}
  	\caption{Time domain representation of 2 bunches  with 0.5 nsec bunch length($\ \sigma_{s}$) and 5 (top trace), 10 (mid) and 20 nsec (bottom) bunch distance and frequency-domain information of the same. }
    \label{fig:TFD-multibunches2}
\end{figure}

\subsection{Spectra of irregular bunch patterns}

In most previous examples we have looked at a continuous sequence of bunches, which is a valid description only for a few continuously operating linear accelerators.
In most circular accelerators one needs to leave several Rf-buckets empty in order to allow injection or extraction kickers to change their deflecting field during the time gap without bunches.
\clearpage
\begin{figure}[!ht]
  \begin{center}
    \includegraphics[width=0.75\textwidth]{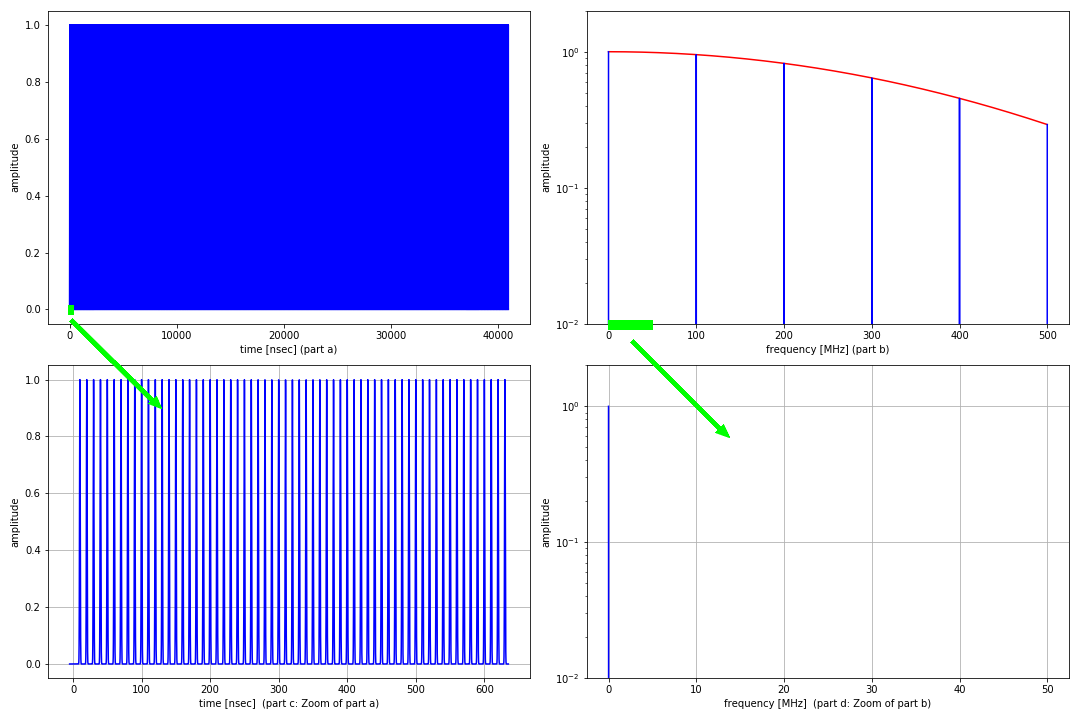}
 \end{center}
  	\caption{Time- and frequency-domain representation of a continuous sequence of bunches (10 nsec bunch distance) in time domain(top left part a) and frequency domain (top right part b). The lower graphs show as zoom the first 64 bunch samples (part c) and a zoom into the lower frequency range (part d).}
    \label{fig:TFD-bunchnogap}
\end{figure}
\begin{figure}[!ht]
  \begin{center}
    \includegraphics[width=0.75\textwidth]{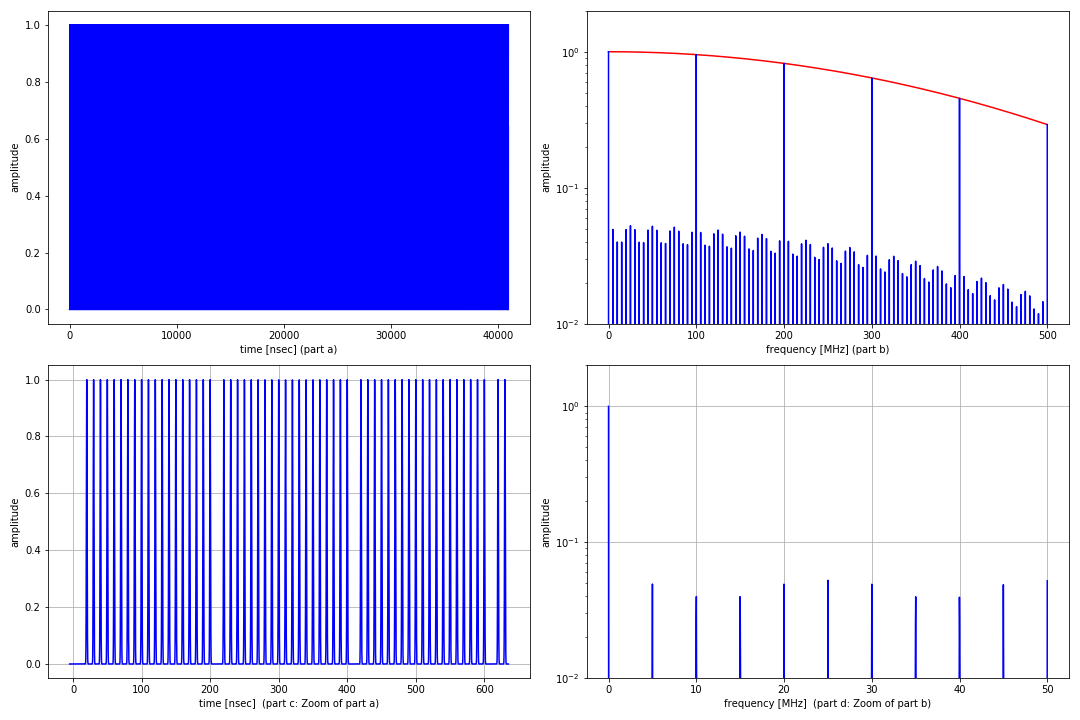}
 \end{center}
  	\caption{Small circular accelerator
  	(20 RF buckets of 10 nsec distance, 19 buckets filled with bunches and one bucket left free). Time- and frequency-domain representation (top left part a) and frequency domain (top right part b). The lower graphs show as zoom the first 64 bunches (part c) and a zoom into the lower frequency range (part d).}
    \label{fig:TFD-bunchgap}
\end{figure}
In order to show the spectral information of bunches in a circular accelerator with a
beam gap we compare the spectrum of an accelerator filled with consecutive bunches of 10 nsec bunch distance (\Fref{fig:TFD-bunchnogap}) with the situation, when a gap of one out of twenty RF buckets is not filled with beam (\Fref{fig:TFD-bunchgap}). This time the vertical scale in frequency-domain is chosen logarithmically over a small range of two decades.

In the \Fref{fig:TFD-bunchnogap} we see a "clean" spectrum with the expected spectrum of  a line at the fundamental bunch frequency (100 MHz) and its harmonics.
In \Fref{fig:TFD-bunchgap} we see many additional spectral lines coming up with a distance of 5 MHz. These additional lines are called \textbf{revolution harmonics}. How can we explain the appearance of these lines?

Again we make use of the convolution theorem to understand the appearance of these additional spectral lines. We describe the beam signal of the \Fref{fig:TFD-bunchgap} with a continuous beam \textbf{multiplied} with a rectangular wave function, which has a value of "1" during the time when there is beam in the accelerator and "0" during the bunch gaps. In our specific case the rectangular multiplicative function has a repetition frequency of 20 (bunches) * 10 nsec (bunch-distance) = 200 nsec or in other words a revolution frequency of the beam of 1/200 nsec = 5 MHz.
The convolution theorem gives now that the Fourier-transform of the product is the convolution of the two individual Fourier transforms. So we need to convolute the Fourier transform of the continuous beam (this gives lines every 100 MHz) with the Fourier transform of a rectangular function of 5 MHz, which gives a fundamental line at 5 MHz and its harmonics with decreasing amplitude.

For students interested in the mathematics, here a short primer of a  quantitative treatment:
\begin{figure}[!ht]
  \begin{center}
    \includegraphics[width=0.95\textwidth]{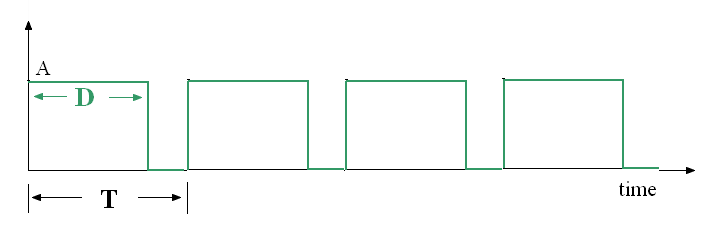}
 \end{center}
  	\caption{asymmetric rectangular wave}
    \label{fig:TFD-rectwave}
\end{figure}
\Fref[b]{fig:TFD-rectwave} illustrates the modulating function $mod(t)$ from above with an amplitude $A$, a period length $T$ and a duty cycle $D$.
Since this function is periodic, we get the Fourier transform by computing the coefficients of the Fourier-Series as:
\begin{equation}
c_k\ =\ \frac{1}{T}\int_0^T mod(t)e^{-jk\omega_0 t}dt\qquad\rm{with}\ \omega_0= \frac{2\pi}{T}
\end{equation}
After some transformation this reads
\begin{equation}
c_k\ =\ A\cdot D\cdot \frac{\sin{k\pi D}}{k\pi D}\cdot e^{-jk\pi D}
\label{eq:TFD-mod_ck}
\end{equation}
For the magnitude of the Fourier coefficient $c_k$ we can ignore the last term of 
\ \Eref{eq:TFD-mod_ck}, since it is just a phase of modulus 1.
For a symmetric rectangular wave ($D=0.5$)\ \Eref{eq:TFD-mod_ck} results in only 
\textbf{odd harmonics} with values $c_{k'}\propto 1/k', k'=2k-1$.
For a small duty cycle like in our \Fref{fig:TFD-bunchgap} (D=0.05) we can approximate the sin-function by its argument and hence get almost constant values for the coefficients
$c_k$. This explains why we see in the frequency domain of \Fref{fig:TFD-bunchgap} all harmonics of the rectangular modulation wave with almost equal amplitude.
\newpage

\subsubsection{A real life example (CERN SPS)}

We conclude this section on spectra of irregular filled accelerators with a measurement example on a real accelerator.

\begin{figure}[!ht]
    \centering
       \includegraphics[width=0.5\textwidth]{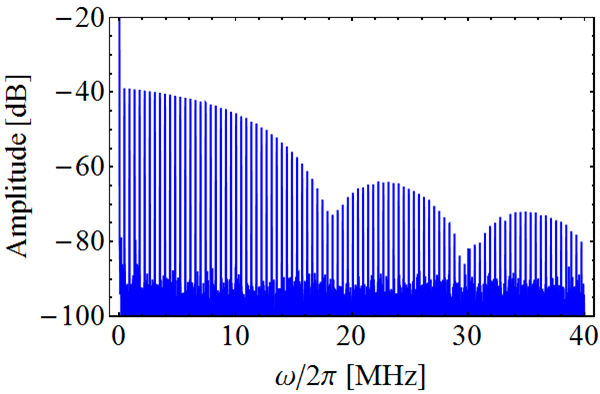} 
        \caption{\label{fig:TFD-P23}Spectrum of a single bunch in a circular accelerator}
\end{figure}
\begin{figure}[!ht]
        \centering
        \includegraphics[width=0.5\textwidth]{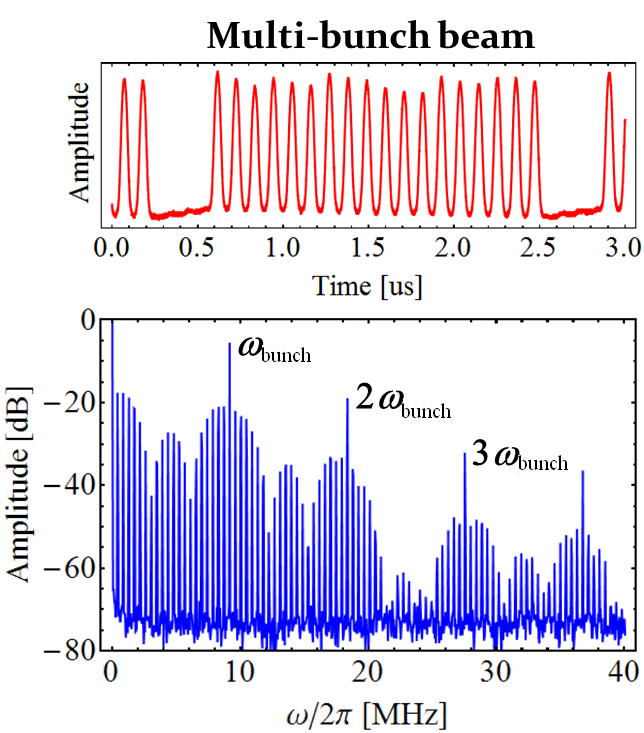} 
        \caption{\label{fig:TFD-P24}Spectrum of several bunches including a beam gap in the same accelerator as in \Fref{fig:TFD-P23}}
\end{figure}
\Fref[b]{fig:TFD-P23} shows the spectrum of a single bunch with a revolution frequency $f_{rev}$ of 43.45 kHz. We see many revolution harmonics all spaced by 43.45 kHz.  From the non-Gaussian envelope of the revolution harmonics we conclude that the bunch shape is "somewhat between triangular and parabolic".\\
\Fref[b]{fig:TFD-P24} shows the beam-spectrum of the same accelerator filled with 18 bunches including a beam gap. In this configuration the bunch distance is about 10.8 $\mu$sec corresponding to a bunch frequency of 9.3 MHz.
Now the most dominant spectral lines are the bunch frequency and its harmonics and in-between the revolution harmonics.
\clearpage
\subsubsection{Short summary}
\vspace{5mm}
\begin{importantbox}
In a circular accelerator filled with several bunches at \textbf{equal time distance} $\Delta T$, only spectral lines at the bunch frequency $f_{bunch} =1/\Delta T$ and harmonics are observable. The amplitude envelope of these lines is given by the Fourier transform of the time domain longitudinal particle distribution (bunch shape).
As soon as the regular bunch pattern gets modulated by an injection/extraction gap
we see around each bunch frequency line so called revolution harmonics, which appear at every multiple of the revolution frequency $f_{rev} = 1/T_{rev}$. The amplitudes of these revolution harmonics are given by the Fourier transform of the bunch pattern plus the additional modulation by the bunch spectrum envelope and by the intensity distribution of the individual bunches.
\end{importantbox}
\begin{figure}[!ht]
  \begin{center}
    \includegraphics[width=0.9\textwidth]{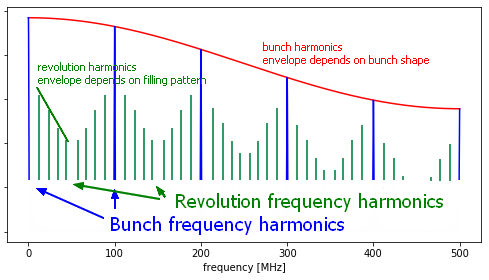}
     \caption{Illustration of a multi-bunch spectrum}   
 \label{fig:TFD-harmonics} 
  \end{center} 
\end{figure}
\newpage

\section{Signals of a single oscillating bunch}
\label{sec:SBO}

\subsection{Transverse Bunch Oscillations (Betatron Oscillations)}

\noindent
Up to this point the bunches were stable on their trajectory around the accelerator. There were no transverse nor longitudinal oscillations around the equilibrium position. Such oscillations will be the subject of the this section.

In both transverse planes the bunches are focused with the help of the lattice quadrupoles. Any constant and static additional deflection will result in a change of the closed orbit, whereas any momentary deflection (for example injection kick) will lead to  a transverse oscillation. As long as no other focusing elements play a role and since the focusing force of the quadrupoles is linear, the bunches will perform a harmonic oscillation around the closed orbit.

The observation of this harmonic oscillation can be done in various ways  \cite{bib:TFD-rhodri2}, in the simplest case by measuring the induced signal into one isolated electrode in the vacuum chamber. This can be for example one electrode of a beam-position monitor. On such an electrode one measures the dipole moment of the bunch charge distribution, which means a signal proportional to the bunch intensity and modulated by the distance of the bunch centroid to the electrode. \Fref[b]{fig:TFD-BPM_AM} illustrates the resulting turn by turn amplitude modulation of the (Gaussian) bunch signals due to the transverse oscillation.

\begin{figure}[!ht]
  \begin{center}
    \includegraphics[width=0.8\textwidth]{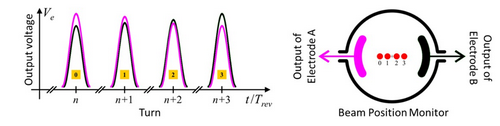}
    \caption{Detecting transverse beam oscillations on a single electrode of a beam position monitor. The oscillation information is superimposed as a small modulation on a large intensity signal.}
    \label{fig:TFD-BPM_AM}
 \end{center}
\end{figure}
 
In order to get the frequency domain representation of such betatron oscillations we   refer again to the convolution theorem (see \Eref{eq:TFD-convolution} and the following section on amplitude modulation).
It is very instructive to do the explicit maths for this:\\

\noindent
Again we write the time domain oscillation $z(t)$\ of a single bunch as a convolution term of equidistant Dirac pulses $x(t)$ with Gaussian pulses $y(t)$. The revolution time is $\Delta t$.
\begin{equation}
z(t)= y(t)\ \circledast\ x(t)\quad
\rm{with}\quad x(t) = \sum_{n=-\inf}^{\inf} \delta (t-n\Delta t)
\quad \rm{and}\quad
y(t)\ =\ \frac{1}{\sqrt{2\pi}\sigma_t}\cdot e^{-\frac{t^2}{2 \sigma_t^2}}
\end{equation}
We introduce a small harmonic amplitude modulation at the betatron tune frequency 
$\omega_q=2\pi qf_{rev};$\\$(A<<1)$, which yields:
\begin{equation}
\begin{split}
z(t) & = y(t)\ \circledast\ \left[ (1+A cos(\omega_q t)) \sum_{n=-\inf}^{\inf} \delta (t-n\Delta t) \right]\\
& = y(t)\ \circledast\ \left[ \sum_{n=-\inf}^{\inf} \delta (t-n\Delta t)+ 
\frac{1}{2}A(e^{j\omega_q t} + e^{-j\omega_q t}) \sum_{n=-\inf}^{\inf} \delta (t-n\Delta t) \right]\\
& = y(t)\ \circledast\ \left[ x(t) +\frac{1}{2}Ae^{j\omega_q t}x(t)+\frac{1}{2}Ae^{-j\omega_q t}x(t)\right]
\end{split}
\end{equation}
The Fourier transform follows from the convolution theorem:
\begin{equation}
\begin{split}
\mathcal{F} \left[ z(t) \right]\ & =\ \mathcal{F} \left[ y(t) \right]
\ \circledast\ \mathcal{F}\left[  x(t) +\frac{1}{2}Ae^{j\omega_q t}x(t)+\frac{1}{2}Ae^{-j\omega_q t}x(t)\right]\\
 & =\ \mathcal{F} \left[ y(t) \right]\ \circledast\ 
\left[ {\color{red}\mathcal{F} \left[ x(t) \right]}\ +
\frac{A}{2}{\color{cyan}\mathcal{F} \left[ e^{j\omega_q t}x(t)\right]}\ +
\frac{A}{2}{\color{blue}\mathcal{F} \left[ e^{-j\omega_q t}x(t)\right]} \right]\\
&=\quad e^{-\frac{\omega^2}{2 {(1/\sigma_t)}^2}}
{\color{red}\left[ \sum_{n=-\inf}^{\inf} \omega_{rev}\delta (\omega - k\omega_{rev}) \right]}\\
&+\ \frac{A}{2}e^{-\frac{\omega^2}{2 {(1/\sigma_t)}^2}}
{\color{cyan}\left[ \sum_{n=-\inf}^{\inf} \omega_{rev}\delta ((\omega -\omega_q) - k\omega_{rev}) \right]}\\
&+\ \frac{A}{2}e^{-\frac{\omega^2}{2 {(1/\sigma_t)}^2}}
{\color{blue}\left[ \sum_{n=-\inf}^{\inf} \omega_{rev}\delta ((\omega +\omega_q) - k\omega_{rev}) \right]}
\end{split}
\end{equation}
\begin{figure}[!ht]
  \begin{center}
    \includegraphics[width=0.92\textwidth]{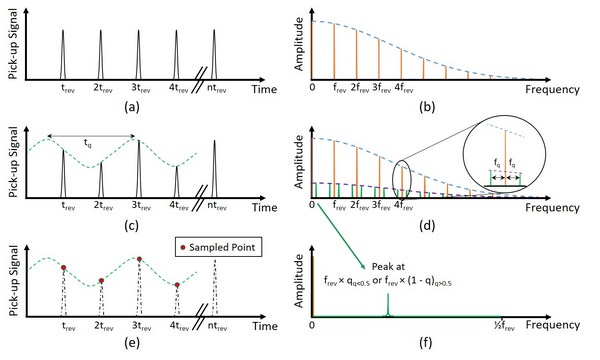}
    \caption{Time and frequency domain representation for the observation of transverse bunch oscillations at a single location on the circumference of the accelerator.
    (a\&b) continuous bunch passage without oscillation; (c\&d) continuous (50\% )amplitude modulation due to an oscillation; (e\&f) sampled once per revolution (figure taken from \cite{bib:TFD-rhodri2}).}
    \label{fig:TFD-tunemeasurement}
 \end{center}
\end{figure}\begin{importantbox}
The result represents the well known series of revolution harmonics with side-bands at $\omega\pm\omega_q$. The envelope function is again a Gaussian function dependent on the bunch length, while the amplitude of the side-bands depend on the modulation depth.
It is important to note that the information about transverse oscillations is \textbf{identically} available around each revolution harmonics. For example in the case of tune measurements it is \textbf{not} important around what revolution harmonics one tries to measure the tune.
\end{importantbox}

\Fref[b]{fig:TFD-tunemeasurement} nicely summarizes the result. The (a\&b) part corresponds to \Fref{fig:TFD-harmonics} in the case of one bunch. In that case bunch frequency and revolution frequency are the same.(c\&d) show the appearance of the betatron side-bands around each revolution harmonics. Finally (e\&f) show the reduced spectrum, which appears if we measure only one position sample per revolution period. Due to aliasing \Sref{sec:aliasing}) the only information we get is a single betatron side-band below the Nyquist-Shannon limit of half the sampling frequency. This form of acquisition is referred to as "Base-band acquisition".

\subsection{Longitudinal Bunch Oscillations (Synchrotron Oscillations)}

In the longitudinal plane the focusing is achieved by the RF system. As explained in \cite{bib:TFD-frank} the bunches perform so called synchrotron oscillations around the stable phase angle of the RF system.
When measuring this oscillation one does not observe like in the case of transverse oscillations an amplitude modulation of the signal, but on the contrary the bunches will pass during the oscillation a little earlier or later through the sensor. This produces a \textbf{phase-modulated} signal in the sensor. The sensor signal we can describe with:
($\beta$ is called the modulation index).
\begin{equation}
s(t)\ =\ A \cdot cos \left[2\pi f_c t+ \beta sin(2\pi f_m t)\right] 
\label{eq:TFD-PM}
\end{equation}
The mathematics of computing the Fourier transform of \Eref{eq:TFD-PM} is somewhat involved, since one has to solve some nifty integrals.
After "some time on a sunny afternoon" we find
\begin{equation}
\mathcal{F}\left[ s(t)\right]\ =\ A \cdot\sum_{n=-\inf}^{\inf} J_n(\beta)
\left[\delta (f - f_c - nf_m) + \delta (f + f_c + nf_m)\right] 
\label{eq:TFD-Four-PM}
\end{equation}
where $J_n(\beta)$ denotes the n-th order Bessel-function of the first kind:
\begin{equation}
J_n(\beta))\ =\ \frac{1}{2\pi}\int_{-\pi}^{\pi}e~^{\left[ \beta sin(x) - nx\right]}dx
\end{equation}

\begin{figure}[!ht]
    \centering
    \begin{minipage}{0.32\textwidth}
        \centering
        \includegraphics[width=0.95\textwidth]{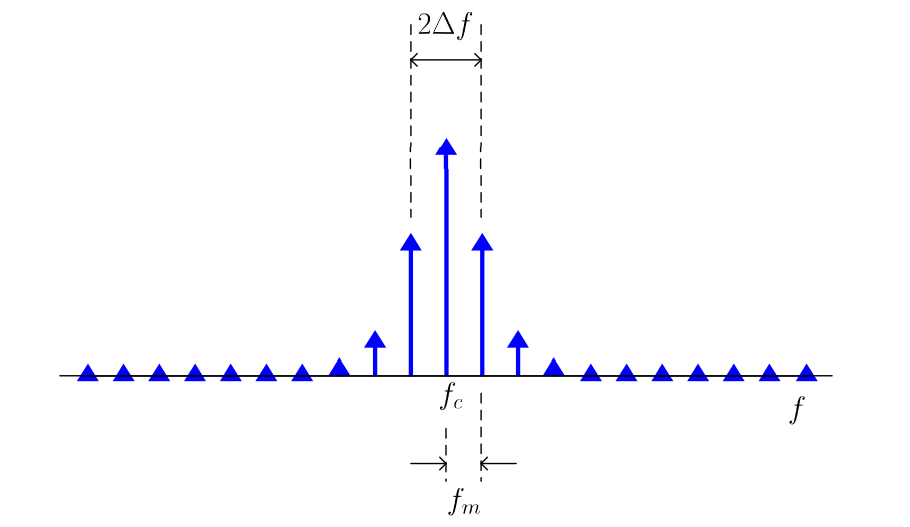} 
        \label{fig:TFD-PM2}
    \end{minipage}\hfill
    \begin{minipage}{0.32\textwidth}
        \centering
        \includegraphics[width=0.95\textwidth]{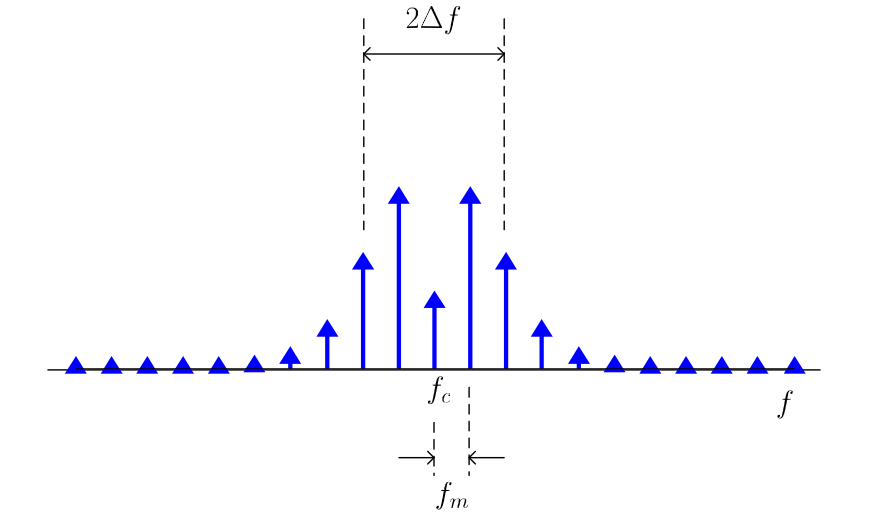} 
        \label{fig:TFD-PM3}
    \end{minipage}
        \begin{minipage}{0.32\textwidth}
        \centering
        \includegraphics[width=0.95\textwidth]{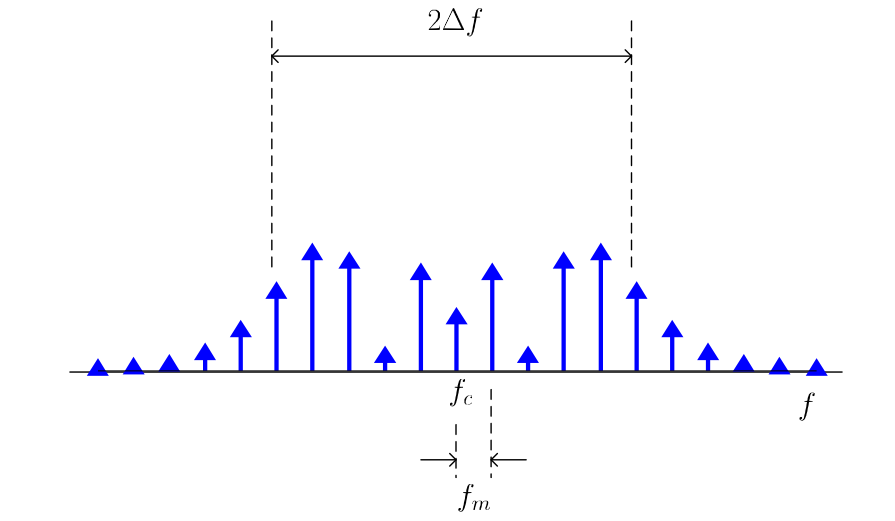} 
        \label{fig:TFD-PM4}
    \end{minipage}
    \caption{Fourier-spectrum (see \Eref{eq:TFD-Four-PM}) of a phase modulated carrier for different modulation-indices. From left to right:$\beta =1, 2, \rm{and} 5$.}
        \label{fig:TFD-PMfigs}
\end{figure}
\Fref[b]{fig:TFD-PMfigs} shows the computed spectra in case of phase-modulation for three different modulation indices. Depending on the modulation index we see many side-bands appearing around the carrier-frequency $f_c$. All Side-bands appear at integer multiples of the modulation frequency $f_m$. We can also see that for a small modulation index (narrowband phase-modulation) we obtain a spectrum similar to amplitude modulation, i.e. two major side-bands left and right of the carrier.\\
It is also interesting to note that in a proton synchrotron the stable phase angle is close to the zero-crossing of the RF voltage, so only a small phase modulation will occur in case of synchrotron oscillations (due to high slope of the Rf voltage). On the contrary in an electron synchrotron the stable phase angle is often close to the crest of the RF voltage and hence larger phase modulations will occur with more side-bands.
\newpage
\subsection{Coupled bunch oscillations}

So far we have only looked at oscillations in one plane independent on the two other planes. But the oscillations in both transverse planes can easily get coupled through accelerator imperfections \cite{bib:TFD-JoergW}, mainly through rotated quadrupoles.
But also the longitudinal plane can couple into the transverse planes, for example through non vanishing dispersion at the location of the RF cavities.

Since the coupling process is mainly a linear superposition, the mathematical treatment is the same as in the sections above, we will as result simply measure with a single sensor many  tune-related side-bands having their origin at the horizontal tune $q_h$, the vertical tune $q_v$ or the synchrotron tune $q_s$.\\
\begin{figure}[!ht]
  \begin{center}
    \includegraphics[width=0.95\textwidth]{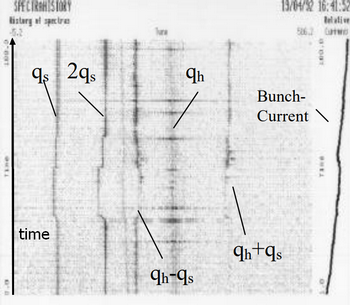}
    \caption{Baseband tune spectra recorded as spectrogram during the injection of particles into the CERN LEP storage ring (1989 - 2000).}
    \label{fig:TFD-tunehistory}
 \end{center}
\end{figure}
\Fref[b]{fig:TFD-tunehistory} shows a measurement example from "the good old days" (LEP 1992 \cite{bib:TFD-schmickl97} ), when due to synchro-betatron coupling several synchrotron side-bands were visible in the horizontal plane. Please note also the signals at the tunes $q_h\pm q_s$.
Time in this spectrogram goes from bottom to top. The sudden changes in the synchrotron tune are due to manipulations on the RF system, i.e. changing the total RF voltage.
\newpage

\section{Signals of multiple oscillating bunches}
\label{sec:TFD-mbmp}

In the previous \Sref{sec:SBO} we have treated oscillations of a \textbf{single} bunch in the transverse and longitudinal plane. The situation gets really interesting if we have a more complex multi-bunch filling scheme.
If there was no "cross-talk" between bunches, then in a multi-bunch filling scheme each bunch would  incoherently perform its own oscillation and we would measure signals as described in the previous section. 

But in a real accelerator the motion of the individual bunches gets coupled together through wake-fields induced in the vacuum chamber or in particular into RF cavities.
\cite{bib:TFD-KevinLi}.
Assuming equal intensities of the bunches, the bunches  will all oscillate at the same frequency (at the betatron tune $\nu$), but with different phases and amplitudes. We describe all possible bunch motion by the linear  \textbf{superposition of modes of oscillation}, where each mode is characterized by a bunch-bunch phase difference of
\begin{equation}
\Delta\Phi = m\frac{2\pi}{M}
\label{eq:TFD-dphi}
\end{equation}
where m is the multi-bunch mode number (m=0, 1, ...,M-1). Each multi-bunch mode is associated to a characteristic set of frequencies
\begin{equation}
\omega=pM\cdot\omega_0\ \pm\ (m+\nu)\cdot\omega_0\ =\ p\cdot\omega_{RF}\ \pm\ (m+\nu)\cdot\omega_0
\label{eq:TFD-dphi2}
\end{equation}
where p is an integer number ($-\inf<p<\inf$), $\omega_0$ is the revolution frequency,
 $\omega_{rf} = M\cdot\omega_0$ is the bunch repetition frequency (homogeneous filling), and $\nu$ is the tune. The term $\pm(m+\nu)\omega_0$ describes the side-bands around each revolution harmonic$\ pm\cdot\omega_0$.

\subsection{Illustrations of various multi bunch modes}

This is on the first side rather mind boggling, but it will become much clearer with a couple of examples in the following sections. During the course we make use of small video clips, but in this write-up we can only look at snapshots of these videos.
We shall start with the simple case of one bunch as recapitulation and explanation of the graphs.

\subsubsection{Single bunch as recapitulation}

\begin{figure}[!ht]
  \begin{center}
    \includegraphics[width=0.85\textwidth]{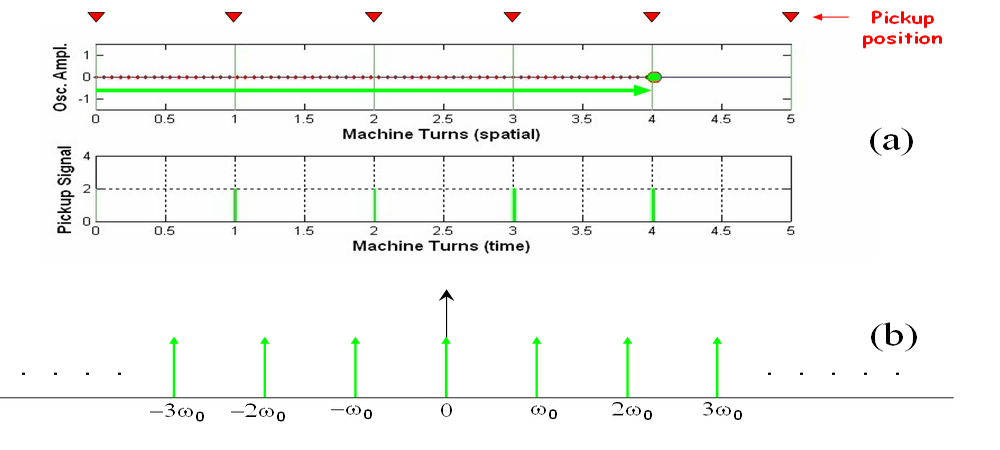}
    \caption{time-domain (a) and frequency-domain (b) information of a single stable bunch over five complete turns.}
    \label{fig:TFD-MultiBOnestable}
 \end{center}
\end{figure}
In the top \textbf{a-part} of \Fref{fig:TFD-MultiBOnestable} the horizontal axis shows time domain information over five complete turns of a circular accelerator, the location of each bunch travelling from left to right is indicated by a green spot in the top trace.
In the lower trace we record the position of this bunch with a green line, each time it passes the pickup for which the location is indicated by the red arrow above.
The lower \textbf{b-part} shows the frequency domain information for an endless sequence of bunch passages. As expected we see the series of harmonics of the revolution frequency $f_{rev}$. For simplicity all harmonics are drawn with the same amplitude as green arrows.\\
\begin{figure}[!ht]
  \begin{center}
    \includegraphics[width=0.85\textwidth]{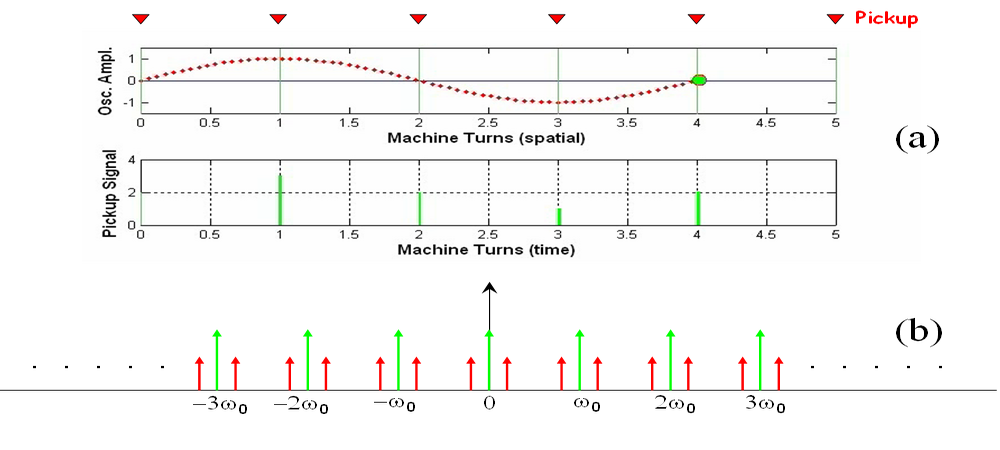}
    \caption{time-domain (a) and frequency-domain (b) information of a single bunch over five complete turns oscillating at the betatron tune.}
    \label{fig:TFD-MultiBOneunstable}
 \end{center}
\end{figure}
\Fref[b]{fig:TFD-MultiBOneunstable} shows the same bunch, this time oscillating at the betatron tune ($q = 0.25$). As expected we see in the b-part the upper and lower side-bands (red arrows) around each revolution harmonics (green arrows).

\subsubsection{Ten bunches - stable}
\begin{figure}[htpb]
  \begin{center}
    \includegraphics[width=0.85\textwidth]{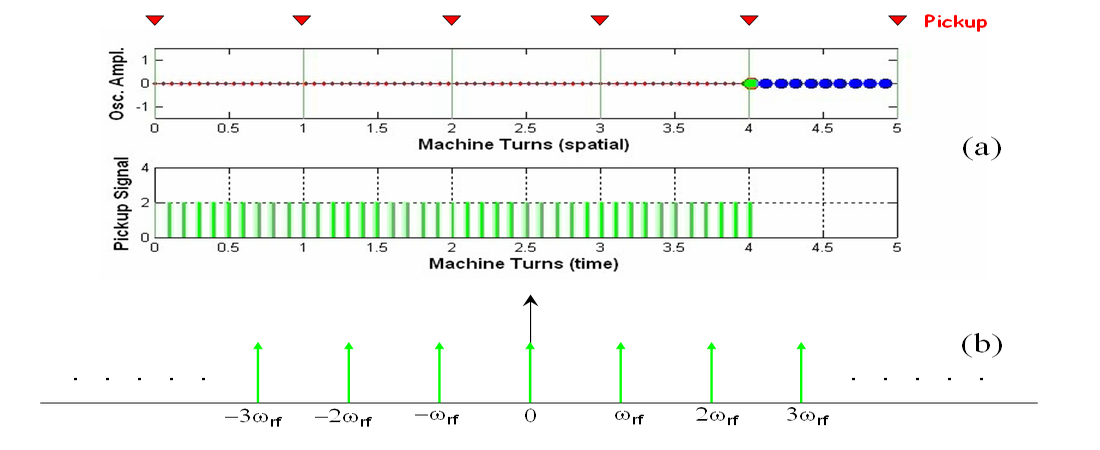}
    \caption{time-domain (a) and frequency-domain (b) information of ten stable bunches over five complete turns. This figure is equivalent to \Fref{fig:TFD-MultiBOnestable}  with the difference that the frequency of the harmonics ($n\cdot\omega_{rf}$) is ten times higher.}
    \label{fig:TFD-MultiBTenStable}
 \end{center}
\end{figure}

We repeat the exercise for \textbf{ten} equidistant bunches in \Fref{fig:TFD-MultiBTenStable}. We assume a harmonic number of h=10, so each Rf-bucket is filled with a bunch. Following the nomenclature of 
\Sref{sec:bunchpatterns} the bunch repetition frequency $\omega_{rep}$ and the RF-frequency $\omega_{rf}\ $ are the same. Therefore in the b-part the (angular) RF frequency 
$\omega_{rf}\ $ is used as horizontal axis.
Since the bunches do not oscillate, we do not see any betatron side-bands in the b-part of \Fref{fig:TFD-MultiBTenStable}. The only difference compared to \Fref{fig:TFD-MultiBOnestable} is that the frequency is ten times higher, so we see the harmonics at multiples of the bunch repetition frequency $\omega_{rf}$.
\subsubsection{Ten bunches - oscillation mode 0}
\begin{figure}[htpb]
  \begin{center}
     \includegraphics[width=0.95\textwidth]{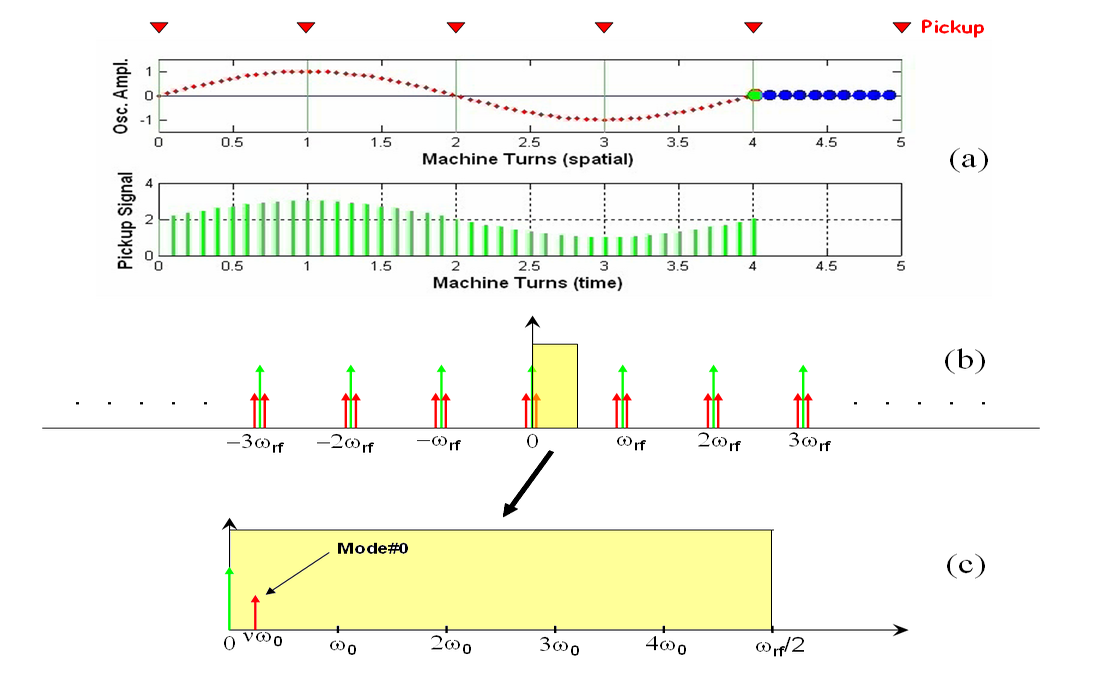}
     \caption{time-domain (a) and frequency-domain (b) information of ten bunches
      oscillating at the betatron tune over five complete turns. All bunches oscillate with the same phase, i.e. the multi-bunch mode number is\\ m = 0.
    The lower graph (c) displays as zoom the baseband information of (b) depicted as    yellow rectangle.}
    \label{fig:TFD-MultiBTenMode0}
 \end{center} 
\end{figure}

Finally in \Fref{fig:TFD-MultiBTenMode0} we display the first mode of multi-bunch oscillations. We have all bunches oscillate at the same frequency (the betatron tune $q=0.25\ $) with the \textbf{same} oscillation phase or $\Delta\Phi = 0$ in 
\Eref{eq:TFD-dphi2}. Following \Eref{eq:TFD-dphi} this corresponds to the multi bunch mode number m=0. In literature this mode of oscillation is called \textbf{$\sigma$-mode}.
As described before the spectral information displayed in the b-part is redundant around each RF-frequency harmonic, so we will in the following examples concentrate on the baseband information, which is displayed as zoom in the c-part of \Fref{fig:TFD-MultiBTenMode0}.
The same spectral information of \Fref{fig:TFD-MultiBTenMode0}c is obtained by measuring \textbf{one} bunch in a \textbf{ten times smaller accelerator} oscillating at a \textbf{ten times smaller betatron tune}, compare \Fref{fig:TFD-MultiBOneunstable}b. The important difference is that due to the large number of bunches the betatron wave is much finer sampled.
\newpage
\subsubsection{Ten bunches - oscillation mode 1}
\begin{figure}[!ht]
  \begin{center}
    \includegraphics[width=0.95\textwidth]{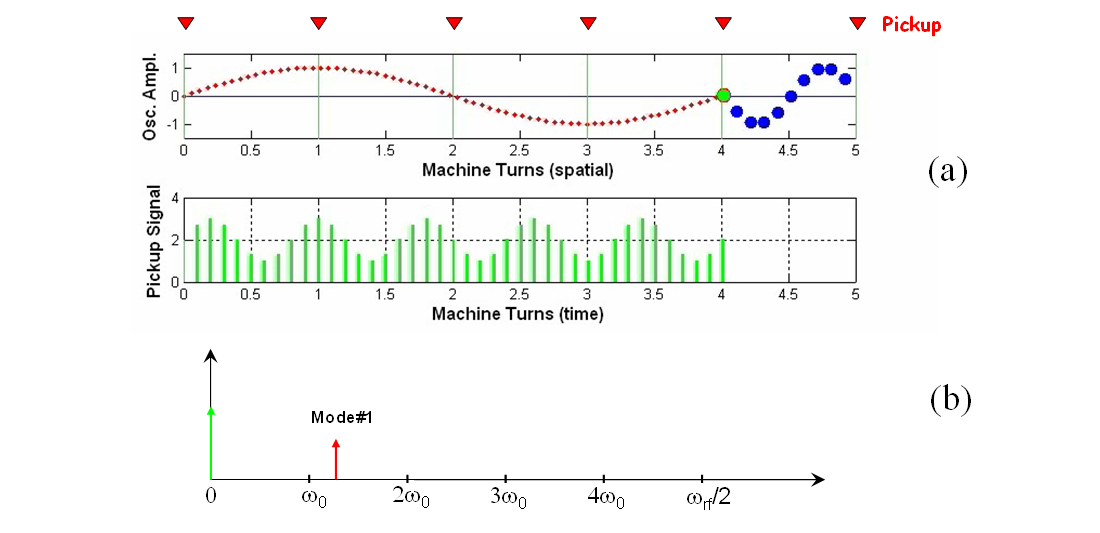}
    \caption{time-domain (a) and baseband frequency-domain (b) information of ten bunches
    oscillating at the betatron tune over five complete turns. The bunches oscillate with a phase advance $\Delta\Phi = 2\pi /10$, i.e. the multi-bunch mode number is m = 1.}
    \label{fig:TFD-MultiBTenMode1}
 \end{center} 
\end{figure}
With multimode number \textcolor{red}{m=1} the bunches oscillate with a relative phase advance of\\
$\Delta\Phi = \textcolor{red}{1}\cdot 2\pi /10$ relative to each other. Through this relativ phase advance the sampling of the betatron wave by the ten bunches make it look like a single oscillation
at a higher frequency. Hence the m=1 mode appears as side-band of the first revolution harmonic $\omega_0$.(\Fref{fig:TFD-MultiBTenMode1}b).
\subsubsection{Ten bunches - oscillation mode 2, 3 and 4}
\begin{figure}[!ht]
    \centering
    \begin{minipage}{0.32\textwidth}
        \centering
        \includegraphics[width=0.95\textwidth]{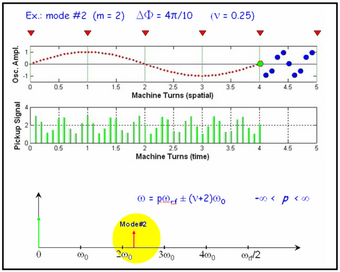} 
        \caption{mode = 2}
        \label{fig:TFD-MultiB_Mode2 }
    \end{minipage}\hfill
    \begin{minipage}{0.32\textwidth}
        \centering
        \includegraphics[width=0.95\textwidth]{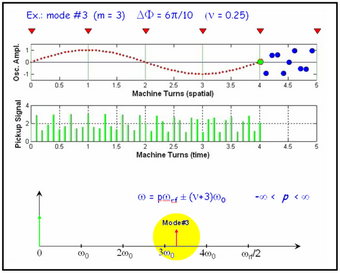} 
        \caption{mode = 3}
        \label{fig:TFD-MultiB_Mode3 }
    \end{minipage}
        \begin{minipage}{0.32\textwidth}
        \centering
        \includegraphics[width=0.95\textwidth]{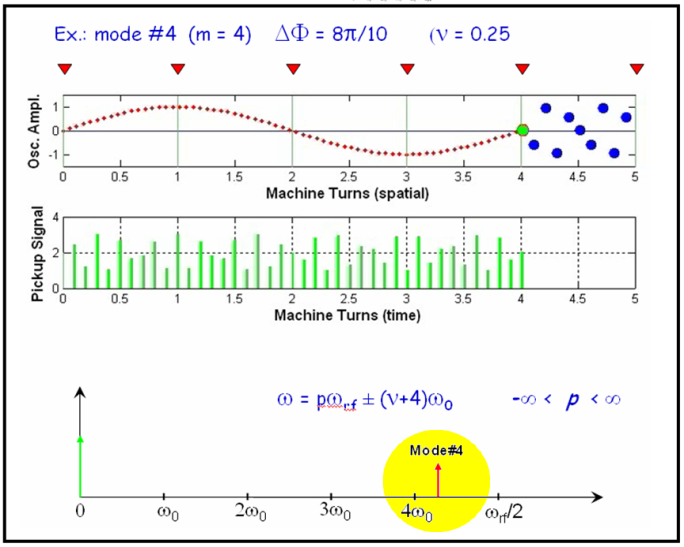} 
        \caption{mode = 4}
        \label{fig:TFD-MultiB_Mode4 }
    \end{minipage}
\end{figure}
In the higher modes the phase advance between the oscillation of the bunches is larger and larger. Hence the sampled betatron wave appears to be at a higher and higher frequency and shows up in the spectrum as the upper side-band of the higher revolution harmonics.
(yellow circles in the three figures).
\subsubsection{Ten bunches - oscillation mode 5}
\begin{figure}[!ht]
  \begin{center}
    \includegraphics[width=0.95\textwidth]{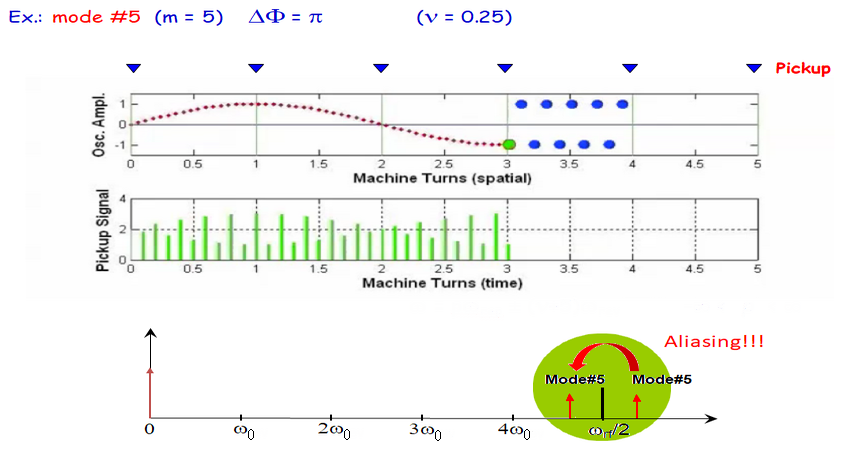}
    \caption{time-domain (a) and baseband frequency-domain (b) information of ten bunches
    oscillating at the betatron tune over five complete turns. The bunches oscillate with a phase advance $\Delta\Phi =\textcolor{red}{5}\cdot 2\pi /10 =\pi$, i.e. the multi-bunch mode number is m = 5.}
    \label{fig:TFD-MultiBTenMode5}
 \end{center} 
\end{figure}
This mode is called in literature due to the very specific phase advance of $\Delta\Phi =\pi$ between consecutive bunches also the \textbf{$\pi$-mode} of oscillation. Following bunches oscillate with opposite phases and this leads to the highest frequency components of the sampled betatron wave. Following the logic of the previous sections, this would lead to the upper side-band of the revolution harmonics at $\omega = \omega_{RF}/2+\nu\cdot\omega_0$. But this frequency is above half the sampling frequency $\omega_{RF}/2$! Hence aliasing will occur (see \Sref{sec:aliasing}) and we measure the mirror frequency
$\omega_{RF}-\left[\omega_{RF}/2\ \textcolor{red}{\textbf{+}}\ \nu\cdot\omega_0\right] =\omega_{RF}/2\ \textcolor{red}{\textbf{-}}\ \nu\omega_0$, which is the lower side-band.
The situation is shown in \Fref{fig:TFD-MultiBTenMode5} and the effect of aliasing is highlighted with the green circle.
\subsubsection{Ten bunches - oscillation mode 6-9}
The following higher modes are shown for completeness; they do not bring much new information. One clearly sees that the betatron wave gets sampled at even higher frequencies, but due to aliasing the side-bands appear now in reverse order as lower side-bands of the corresponding revolution harmonics. 
\begin{figure}[!ht]
    \centering
    \begin{minipage}{0.49\textwidth}
        \centering
        \includegraphics[width=0.95\textwidth]{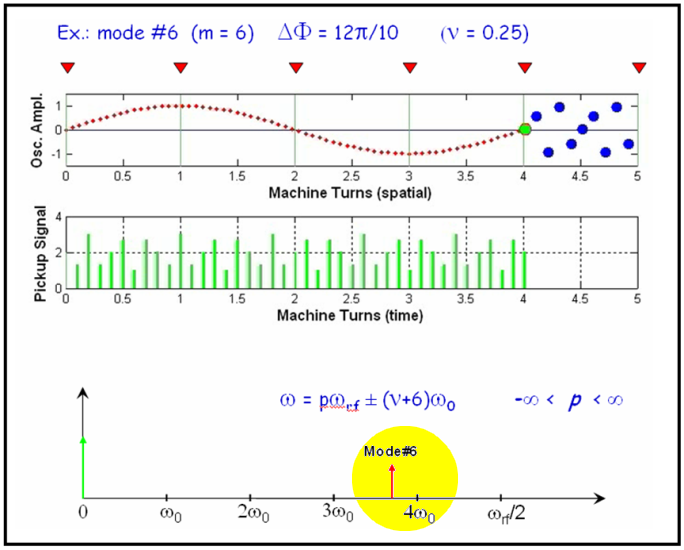} 
        \caption{mode = 6}
        \label{fig:TFD-MultiB_Mode6 }
    \end{minipage}\hfill
    \begin{minipage}{0.49\textwidth}
        \centering
        \includegraphics[width=0.95\textwidth]{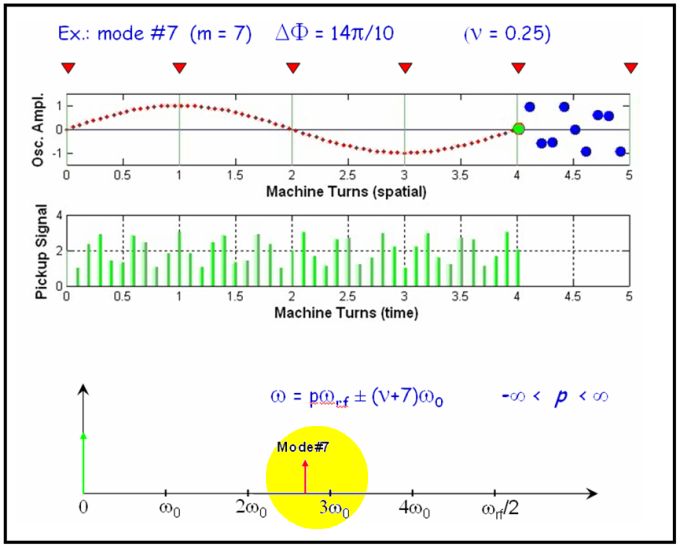} 
        \caption{mode = 7}
        \label{fig:TFD-MultiB_Mode7 }
    \end{minipage}
\end{figure}
\begin{figure}[!ht]
    \centering
    \begin{minipage}{0.49\textwidth}
        \centering
        \includegraphics[width=0.95\textwidth]{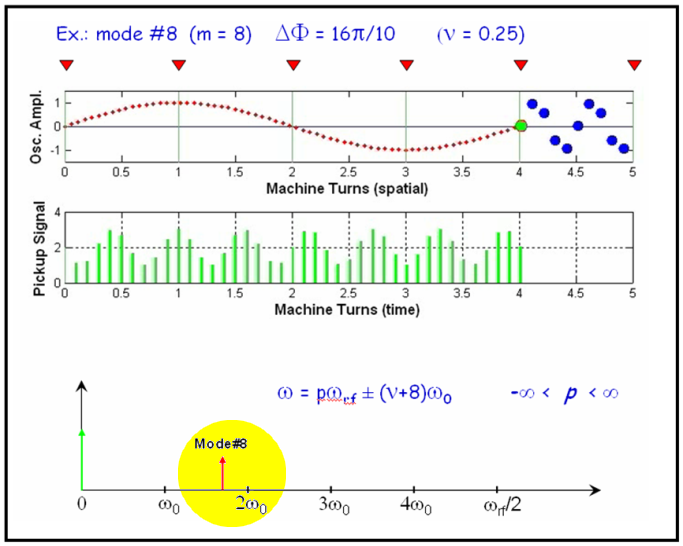} 
        \caption{mode = 8}
        \label{fig:TFD-MultiB_Mode8 }
    \end{minipage}\hfill
    \begin{minipage}{0.49\textwidth}
        \centering
        \includegraphics[width=0.95\textwidth]{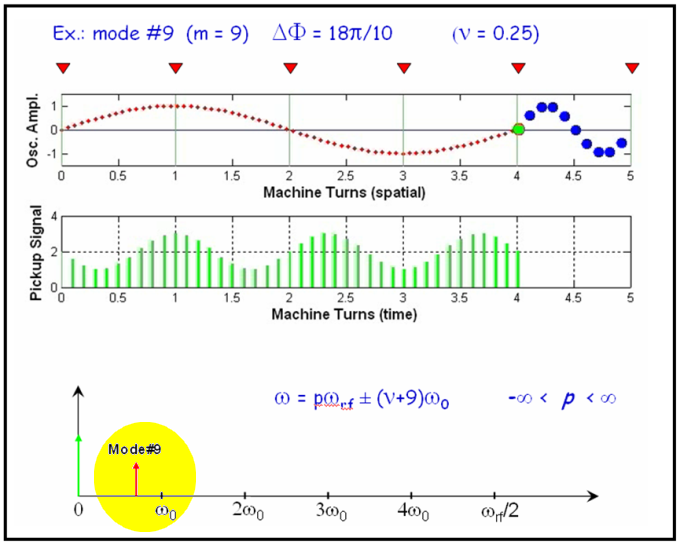} 
        \caption{mode = 9}
        \label{fig:TFD-MultiB_Mode9 }
    \end{minipage}
\end{figure}
\subsubsection{Ten bunches - oscillation mode summary}
\begin{figure}[!ht]
  \begin{center}
    \includegraphics[width=0.95\textwidth]{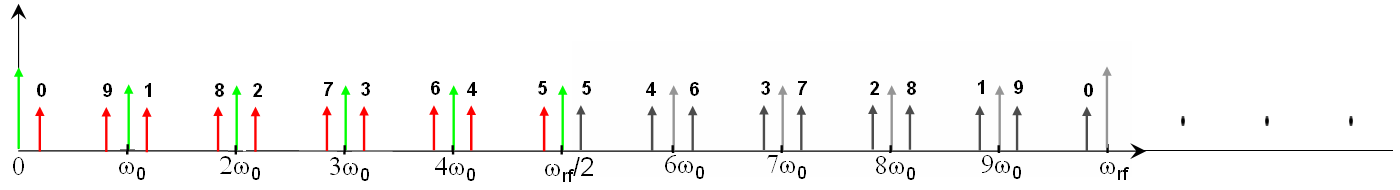}
    \caption{Summary of all side-bands with their corresponding multi-bunch oscillation mode numbers. In color: Baseband spectral lines.}
    \label{fig:TFD-MultiBsummary}
 \end{center} 
\end{figure}

The appearance of all multi-bunch oscillation modes can be summarized in \Fref{fig:TFD-MultiBsummary}. The lower mode numbers appear as upper side-bands of the revolution harmonics up to half the sampling frequency (= RF frequency), then the higher mode numbers appear as lower side bands in reverse order. It should also be mentioned, that of course all these side-bands repeat around all higher harmonics of the RF frequency and they can be used for measurements equally well (using down-mixing equipment for example).
\subsection{Multi-Bunch Oscillations: Relevance}
\begin{figure}[!ht]
  \begin{center}
    \includegraphics[width=0.5\textwidth]{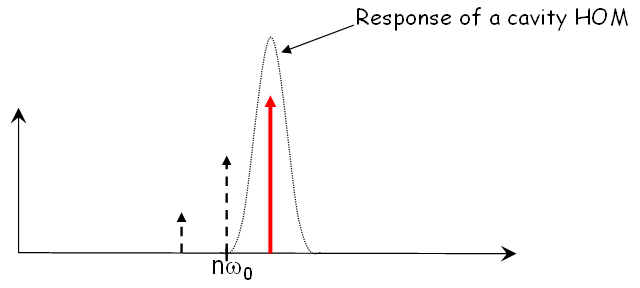}
    \caption{Situation to be avoided in the design of a particle accelerator: The resonance frequency of a high-order-mode of a cavity (HOM) coincides with a multi-bunch oscillation mode.}
    \label{fig:TFD-MultiBHOM}
 \end{center} 
\end{figure}
One might ask for what reason so much effort is put into understanding and classifying these multi-bunch oscillation modes?
The real particle beam will of course oscillate at any linear combination of these modes and hence we can see from \Fref{fig:TFD-MultiBsummary} that a dense "carpet" of possible oscillating frequencies covers the whole spectral range. This is in particular important for larger accelerators, since with a long revolution time the revolution frequency is low and therefore the distance between the revolution harmonics becomes very small. For example in the LHC with $T_{rev}= 89\mu sec$ the distance between revolution harmonics is only about 11 kHz. When designing accelerator components one must make sure that there are no narrow band impedance sources (like RF cavities) introduced, where High-order-mode (HOM) resonances have the frequency of one of the revolution harmonics.
Otherwise the particle beams will be stimulated by the presence of this resonator to become unstable at exactly this multi-bunch oscillation mode.
The situation to avoid is depicted in \Fref{fig:TFD-MultiBHOM}.

\subsection{Multi Bunch Oscillations: Measurement Examples}
\begin{figure}[!ht]
    \centering
    \begin{minipage}{0.49\textwidth}
        \centering
        \includegraphics[width=0.95\textwidth]{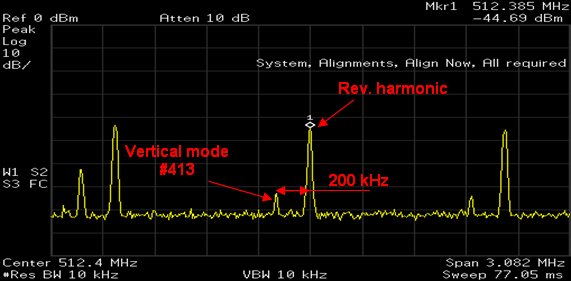} 
        \caption{Spectral line at 512.185 MHz. Lower side-bands of 2 $f_{RF}$, 200 kHz apart from the 443 rd revolution harmonic: vertical mode number 413}
        \label{fig:TFD-MultiBEXver}
    \end{minipage}\hfill
    \begin{minipage}{0.49\textwidth}
        \centering
        \includegraphics[width=0.95\textwidth]{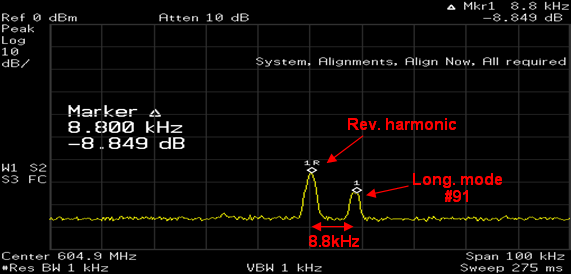} 
        \caption{Spectral line at 604.914 MHz. Upper side-band of  $f_{RF}$, 8.8 kHz apart from the $523^{rd}$ revolution harmonic: longitudinal mode number 91}
        \label{fig:TFD-MultiBEXlong}
    \end{minipage}
\end{figure}

Last not least two measurement examples are shown; one in the vertical plane, one from the longitudinal plane (\Fref{fig:TFD-MultiBEXver} and \Fref{fig:TFD-MultiBEXlong}). Both measurement were made at the ELETTRA synchrotron with the following parameters:
$f_{RF}$ = 499.654 MHz, bunch spacing $\simeq 2nsec$, 432 bunches, revolution frequency $f_0$=1.15 MHz, $\nu_{VER}$ = 0.17 (= 200 kHz), $\nu_{long}$ = 0.0076 (=8.8 kHz).


\section{Analysis of Stationary Spectra}

In many cases the physics process which we want to observe is stationary, which means that we get the same spectral information no matter at what moment we take our measurement samples. Our interest can be in observing the multitude of spectral lines or it could be that we want to measure the frequency of one spectral line with the highest possible accuracy. The following section will have a closer look at the frequency resolution of a measured spectrum.

\subsection{Methods to improve the frequency resolution}
\label{sec:TFD-resolution}

Referring back to \Sref{sec:aliasing} we know that the frequency resolution
$\Delta f$ of a DFT of $N$ samples measured at a sampling frequency $f_s$ is
$\Delta f = 2\cdot f_s/N$. So the most straightforward method to improve the frequency resolution is to increase the number of samples. But this is not always possible, sometimes limited by limited amount of storage capacity in the front-end sampler or by the fact that the observed physical process only lasts for a limited amount of time.
So we should be looking at ways to improve the frequency resolution by keeping the number of samples finite and constant.

\subsubsection{Effect of data windowing: spectral broadening}
\label{sec:windowing}

\Fref[b]{fig:TFD-P31} illustrates what actually happens at the moment when we limit our measurement of a continuous periodic signal to a finite set of samples.
\begin{figure}[!ht]
  \begin{center}
    \includegraphics[width=0.95\textwidth]{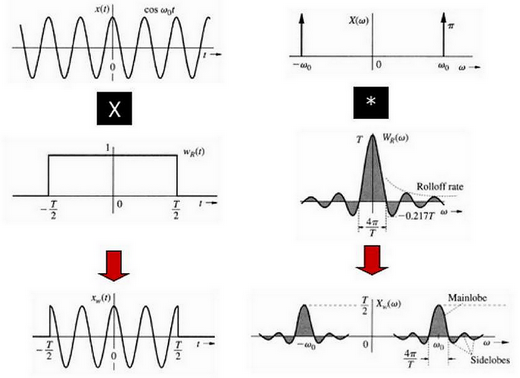}
    \caption{effect of data windowing on the spectrum of a continuous sin-wave}
    \label{fig:TFD-P31}
 \end{center} 
\end{figure}

The process of "looking at the continuous input signal" only for a given time span is like looking through a time-window, or applying a time-window function to the input data.
Therefore this process is called "windowing". Mathematically we can express this in time-domain as the multiplication of the continuous input signal with the window function, which has the value "0" outside the observation time and the value "1" during the observation. Again the convolution theorem (see \Sref{sec:TFD-convolution}) tells us how to obtain the effect of data windowing in frequency domain: The resulting spectral information is the convolution of the spectrum of the continuous input signal with the spectrum of the data window. Going back to \Fref{fig:TFD-P31} we have as continuous input signal a sine wave, the Fourier transform we know is a single line at the frequency of the sine wave. The Fourier transform of a rectangular pulse is basically a sinc-function (see \Eref{eq:TFD-mod_ck} in section \Sref{sec:TFD-mbmp}). Hence the resulting spectral information after windowing is a broadened spectral line, often also called
"spectral leakage". If the length of the data window is $T$ seconds, the width of the main lobe of the sinc function is $\frac{4\pi}{T} [sec^{-1}]$. Obviously, by measuring for a longer time $T$, the effect of the data window disappears until we get for an infinitely long observation time again the single spectral line of the sine wave.

\begin{figure}[!ht]
  \begin{center}
    \includegraphics[width=0.8\textwidth]{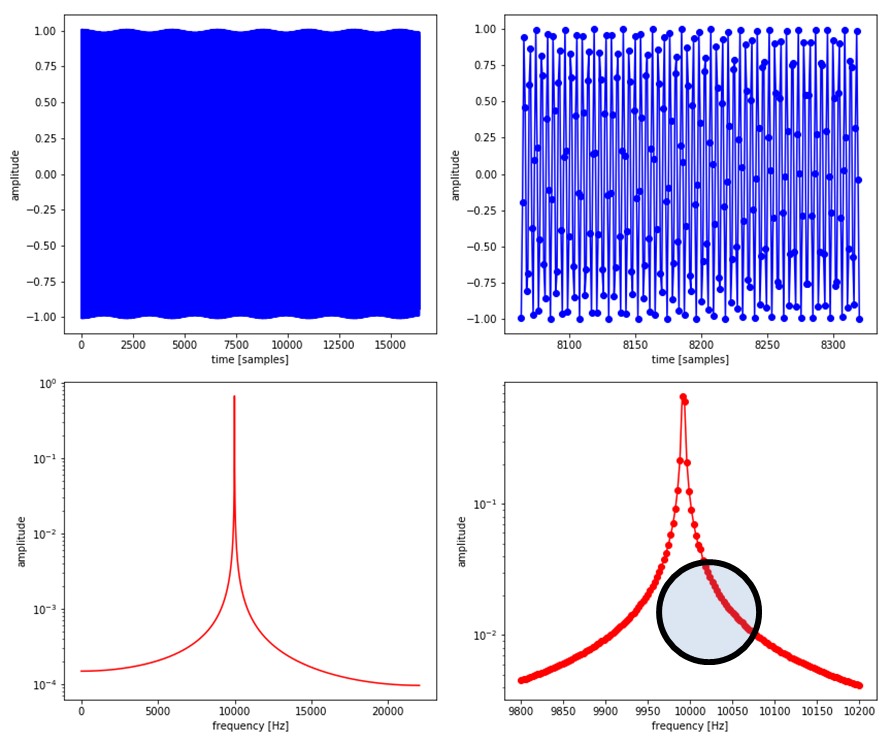}
    \caption{observation of two signals (one large amplitude, one small amplitude) through a rectangular data window}
    \label{fig:TFD-P32}
 \end{center} 
\end{figure}
\subsubsection{Effect of data windowing: masking of small signals}

The above case of having a single sine wave as input signal is certainly a very specific case. A more important aspect of data windowing becomes visible in case we have more than one spectral line in our measurement. we shall examine a signal composed of two spectral lines:
\begin{align}
A\ &=\ A_1\sin{2\pi\omega_1 t} + A_2\sin{2\pi\omega_2 t}\notag\\
A_1\ &=\ 1;\ A_2\ =\ 0.01\quad\omega_1= 9990;\ \omega_1= 10010
\end{align}

\Fref[b]{fig:TFD-P32} illustrates the case for a dominant large amplitude signal $A_1$ and a second signal at a slightly higher frequency with only 1\% signal amplitude 
$A_2$. On the left side of \Fref{fig:TFD-P32} we see the full data set (16384 samples) and the resulting Fourier transform, on the right side a zoom in time-domain as well as in frequency domain. The black circle in the spectrum indicates the area in which we would expect to see signal peak from the 1\%-component at 10010 [Hz]. But the expected information is completely masked by the spectral leakage of the signal at 9990[Hz] due to the data windowing. In time domain the small amplitude modulation is still visible.

\subsubsection{Various data window shapes}

How can the situation be improved? We would like to have less spectral leakage. So basically we are looking for a data window, for which the Fourier transform has a more narrow main lobe. The so far used "rectangular data window" suffers very much from the steep transition "off" directly to the "100\% on". So in the past people have tried several "multiplicative functions" with smooth transitions from "off" to "on".

\begin{figure}[!ht]
  \begin{center}
    \includegraphics[width=0.8\textwidth]{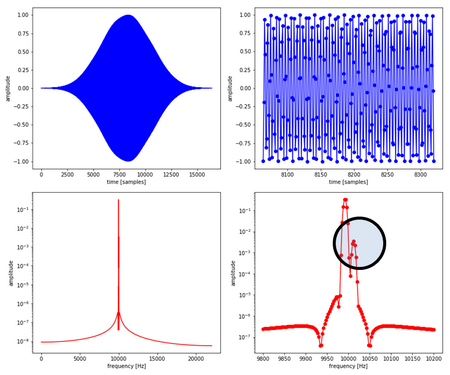}
    \caption{observation of two signals (one large amplitude, one small amplitude) through a so called Blackman-Harris data window}
    \label{fig:TFD-P33}
 \end{center} 
\end{figure}
\Fref[b]{fig:TFD-P33} makes use of such a smooth transition by applying a data window of the following form (Blackman-Harris window):
\begin{align}
w(n) &= a_0 - a_1\cos{\frac{2\pi n}{N}} + a_2\cos{\frac{4\pi n}{N}} - a_3\cos{\frac{6\pi n}{N}}\notag\\
a_0 &= 0.35875\quad a_1=0.48829 \quad a_2=0.14128 \quad a_3=0.01168
\end{align}
On the left side of \Fref{fig:TFD-P33} we see again the full input data set (multiplied with the data window) and its Fourier transform. On the right side we see again a zoom
in time- and frequency domain. Now the spectral leakage is much smaller, the dynamic range around the peak is extended and we clearly can resolve the spectral information coming from the small amplitude component.

\begin{figure}[!ht]
  \begin{center}
    \includegraphics[width=0.8\textwidth]{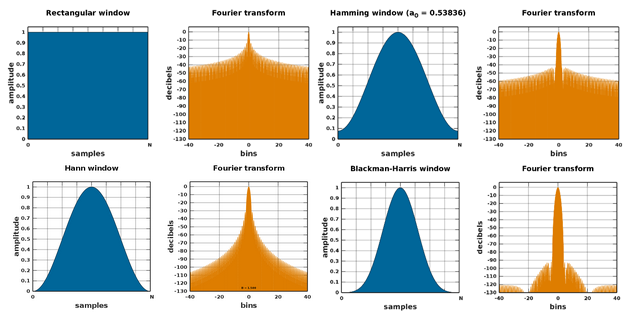}
    \caption{Different data windows and their Fourier transform taken from
    \cite{bib:TFD-datawindows}}
    \label{fig:TFD-P34}
 \end{center} 
\end{figure}

In literature one can find many different families of data windows, all with slightly different optimization criteria (see for example \cite{bib:TFD-datawindows}).
The above used Blackman-Harris window is part of the Hamming family, in which one tries to minimize side-lobe levels by adding more shifted sinc functions.
One thing should be clear, besides all beneficial effects of various data windows, all have one common problem: amplitude loss. Since we multiply in particular the first and last data samples with a weight close to zero, we are losing total signal power in our data sample. In particular cases, when frequency resolution is not the major concern, but for instance signal to noise ratio is a major concern, one better takes a rectangular data window. Modern spectrum analysers have a multitude of data windows implemented in their software, so one should try the one that suits best the needs.
For comparison four different data windows and their Fourier transform have been taken from \cite{bib:TFD-datawindows} and are shown in \Fref{fig:TFD-P34}.

\subsubsection{Interpolation between frequency bins}

In the previous sections we were mainly concerned with spectral leakage due to different data windows and did not answer the question on how to obtain the most accurate frequency measurement. This is in particular relevant for small number of data samples $N$, since the Fourier spectrum only contains N/2 values in the frequency range $[0...f_s/2]$.
So what is the right frequency? Is it the frequency of the bin with the highest amplitude?
No, we can do better:
\begin{figure}[!ht]
  \begin{center}
    \includegraphics[width=0.6\textwidth]{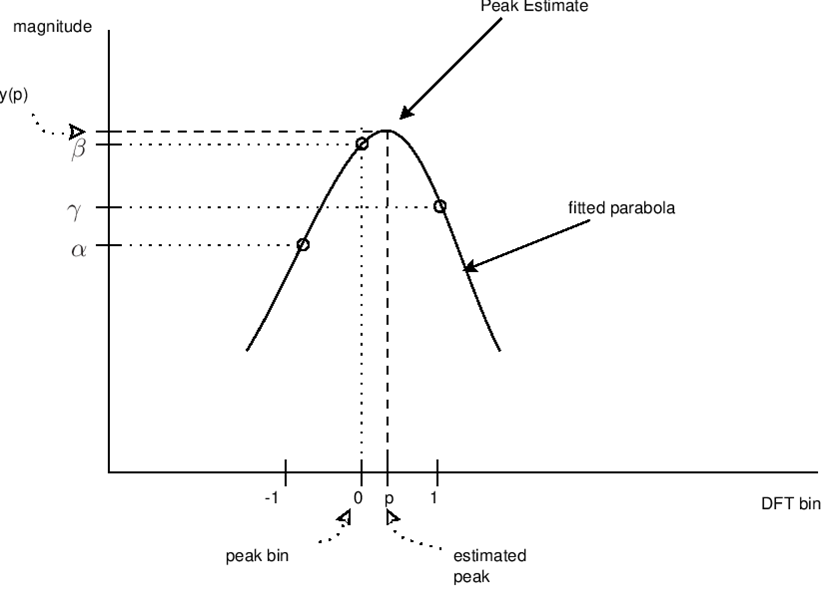}
    \caption{Interpolation around the amplitude maximum of a spectral line}
    \label{fig:TFD-P35}
 \end{center} 
\end{figure}
\Fref[b]{fig:TFD-P35} illustrates the situation. In the shown example there are three measured amplitudes, the middle one with the largest amplitude. But since the measurement on the right side of the peak value is higher than on the left side, the measured frequency will be slightly higher than the value of the bin with the maximum amplitude.
We can obtain a higher accuracy by interpolation between the three points.
We will limit the discussion to "three point interpolation methods", for which we distinguish between a simple parabolic interpolation between the three measured values or the so called "Gaussian interpolation", for which we interpolate with a parabola the log-values of the measured spectral values. Interpolations with more points can also be used, but one has to be careful for example with the purity of the spectral information and make sure that there is no signal from neighbouring frequency lines influencing too much the off-peak measurements.

\begin{figure}[!ht]
  \begin{center}
    \includegraphics[width=0.95\textwidth]{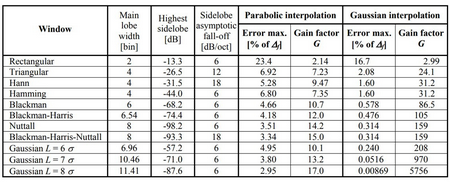}
    \caption{Overview of the major data windows, their characteristics and the possible interpolation gain $G$ for accurate frequency determination. (Taken from\cite{bib:TFD-gasior3})}
    \label{tab:TFD-T1}
 \end{center} 
\end{figure}
\begin{figure}[!ht]
  \begin{center}
    \includegraphics[width=0.8\textwidth]{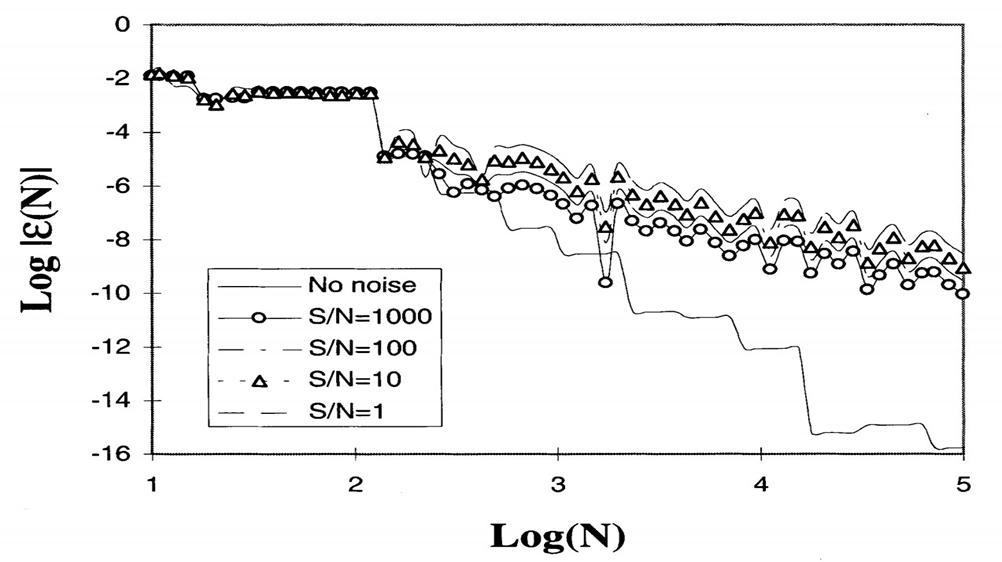}
    \caption{DFT frequency resolution $\epsilon (N)$ as function of the number of samples $N$ for various signal to noise (S/N) ratios. Plot taken from \cite{bib:TFD-bartolini} }
    \label{fig:TFD-36}
 \end{center} 
\end{figure}

As we have seen in the previous sections, the shape of the spectrum around each line depends on the data window that we are using. In consequence the gain in accuracy that we can get will depend on the choice of the window.
The details are given in \Tref{tab:TFD-T1} in which the data windows are characterized by the main lobe width, the highest side lobe level and side lobe asymptotic fall-off. The maximum residual interpolation error is given a percentage of the spectrum bin spacing
$\Delta f$. A number, which nicely quantifies the possible gain in accuracy by interpolation is given through the gain factor $G$, which is defined as:
\begin{equation}
\rm{Gain\ factor}\ G:=\frac{\Delta f}{2\cdot \rm{error\ max}}
\end{equation}
At the end a word of warning. \Fref[b]{tab:TFD-T1} shows for some data windows in particular for the Gaussian extrapolation very large gains in accuracy. One should really stress that such accuracy can only be achieved if the signal to noise ratio is very high and if there are no neighbouring spectral lines entering the game.
The best frequency resolution in a typical accelerator application (tune measurement) have been studied in \cite{bib:TFD-bartolini}. In particular the gain in resolution by interpolation has been studied for different signal to noise ratios.
The result is shown in\Fref{fig:TFD-36} in which in double logarithmic scale the frequency resolution $\epsilon (N)$ is plotted against the number of samples $N$.

We clearly see that in the case of a small signal to noise (S/N) ratio the scaling with $N$ is close to the simple $1/N$ scaling one obtains by taking as frequency the value of the bin with the highest amplitude down to a $1/N^2$ scaling by interpolation and a good S/N ratio.

\begin{importantbox}
By taking as measured value for the frequency of a signal simply the frequency of the bin with the highest amplitude we get a frequency resolution $\Delta f\propto 1/N$ ($N$ measurement samples). By using proper data windowing and interpolation we can get close to a frequency resolution scaling of $\Delta f\propto 1/N^2$
\end{importantbox}

\subsubsection{Some do even better: data fitting with the NAFF algorithm}

So far we have made no assumption on the time domain function, which we want to analyse.
But in some cases we know based on modelling the shape of the time domain function.
In that case we can use the measured samples and perform a fitting algorithm in order to obtain a very precise estimate of the model parameters.

Today we find this method implemented as NAFF algorithm (:= Numerical Analysis of Fundamental Frequencies) and it goes back to work of J.~Laskar \cite{bib:TFD-Laskar}, who called the method FMA (:= Frequency Map Analysis). The NAFF code was developed by J.~Laskar et al. in the Nineties as a method to analyse chaotic dynamical systems by the Numerical Analysis of the Fundamental Frequencies in the sampled signal.
\begin{figure}[!ht]
  \begin{center}
    \includegraphics[width=0.9\textwidth]{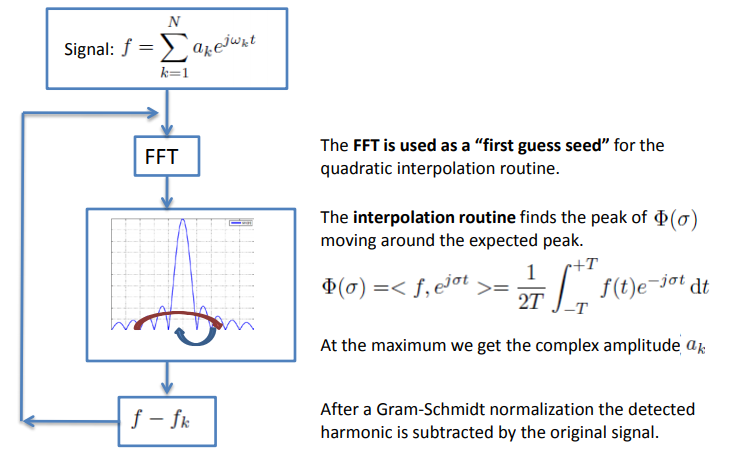}
    \caption{Illustration of the fitting process used in the NAFF algorithm. Graphics taken from \cite{bib:TFD-biancacci} }
    \label{fig:TFD-37}
 \end{center} 
\end{figure}
With the help of \Fref{fig:TFD-37} we can describe the NAFF algorithm in the following way:
\begin{itemize}
\item{
Assume a model function for the observed signal. In the best case a sum of few individual spectral lines.
}
\item{
Obtain the dominant spectral line by a FFT algorithm. Use this single spectral line (and the side lobes depending on the data window) as seed for a fitting algorithm.
}
\item{During the fitting vary slightly the frequency of the fundamental line until maximum overlap with the measured data is found.
}
\item{Subtract the spectrum of the most fundamental line from the measurement and proceed in a following iteration with the next higher larger spectral components.
}
\end{itemize}
So comparing this method with the previously outlined "three-point extrapolation" one clearly sees that more data points are involved in obtaining the best estimate for the frequency. On the downside this increases of course the sensitivity to additional noise in the data. Theoretically one can obtain a $1/N^4$ scaling with the NAFF algorithm, but such excellent performance can only be achieved for noise free data, which in general is only available in computer simulations. For measurements in an accelerator environment with ordinary signal to noise ratios (40...60 dB), also the NAFF algorithm does not yield a better performance than $1/N^2$ scaling.

\subsection{Special cases without spectral leakage}

\subsubsection{Rational ratio between measured frequency and sampling frequency}

\begin{figure}[!ht]
  \begin{center}
    \includegraphics[width=0.95\textwidth]{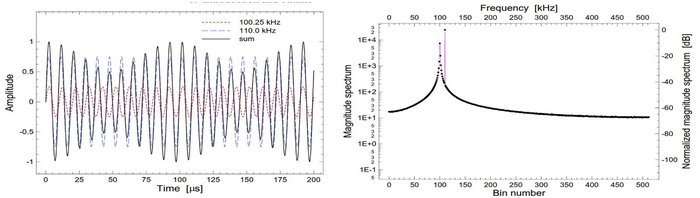}
    \caption{time and frequency data of two signals (100.25 kHz and 110 kHz) sampled at 1024 kHz}
    \label{fig:TFD-3839}
 \end{center} 
\end{figure}

\Fref[b]{fig:TFD-3839} shows as an example two sampled sine waves with frequencies of 100.25 kHz (red dotted line) and 110 kHz (blue dashed line) sampled at 1024 kHz.
In the spectrum we see the lower frequency with the expected spectral leakage from a rectangular data window, whereas the higher frequency appears as single line!
How can this be explained?
\begin{figure}[!ht]
  \begin{center}
    \includegraphics[width=0.9\textwidth]{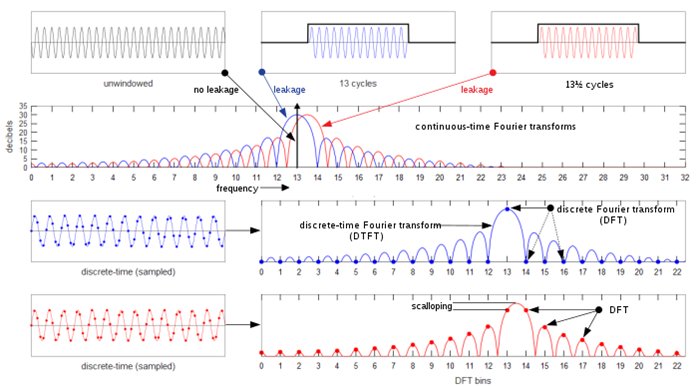}
    \caption{Graphical representation of the sampling and DFT of a continuous sine wave (black graphs) captured \textbf{exactly} over 13 full cycles of the sine wave (blue graphs) or captured over 13.5 cycles (red graphs)}
    \label{fig:TFD-40}
 \end{center} 
\end{figure}
\Fref[b]{fig:TFD-40} gives the answer in a graphical way. In order to make the effect visible much fewer data points are used.

A continous sine wave is shown with its theoretical Fourier transform as black graphs in the figure.
In blue we show the same sine wave captured exactly over 13 full cycles whereas the red graphs show the sine wave sampled over 13.5 cycles.
For the blue and red case we show the results of the discrete-time Fourier transform (DTFT), which are identical in shape, but in the "blue" case of sampling exactly 13 full cycles only one point of the DTFT (the maximum) has a value different from zero, whereas the other values correspond exactly to the "0" values of the side-lobe sinc-function.
In the red case the spectrum has the same shape, but here the DTFT points fall onto the maxima of the side-lobes of the sinc-function.

This explains the phenomenon described in the previous \Fref{fig:TFD-3839}.
If the measured frequency and the sampling frequency have a fixed ration, notably the ratio of two natural numbers, then the sampling and Fourier transformation will happen without any spectral leakage and artefacts.
In the given example of \Fref{fig:TFD-3839}, the second wave has a fixed ratio of 110/1024 with the sampling frequency and can hence be measured without spectral leakage.

\subsubsection{I-Q sampling}

The example shown above looks somehow artificial, since it was by pure chance that one of the signals to be measured had a rational ratio with the sampling frequency, hence we could measure it without spectral leakage.

But in the accelerator domain we very often know exactly the frequency of a signal to be measured, we are only interested in measuring exactly the amplitude and phase of the signal. Very often accelerator signals have a fixed ratio with the RF-frequency, which determines the time structure of all bunched beams.

The most commonly used technique to measure amplitude and phase of a signal with known frequency is called I-Q sampling and it is explained as follows:

We describe the signal to be measured as 
\begin{equation}
y(t)\ =\ \textcolor{blue}{A}\cdot\sin (\omega t +\textcolor{blue}{\phi_0})
\label{eq:TFD-IQ1}
\end{equation}
$\textcolor{blue}{A, \phi}$ to be measured.
We rewrite \Eref{eq:TFD-IQ1} as
\begin{align}
y(t)\ &=\ \textcolor{blue}{A}\cos\textcolor{blue}{\phi_0}\cdot\sin (\omega t) +
\textcolor{blue}{A}\sin\textcolor{blue}{\phi_0}\cdot\cos (\omega t)\notag\\
y(t)\ &=\ I\cdot\sin (\omega t) + Q\cdot\cos (\omega t)
\label{eq:TFD-IQ2}
\end{align}
with $I$ defined as "Inphase"-component and $Q$ as "Quadrature" component ($90^0$) out of phase. It is now sufficient to sample the signal to be measured at \textbf{four} times the signal frequency $\omega/2\pi$ producing periodically four samples $I,Q,-I,-Q$.
From the measured values of $I$ and $Q$ the amplitude and phase of the signal can be computed exactly using:
\begin{equation}
\textcolor{blue}{A}=\sqrt{I^2+Q^2}\qquad
\textcolor{blue}{\phi_0}= \arctan (\frac{Q}{I})
\label{eq:TFD-IQ3}
\end{equation}
\begin{figure}[!ht]
  \begin{center}
    \includegraphics[width=0.7\textwidth]{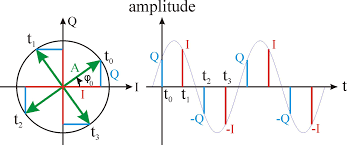}
    \caption{Illustration of I-Q sampling}
    \label{fig:TFD-41}
 \end{center} 
\end{figure}
\Fref[b]{fig:TFD-41} illustrates the process of I-Q sampling. Since the phase between the signal to be measured and the sampling clock is not known, 4 samples per period are necessary. If the phase was exactly known or could be set to "zero", then sampling the input signal exactly at its maximum with one sample per cycle would be sufficient.
But this condition can normally not be met technically, so I-Q sampling with 4 samples per cycle has become a standard.

\subsection{vector network analysis (VNA)}
\label{sec:TFD-VNA}

This section is included as a small outlook to multi-particle dynamics of a whole particle beam. So far we have assumed that all particles in an observed system oscillate with the same frequency and we have treated the beam as a single macro particle oscillating at a particular frequency (betatron tune or synchrotron tune). We have then focused on methods on how to determine this oscillation frequency with the highest possible resolution.

\begin{figure}[!ht]
  \begin{center}
    \includegraphics[width=0.8\textwidth]{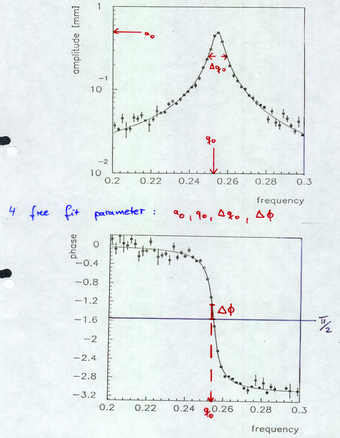}
    \caption{VNA analysis of the LEP particle beam (transverse plane), taken from the LEP-logbook}
    \label{fig:TFD-42}
 \end{center} 
\end{figure}

A real particle beam behaves differently. Particles have different momenta, they experience in consequence different forces and they oscillate with different frequencies.
So even by using observation methods without any spectral leakage, we will experience that a real particle beam has a certain spread in resonance frequencies.
In order to measure the spread in resonance frequencies of a particle beam a method called vector network analysis was performed. The measurement setup consists of a fast magnet, which can excite the beam at frequencies close to its resonance plus a position monitor, measuring the response of the beam. The measurement starts at a frequency well below the resonance, measures the beam response and then continuous by increasing each time the excitation frequency in small steps.

One result from the "early days of LEP (1991)" is shown in \Fref{fig:TFD-42}. The individual measurement points (amplitudes and phases) have been fitted with a resonance curve in order to determine precisely the medium resonance frequency $q_0$, the width of the resonance $\Delta q_0$ and the phase shift $\Delta \Phi$ between the beam exciter and observation point. 

\section{Analysis of Non-Stationary Spectra}

Up to here we have made a very significant assumption: The phenomenon that we want to observe is \textbf{stationary} during the measurement time! This is an important assumption and in the case of very long measurements (i.e $10^6$ samples to get a high frequency resolution) not a very realistic assumption. Furthermore in many cases the measurement serves to monitor beam parameters during dynamic changes of an accelerator (beam acceleration) and under such conditions observed signals are rarely stationary in time.
\begin{figure}[!ht]
  \begin{center}
    \includegraphics[width=0.95\textwidth]{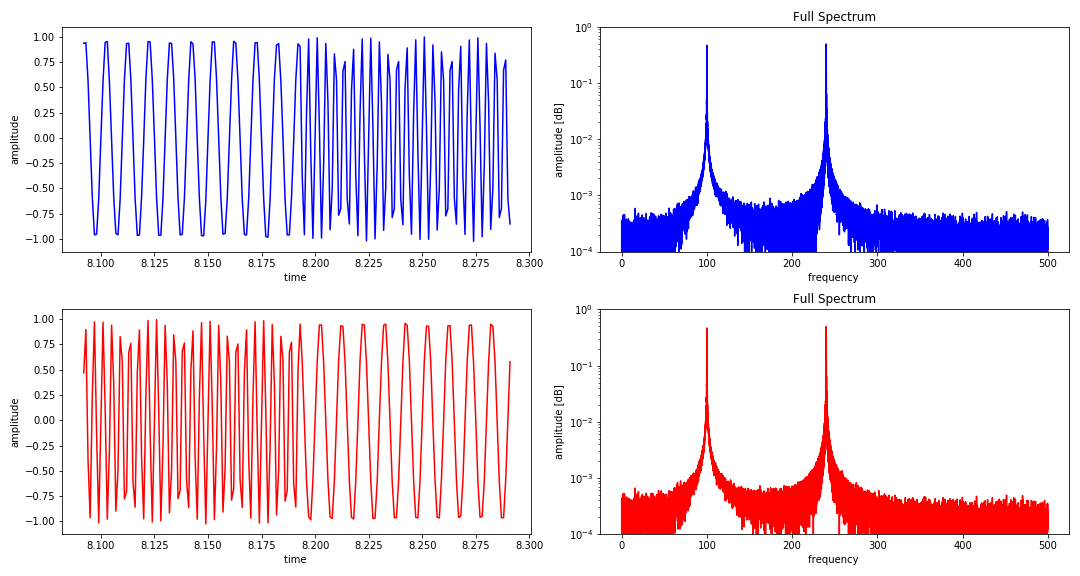}
    \caption{Simple example of an FFT (right) for two consecutive sine-waves in time domain:
    in blue: 100 Hz signal first, then 250 Hz signal; in red: 250 Hz signal first, then 100 Hz signal}
    \label{fig:TFD-43}
 \end{center} 
\end{figure}
\Fref[b]{fig:TFD-43} shows a simple constructed example: The first half of the observation time we have a continuous sine wave, followed by another continuous sine wave in the second half.
In the first case (blue) the lower frequency (100 Hz) is sent first, in the second case (red), the higher frequency (250 Hz) is sent first.
The resulting FFT in both cases is identical, which clearly demonstrates the fact that with a simple FFT we have no time resolution for varying spectral content.

\subsection{Overview of presented tools}

So how do we proceed when the spectral content varies in time?

The most frequently used tool to explore non stationary signals are the so called spectrograms. In that case the observed time period $\Delta T$ is divided into several shorter slices (of equal length) and the data in each slice is subject to a Fourier transform. The resulting sequence of spectra is either displayed as two-dimensional (2D) projection, in which the amplitude information is given by a color code or by a three-dimensional (3D) display often also called "waterfall"-display.
More details are given in \Sref{sec:TFD-spectrogram}.
A single and global Fourier-analysis of an input has no time resolution, whereas  a Spectrogram gives spectral information at regular time intervals. The length of the individual slices determines the frequency resolution that can be obtained. 
Since the time intervals of equal length, the frequency resolution is the same in every slice of the analysis.
\begin{figure}[!ht]
  \begin{center}
    \includegraphics[width=0.95\textwidth]{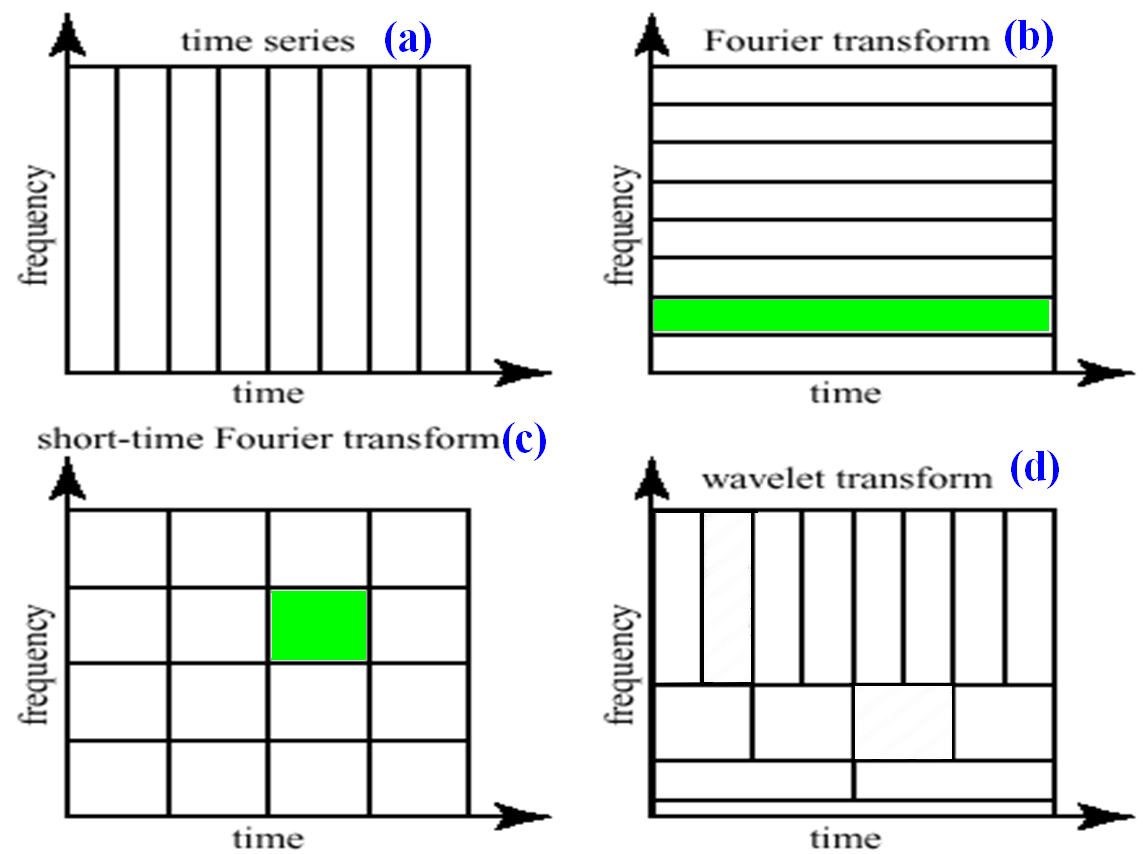}
    \caption{Comparison of the time/frequency resolution of several methods}
    \label{fig:TFD-54}
 \end{center} 
\end{figure}

In most cases this is sufficient, but in specific cases at a short moment in time something interesting might happen and then the frequency resolution of a spectrogram will not be adequate. In such a case a wavelet-analysis can be used.
In a wavelet-analysis the elementary sine-and cosine-functions of a Fourier-transform are replaced by two-dimensional wavelet-functions, yielding a dynamically adapting frequency resolution (see \Sref{sec:TFD-wavelets}).
In \Fref{fig:TFD-54} the frequency resolution, which can be obtained from various methods
is graphically explained. A simple time series of data has of course no frequency information (case a). A Fourier transform of the whole time series gives a frequency resolution, but within the data set no further time resolution is obtained (case b).
A short-time Fourier transform gives time and frequency resolution (case c). Since to first order the frequency resolution is inversely proportional to the length of the time sample, the surface of each rectangle in case b and c is the same (marked in green). Frequency resolution has to be traded in for time resolution. In a wavelet transform the time/frequency resolution depends on the details of the transform. In general high frequencies can be analysed with higher time resolution than low frequencies (case d).

In many cases the spectral information is dominated by a single resonance line ( i.e. betatron- or synchrotron-tune in accelerators) and we only want to know how this quantity changes with time. In such a case we can used a technique called "Phase-locked loop",, which tracks the main resonance as a function of time (see \Sref{sec:TFD-PLLtracking}).

Last not least in some cases we want to observe very short lived phenomena for which we have in a circular accelerator only a few samples ($N$), since we can measure only once per beam revolution. In such cases we can use multiple measurement sensors distributed around the azimuth of the accelerator ($M$ monitors), i.e. we sample in time and space, and by a transformation convert this data-set into a data-set, which looks like recorded at a single point with $N\cdot M$ samples (see \Sref{sec:TFD-TimeandSpaceSampling}).

\subsection{Spectrograms}
\label{sec:TFD-spectrogram}

\begin{figure}[!ht]
  \begin{center}
    \includegraphics[width=0.4\textwidth]{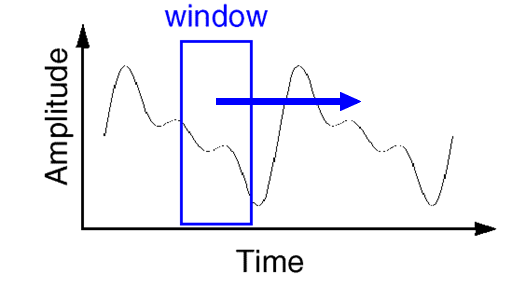}
    \caption{Illustration of a Short Time Fourier Transform}
    \label{fig:TFD-44}
 \end{center} 
\end{figure}

\begin{figure}[!ht]
  \begin{center}
    \includegraphics[width=0.95\textwidth]{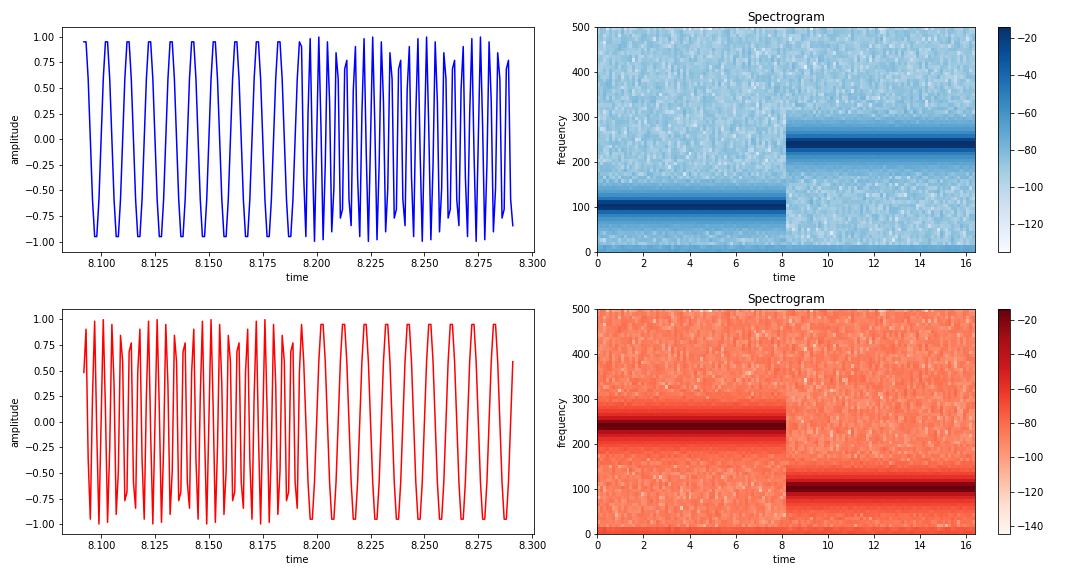}
    \caption{Time domain functions as in \Fref{fig:TFD-43}, but spectral content displayed as Spectrogram}
    \label{fig:TFD-45}
 \end{center} 
\end{figure}
Spectrograms are going back to work of Dennis Gabor in the year 1946 \cite{bib:TFD-gabor}.
In the Gabor-transform or the so called " Short Time Fourier Analysis (STFT)",
a short Gaussian data window is moved along the measured data stream and for each position of the data window the Fourier transform is computed. This way the analysis is always most sensitive to the spectral content at the position of the data window.
Mathematically the Gabor transform $G(\tau ,\omega)$ of the time domain function $x(t)$ can be written as:
\begin{equation}
G(\tau ,\omega)=\int_{\infty}^{\infty} x(t)\cdot  e^{-\pi (t-\tau )^2}\cdot e^{-j\omega t} dt
\label{eq:TFD-gabor}
\end{equation}
As we can see in \Eref{eq:TFD-gabor} we have the well known Fourier transform of the signal $x(t)$ multiplied with a Gaussian-shaped time window (center-time equals to $\tau $).
We can easily see that the width of the Gaussian-shaped time window determines the time resolution of the Gabor analysis. A short data window gives only weight to a small fraction of the time-function $x(t)$, hence giving a good time resolution , but at the price of few residual data points. Enlarging the data window gives more signal, but smaller time resolution.
The window function width can also be varied to optimize the time-frequency resolution trade-off for a particular application by replacing the 
exponent $-\pi (t-\tau )^2$ with $-\pi \alpha (t-\tau )^2$ for some chosen $\alpha $.

A spectrogram could now be simply a display of the individual power spectra of the STFT in a 3-D display or in a 2-D coloured contour plot. On the horizontal axis we have the window centre $\tau $ and on the vertical axis the frequency of the Fourier transform $\omega $.

\begin{figure}[!ht]
  \begin{center}
    \includegraphics[width=0.95\textwidth]{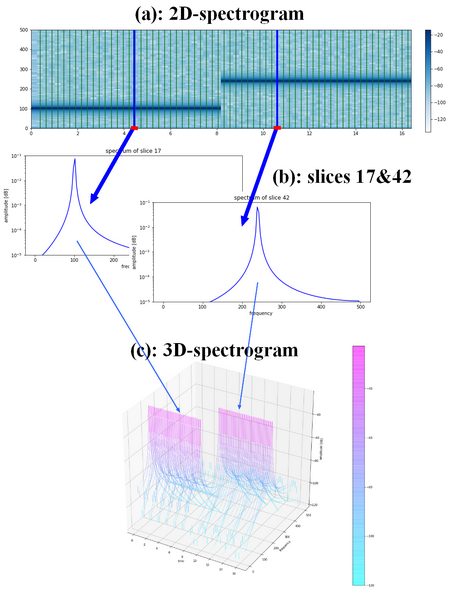}
    \caption{(a): 2D-Spectrogram as in \Fref{fig:TFD-45}, (b): FFT of two individual slices, (c): 3D-Spectrogram (Waterfall-Plot)}
    \label{fig:TFD-46}
 \end{center} 
\end{figure}
For sampled data the sliding data window is replaced by a segmentation of the data into slices and individual Fourier transforms of each slice.

\Fref[b]{fig:TFD-44} illustrates the process of the sliding data window of a STFT (in this case drawn as a rectangle).

\Fref[b]{fig:TFD-45} displays the same time domain signals as in \Fref{fig:TFD-43}, but the spectral content is displayed as 2D-spectrogram, in which the amplitude is encoded as color-map (defined in the color-bar right of the spectrogram). We can clearly see at what moment in time the frequency of the input signal changes the frequency. 

\Fref[b]{fig:TFD-46} illustrates more details of a spectrogram analysis.
In \Fref{fig:TFD-46}a the 2D-spectrogram of \Fref{fig:TFD-45} is repeated. The horizontal axis represents the time of the observation. Each green vertical line in the plot indicates the center of one observation slice. The length of each slice is indicated with the two red bars. In \Fref{fig:TFD-46}b two of the slices are displayed. One clearly sees that due to the shorter length the frequency resolution of these graphs is much less than in \Fref{fig:TFD-43}. Last not least \Fref{fig:TFD-46}c shows the same data as 3D-Spectrogram.

Spectrograms as analysis tool are implemented as complete packages into most modern computing tools. Almost all parameters of the analysis can be given as input. The slices can be adjacent or overlapping. One can apply data windowing to each slice and one can chose the color code for the display. It is worth spending some time in exploring the options "on a rainy afternoon".

\subsection{Wavelet Analysis}
\label{sec:TFD-wavelets}

A generalization of Short-Time-Fourier-Transforms or Spectrograms has been developed due to the need of an analysis tools with time resolution and with variable frequency resolution. 
Wavelet-Analysis goes back to very early work of Alfred Haar (1909)\cite{bib:TFD-Haar}, but the mathematical basis for the use nowadays has been laid by Ingrid Daubechies in the late 1980's \cite{bib:TFD-Daub}. Today wavelet-transform analysis can be found in many software tools like data de-noising and  data compression including the JPEG compression of 2D-data (photos/videos).
For the analysis of data recorded in accelerator domain there are only limited applications, predominantly in the detection of data consistency or filtering.

\bgroup
\def\arraystretch{2}
\begin{table}[!ht]
\centering
\begin{tabular}{|l|l|} 
\hline
\textbf{Transform}  & \textbf{Mathematical Expression}                                                                                                                                                                                                                                                                  \\ 
\hline
Fourier Transform (FT)                                         & $\mathcal{F}_x
\textcolor{red}{ (\omega)}\ = \ \int_{-\infty }^{+\infty } x(t)\cdot \textcolor{red}{e^{-j\omega t}}dt$                                                                                                                                             \\ 
\hline
\begin{tabular}[c]{@{}l@{}}Gabor Transform (STFT)\end{tabular}  & $G(\textcolor{blue}{\tau} ,\textcolor{red}{\omega})=\int_{-\infty}^{\infty} x(t)\cdot \textcolor{blue}
{e^{\alpha \pi (t-\tau )^2}}\cdot \textcolor{red}{e^{-j\omega t}}dt $ \\ 
\hline
\begin{tabular}[c]{@{}l@{}}Wavelet Transform (WT)\end{tabular} & $\Psi_x^{\psi} (\textcolor{blue}{\tau}, \textcolor{red}{s})\ = \frac{1}{\sqrt{s}}\ \int_{-\infty }^{+\infty } x(t)\cdot \psi \left(\frac{t-\textcolor{blue}{\tau}}{\textcolor{red}{s}}\right) dt$                                                                                                                                      \\
\hline
\end{tabular}
\caption{mathematical expressions for FT, STFT and WT. The variables in red color constitute the frequency resolution, the variables in blue color the time resolution.}
\label{tab:TFD-table2}
\end{table}
\egroup
With the help of \Tref{tab:TFD-table2} we can understand the main idea of a wavelet-transform. The table compares the mathematical expressions for all three transforms under discussion. The terms determining the frequency resolution are highlighted in red color, the time resolution is marked in blue color. As we can see the Fourier transform has no further time resolution within the observed function $x(t)$, whereas in the case of a Gabor transform the time resolution is realized by the sliding (Gaussian) data window.
In both cases the function $x(t)$ is analysed by the use of the orthonormal sine- and cosine functions $e^{-j\omega t}$. In the case of the wavelet transform the basis for analysing $x(t)$ are so called "mother-wavelets" $\psi\left(\frac{t-\tau}{s}\right)$, which combine the frequency resolution (scale parameter $s$) and time resolution (scale parameter $\tau$). One can imagine a mother wavelet as a short time limited "bleep" of oscillation, which gets shifted in time by the parameter $\tau$ and stretched in time by 
the parameter $s$. For each instance of a wavelet a correlation with the measured data $x(t)$ is computed. If the correlation is high, this indicates that at this particular time the frequency content of $x(t)$ is close to the frequency content of the wavelet.
Whereas FT and STFT have always sine-and cosine functions as basis, there is a large range of mother wavelets from which one can chose depending on the needs of the application.
A non exhaustive list of wavelets can be found in \cite{bib:TFD-waveletlist}

\begin{figure}[!ht]
  \begin{center}
    \includegraphics[width=0.6\textwidth]{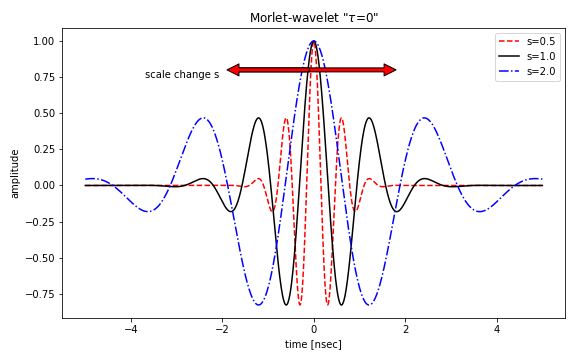}
    \caption{Graphical representation of a Morlet-Wavelet with different scalings $s$}
    \label{fig:TFD-48}
 \end{center} 
\end{figure}

Let us do an example to explain wavelet analysis. First we pick a mother wavelet:
We pick a "Morlet Wavelet".
Its mathematical expression is:
\begin{equation}
\psi_{\rm Morlett}\left(\frac{t-\tau}{s}\right)\ =\ 
e^{-\frac{\left(\frac{t-\tau}{s}\right)^2}{2}}\cdot \cos{\left( 5\left(\frac{t-\tau}{s}\right)\right)}
\label{eq:TFD-morlet}
\end{equation}
with the two free parameters for the scale ($s$) and for the origin ($\tau$).
In the following two figures we show the graph of the Morlet-wavelet. \Fref[b]{fig:TFD-47}
shows the principal shape with $s=1.0$, but with different values of $\tau$ for the origin, whereas \Fref{fig:TFD-48} keeps the origin fixed and varies the scale parameter $s$. And this gives us a direct inside into how the wavelet analysis works:
For every parameter set ($\tau, s$) a correlation coefficient between the data and the wavelet is computed. If the correlation is high, this means that around the time defined by the wavelet center ($\tau$), the data has a high spectral component of the spectral component of the wavelet. Depending on the choice of the wavelet and by varying the scale parameter $s$ different spectral components are probed.
\begin{figure}[!ht]
  \begin{center}
    \includegraphics[width=0.6\textwidth]{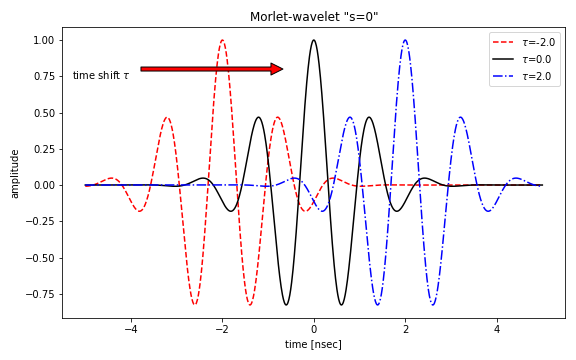}
    \caption{Graphical representation of a Morlet-Wavelet with different origins $\tau$}
    \label{fig:TFD-47}
 \end{center} 
\end{figure}

\begin{figure}[!ht]
  \begin{center}
    \includegraphics[width=0.95\textwidth]{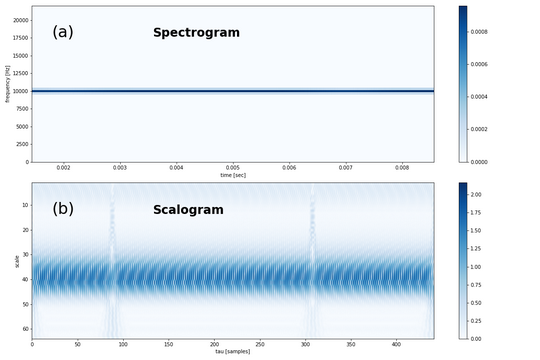}
    \caption{Continous 10 KHz sine wave sampled over some 500 turns with two artificial irregularities created at sample 88 and 308. (a): Spectrogram, (b): Scalogram using a Morlet wavelet}
    \label{fig:TFD-49}
 \end{center} 
\end{figure}

The graphical two dimensional representation of a wavelet-analysis is called "scalogram".
In a scalogram the correlation coefficients of a wavelet-analysis are plotted in a color-code as a two dimensional array with the displacement parameter $\tau$ on the horizontal axis and the scale parameter $s$ on the vertical axis. Some authors convert the scale parameters into "equivalent frequencies", but one should be careful with this:
A scalogram is not yet another way of making a spectrogram, it is a different way of analysing time domain data. The direct physical interpretation of a scalogram is less obvious than for a spectrogram, but its main advantage is the capability to detect irregularities in a data set.
As example \Fref{fig:TFD-49} shows the spectrogram and the scalogram computed for the same input data set. The input is a pure 10 kHz sine wave sampled at 44.1 kHz for some 500 samples. Artificially two samples (88 and 308) are set to zero.
In the spectrogram we clearly see the expected straight line, which indicates the constant 10 KHz sine wave. The spectrogram is computed as a 128 samples based FFT, and for each computation of the FFT the data set is advanced by one sample (so called "sliding FFT").
The frequency resolution is therefore $1/64$. The two spurious data points do not show up in the result.
The scalogram is computed for 64 discrete scales ranging from 1.0 to 8.0.
The scalogram shows some constant correlation coefficients around scale index 40, which corresponds to a scale parameter $s=3.6$, which in turn for a Morlet wavelet corresponds to a frequency of 10 kHz. Contrary to the spectrogram the scalogram clearly shows the location of the two artificial irregularities in the data set.

\subsection{PLL tracking of a resonance}
\label{sec:TFD-PLLtracking}

\begin{figure}[!ht]
  \begin{center}
    \includegraphics[width=0.8\textwidth]{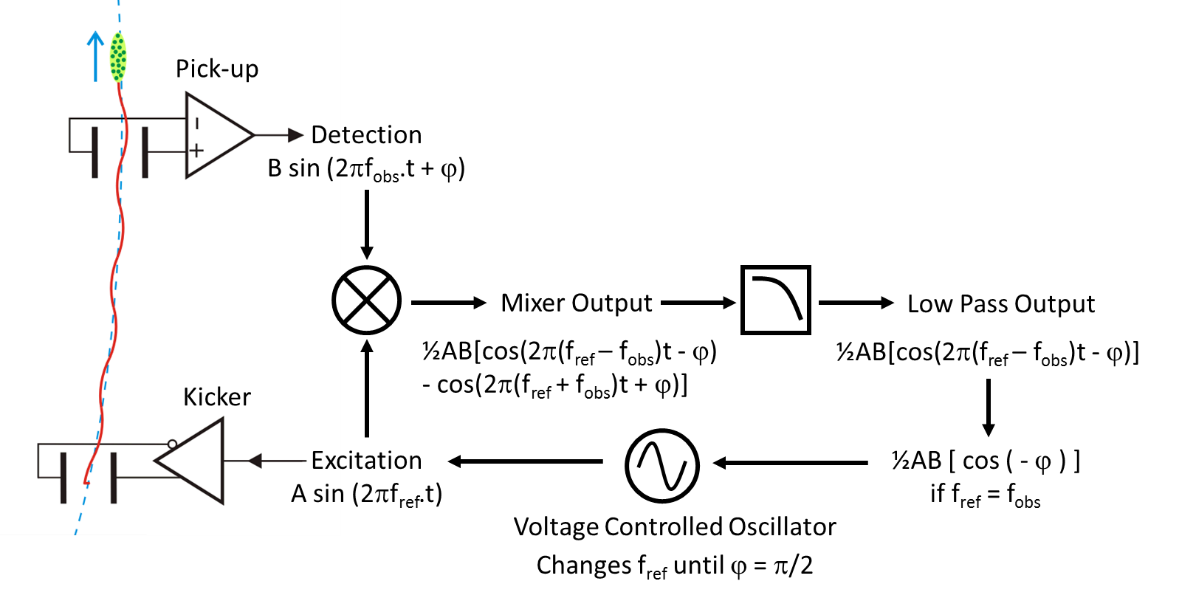}
    \caption{Schematics of a Phase-Locked-Loop Resonance tracker}
    \label{fig:TFD-50}
 \end{center} 
\end{figure}

In many applications of accelerator measurements the spectral information only consists of a single dominant resonance line, notably the betatron tunes in the transverse planes or the synchrotron tune in the longitudinal plane. So if the only required information is how this resonance line changes as a function of time, it is not necessary to compute the full spectral information and to find the resonance frequency from there.
The most direct method is to follow with an external oscillator the frequency of the resonance.
The experimental setup is the same as described for the vector-network analysis in section \Sref{sec:TFD-VNA}. Going back to \Fref{fig:TFD-42} in this section one can see that the phase between the beam exciter signal (continuous sine wave) and the measurement station
changes rapidly by $180^0$ for frequencies just below and above the resonance. This is typical for every resonant system. When the excitation frequency is in resonance, the phase will be just $90^0$. If we now want to follow changes of the resonance of the beam, it will be sufficient to vary the frequency of the exciter by maintaining the excitation phase at $90^0$. Under these conditions the frequency of the exciter is a copy of the resonance frequency of the accelerator.
The electronic circuit, which allows such measurements is shown in \Fref{fig:TFD-50}.
A voltage or numerically controlled oscillator (VCO or NCO) is used to put a harmonic
excitation $A\cdot\sin(\omega t)$ onto the beam. The beam response to this signal is
then observed using a position pick-up. The response is of the form 
$B\cdot\sin(\omega t +\Phi)$, where A and B are the amplitude of the excitation and
oscillation respectively,$\omega$ is the angular excitation frequency and $\Phi$ is the
phase difference between the excitation and the observed beam response. In the phase
detector both signals are multiplied, resulting in a signal of the form
\begin{equation}
\frac{1}{2}\cdot A \cdot B \cdot \cos(-\Phi) - \frac{1}{2}\cdot A \cdot B \cdot 
\cos(2\omega t + \Phi)
\end{equation}
which has a DC component (first term in the equation) proportional to the cosine of the phase difference and a second term varying at twice the excitation frequency. 
The fast varying component can be suppressed with a low pass-filter, hence leaving the DC component, which will be zero when the phase difference between excitation and observation frequencies is 90°. The sign of the component is also correct: When the DC term is negative, the excitation frequency is above the resonance and we need to decrease the excitation frequency. The same principle works for positive values of the DC term, we need to increase the excitation frequency. There fore we can build a closed loop system, where the value of the DC component "drives" the excitation frequency.
Th frequency of the exciter can be examined at any moment, hence a continuous tracking of the resonance is possible. Since this measurement is exclusively based on the phase response, it is usually called "Phase-Locked-Loop (PLL) tracker).

\begin{figure}[!ht]
  \begin{center}
    \includegraphics[width=0.95\textwidth]{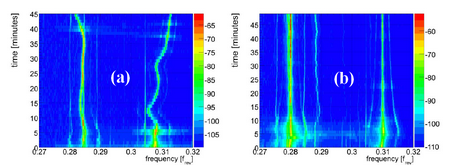}
    \caption{(a): Spectrogram (both planes combined) showing the horizontal and vertical tune during beam acceleration; (b): same measurement with active tune control.}
    \label{fig:TFD-51}
 \end{center} 
\end{figure}

It should be mentioned that \Fref{fig:TFD-50} only shows the principle arrangement of the PLL-circuit. In a real implementation there are additional circuits, which compensate for the phase advance between the beam exciter and the measurement station, which indicate for example if the PLL is locked to the resonance and which control the excitation strength.
The control of the excitation strength is in particular important for accelerators with particle without transverse damping (hadrons). If the beam is stored for longer periods (colliders), any continuous strong excitation will increase the emittance of the beam in a non tolerable way. For leptons this is much less critical, since the radiation damping will immediately damp the effect of the excitation.
A Pll tracker is also very vulnerable in coupling between oscillation planes, or in general if there is more than one resonance within the beam spectrum.
Last not least by running simultaneous PLL tracks to the horizontal and vertical plane for real time tune measurements, a feedback signal to the quadrupole strengths can be generated for on-line control of the betatron tunes. More details on this subject can be found in\cite{bib:TFD-tunecontrol1}.
\Fref[b]{fig:TFD-51} shows measurements made with a PLL tune tracker in the LHC during beam acceleration. Using the technique of a spectrogram as described in the previous section the transverse spectra (horizontal and vertical signals combined) are monitored, whilst the frequencies of the main resonances are tracked with two PLL circuits (red-green high amplitude traces in \Fref{fig:TFD-51}. The same experiment has been repeated with  a real time control loop trimming the strength of the quadrupoles (\Fref{fig:TFD-51}b). One can clearly see that the tunes remain almost constant during this critical machine operation \cite{bib:TFD-tunecontrol2}.
\subsection{Mixed-BPM Sampling in Time and Space}
\label{sec:TFD-TimeandSpaceSampling}

The last section of this write-up addresses a very recent development for a problem in accelerator diagnostics, which is rather frequent: the phenomenon (beam oscillation)  that needs to be observed only lasts for a few machine turns, but the frequency of this oscillation needs to measured quite accurately. Since the beam only passes once per turn through one observation point, only a limited amount of data is available.
All methods described in \Sref{sec:TFD-resolution} can of course be applied, but can we not be smarter?
\begin{figure}[!ht]
  \begin{center}
    \includegraphics[width=0.7\textwidth]{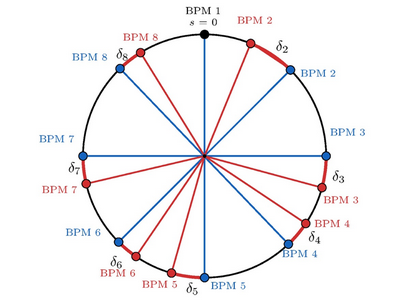}
    \caption{Sketch explaining "Mixed-BPM" sampling in time and space for an example of an accelerator with eight BPM sensors. (taken from \cite{bib:TFD-multiBPM}}
    \label{fig:TFD-52}
 \end{center} 
\end{figure}

Yes, we can: In most circular accelerators a large number $(M)$ of position measurement sensors are implemented, but they are distributed around the ring. The idea is now to acquire data from all sensors synchronized in time over $N$ turns and produce from this a combined data set, which looks like the time-series of one sensor over $M\cdot N$ turns.
The principal problem with this method is that
\begin{itemize}
\item{
the monitors are not positioned in an exactly regular way around the accelerator
}
\item{
the betatron phase advance from one monitor to the other has to be known, if the data has to be recombined into a time series, which mimics equidistant sampling in time.
So the precise knowledge of the accelerator optics functions enters into the recombination process
}
\end{itemize}
\begin{figure}[!ht]
  \begin{center}
    \includegraphics[width=0.7\textwidth]{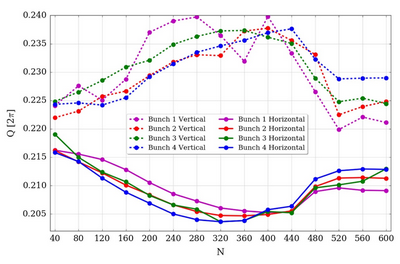}
    \caption{Betatron tunes measured to $10^{-3}$ during the CERN-PS beam injection process (500 machine turns $N$) every 40 turns using the Mixed-BPM method.
    Solid lines indicate the horizontal tunes, dashed lines the vertical tunes of all 4 injected bunches. (taken from \cite{bib:TFD-multiBPM}}
    \label{fig:TFD-53}
 \end{center} 
\end{figure}

In \Fref{fig:TFD-52} the principle of the so called "Mixed-BPM sampling in time and space"
is sketched. In a hypothetical ring the eight installed BPMs at different longitudinal positions are marked with red circles. From the measurements at the eight "red" positions eight new positions have to be calculated as if the BPMs very at equidistant positions.
Those new positions are marked with blue circles. One BPM (BPM1) is set as reference point in black. The seven correction terms $\delta_i$ have to be expressed in betatron phase advance and not in physical distance.
This method of Mixed-BPM analysis has been recently developed and many details can be found in \cite{bib:TFD-multiBPM}. In summary the application of this method gives a net improvement of the frequency resolution based on measurements over only few turns and most errors due to unknown values of the accelerator optics cancel out in the final measurement. Care has to be taken when the tune approaches integer values, in that case errors do not cancel out easily.

An interesting study performed with this method described in the following:
During beam injection into the CERN PS (PS injector synchrotron) a rather large injection bump needs to be applied over 500 turns, which induces a significant change in both betatron tunes. It is important to study the temporal evolution of the betatron tune changes, but a measurement over 500 turns is barely enough to make a single precise tune measurement.
So by applying the Mixed-BPM analysis a window of 40 turns could be moved over the  injection time of 500 tuns and within each window the tune could be measured with a accuracy better than $10^{-3}$. The results of one measurement are shown in
\Fref{fig:TFD-53}. Measurements of this type have been used to adjust the time evolution and the overall strength of the injection bump in order to keep the tune excursions within reasonable limits.

\section{Acknowledgements}

In preparation of the lecture and the write-up I was very grateful to be allowed to reuse lots of visual information from my colleagues. In particular I would like to thank Marco Lonza (INFN) for letting me use more than one time his video animations of multi-bunch modes and all other related graphics. Furthermore I have reused valuable material from Rhodri Jones (CERN), Marek Gasior (CERN), Thibaut Lefevre (CERN) and Heiko Damerau (CERN).

\end{document}